\documentclass[preprint,aps,12pt,showpacs,nofootinbib,tightenlines]{revtex4}
\usepackage{amsmath}
\usepackage{graphicx}
\usepackage{amssymb}

\begin{document}

\title{Consistency and Advantage of Loop Regularization Method Merging with
Bjorken-Drell's Analogy \\
Between Feynman Diagrams and Electrical Circuits}
\author{Da Huang and Yue-Liang Wu}
\email{ylwu@itp.ac.cn}
\affiliation{State Key Laboratory of Theoretical Physics (SKLTP)\\
Kavli Institute for Theoretical Physics China (KITPC) \\
Institute of Theoretical Physics, Chinese Academy of Science,
Beijing,100190, P.R.China}
\date{\today}

\begin{abstract}
The consistency of loop regularization (LORE) method is explored in
multiloop calculations. A key concept of the LORE method is the introduction
of irreducible loop integrals (ILIs) which are evaluated from the Feynman
diagrams by adopting the Feynman parametrization and
ultraviolet-divergence-preserving(UVDP) parametrization. It is then
inevitable for the ILIs to encounter the divergences in the UVDP parameter
space due to the generic overlapping divergences in the 4-dimensional
momentum space. By computing the so-called $\alpha\beta\gamma$ integrals
arising from two loop Feynman diagrams, we show how to deal with the
divergences in the parameter space with the LORE method. By identifying the
divergences in the UVDP parameter space to those in the subdiagrams, we
arrive at the Bjorken-Drell's analogy between Feynman diagrams and
electrical circuits. The UVDP parameters are shown to correspond to the
conductance or resistance in the electrical circuits, and the divergence in
Feynman diagrams is ascribed to the infinite conductance or zero resistance.
In particular, the sets of conditions required to eliminate the overlapping
momentum integrals for obtaining the ILIs are found to be associated with
the conservations of electric voltages, and the momentum conservations
correspond to the conservations of electrical currents, which are known as
the Kirchhoff's laws in the electrical circuits analogy. As a practical
application, we carry out a detailed calculation for one-loop and two-loop
Feynman diagrams in the massive scalar $\phi^4$ theory, which enables us to
obtain the well-known logarithmic running of the coupling constant and the
consistent power-law running of the scalar mass at two loop level.
Especially, we present an explicit demonstration on the general procedure of
applying the LORE method to the multiloop calculations of Feynman diagrams
when merging with the advantage of Bjorken-Drell's circuit analogy.
\end{abstract}

\maketitle

\hfill


\section{ Introduction}

Quantum Field Theory(QFT) is the most successful theory for understanding
the microscopic world in elementary particle physics, nuclear physics and
condensed matter physics. However, when carrying out any calculation beyond
tree level in the framework of perturbation treatment of QFT, one would
encounter the infinities in Feynman integrals, coming from contribution from
large momenta and are usually called ultraviolet (UV) divergences. Thus, the
QFT becomes well-defined only when it can be regularized and renormalized
properly.

Nevertheless, the widely-used regularization methods are known to have some
limitations. For instance, the Pauli-Vallars regularization method is very
useful in the calculation of quantum electrodynamics(QED), but it fails in
non-Abelian gauge theory as it explicitly destroys the non-Abelian gauge
invariance. The dimensional regularization can preserve the gauge symmetry
explicitly and has been useful in the computations for gauge theories, such
as QED and QCD of the standard model\cite{'tHooft:1972fi}. Despite its great
success, it has been known\cite{Bonneau:1989xv, 'tHooft:1972fi} that the
spinor matrix $\gamma _{5}$ and chirality cannot in principle be well
defined in the extended dimensions. Also it has trouble in applying directly
to supersymmetric theories which depend dimension of space-time, and
moreover it cannot keep track of the divergence behaviors (quadratic and
above) of original integrals in the Feynman diagrams. So it is not useful in
some calculations in effective field theories and chiral dynamics where we
need to isolating the quadratic divergences for understanding the dynamical
symmetry breaking and restoration.

Thus it is desirable to develop an alternative new regularization scheme
which possesses the basic properties: being well defined in 4-dimensional
space-time, preserving the gauge symmetry and Lorentz symmetry, keeping
track of the divergent behaviors of original theories, making the pratical
calculations as simple as possible and applicable to both underlying and
effective QFTs as well as supersymmetric and chiral QFTs.

Recently, a new regularization method proposed by one of us\cite%
{wu1,Wu:2003dd} can satisfy all of the properties mentioned above and
checked carefully with explicit calculations for many applications at
one-loop level. For convenience, such new regularization is called the Loop
Regularization since its prescription acts on the so-called irreducible loop
integrals(ILIs)\cite{wu1,Wu:2003dd}. For short, here we may use `LORE' as an
abbreviation of the loop regularization. It has been proved with explicit
calculations at one loop level that the LORE method can preserve non-Abelian
gauge symmetry~\cite{Cui:2008uv} and supersymmetry~\cite{Cui:2008bk}. It can
provide a consistent calculation for the chiral anomaly\cite{Ma:2005md} and
the radiatively induced Lorentz and CPT-violating Chern-Simons term in QED%
\cite{Ma:2006yc} as well as the QED trace anomaly\cite{cui:2011}. It allows
us to derive the dynamically generated spontaneous chiral symmetry breaking
of the low energy QCD for understanding the origin of dynamical quark masses
and the mass spectra of light scalar and pseudoscalar mesons in a chiral
effective field theory\cite{DW}, and also to investigate the chiral symmetry
restoration in a chiral thermodynamic model\cite{HW}. In particular, it
enables us to consistently carry out the quantum gravitational contributions
to gauge theories with asymptotic free power-law running\cite%
{Tang:2008ah,Tang:2010cr,Tang2011}.

It has been analyzed in ref.\cite{wu1} that the LORE method can
straightforwardly be generalized to higher loop calculations with an
explicit demonstration on the general two loop integrals, i.e., the
so-called $\alpha \beta \gamma $ integrals. In fact, our general proof for
the consistency of loop regularization via the $\alpha \beta \gamma $
integrals was just following the same procedure which was adopted by 't
Hooft and Veltman\cite{'tHooft:1972fi} to demonstrate the consistency of
dimensional regularization. Since the LORE method has been realized in four
dimensional space-time without modifying the original theory, its
consistency cannot be proved in the Lagrangian formalism to all orders, thus
it is useful to develop a diagrammatic approach to make such a general
proof. For that, we shall make explicit multiloop calculations to show its
consistency, figuring out a general procedure for practical calculations,
which is a further motivation in our present work. We are going to show in
present paper that the evaluation of the irreducible loop integrals (ILIs)
from Feynman integrals by adopting the ultraviolet divergence-preserving
(UVDP) parametrization naturally leads to the Bjorken-Drell's circuit
analogy between Feynman diagrams and electric circuits. As a consequence,
when merging the LORE method with the Bjorken-Drell's circuit analogy, we
arrive at the interesting observation that there is the one-to-one
correspondence between the divergences of the UVDP parameters and the
subdiagrams of Feynman diagrams, which enables us to extend the procedure to
higher loop Feynman diagrams in a more general and systematic way.

The key concept in the LORE method is the introduction of the ILIs which are
obtained from the Feynman diagrams by using the Feynman parametrization and
the UVDP parametrization. A crucial point in the LORE method is the presence
of two energy scales. They are introduced via the string-mode regulators in
the regularization prescription acting on the ILIs. It has been shown that
the two energy scales play the roles of the ultraviolet (UV) cut-off and
infrared (IR) cut-off to avoid infinities without spoiling symmetries in the
original theory\cite{wu1,Wu:2003dd}. It is then inevitable to encounter the
UV divergence in the UVDP parameter space due to the generic overlapping
divergences. To be more explicit, we carry out a calculation for the general
integrals, the so-called $\alpha \beta \gamma $ integrals, arising from two
loop Feynman diagrams, and show how to deal with the divergences in the UVDP
parameter space by applying the LORE method. By identifying the divergences
in the UVDP parameter space with those in the subdiagrams, we naturally
arrive at the Bjorken-Drell's analogy between Feynman diagrams and electric
circuits, where the UVDP parameters are found to be associated with the
conductance or resistance in electric circuits. A detailed description on
circuit analogy is given in the book by Biorken and Drell\cite{bjor & drel}
in which the circuit analogy was originally inroduced to study the
analyticity properties of Feynman diagrams from the causality requirement.
In our present paper, we observe that the sets of conditions required to
eliminate the overlapping divergent momentum integrals for evaluating the
ILIs is analogous to the conservations of electric voltages in the loop, and
the momentum conservations to the conservations of electric currents at each
vertex. These equations are known as the Kirchhoff's laws in electric
circuits. In particular, it is noticed that the divergence in Feynman
diagrams corresponds to an infinite conductance or zero resistance in
electric circuits. By adopting such an analogy, we perform a detailed
calculation for one- and two-loop Feynman diagrams in the massive scalar $%
\phi ^{4}$ theory, and meanwhile we explicitly demonstrate the general
procedure for applying the LORE method to multiloop calculations of Feynman
diagrams.

We would like to emphasize that our motivation is not just for figuring out
a much simpler regularization scheme, but for finding out whether there
exists in principle a regularization scheme which can overcome some
shortages and limitations in the widely-used regularization schemes.
Meanwhile, we expect that such a regularization scheme must also be
practical and as simple as possible. In fact, for the one loop calculation,
the LORE method is really simple. For the higher-loop calculations, the
procedure and calculation in the LORE method are not as concise as the ones
in the dimensional regularization, but our treatment here makes the
overlapping divergent structure as well as its divergent behavior physically
manifest. To be more precise, the divergence structure for a diagram
includes the overall quadratic or logarithmic divergence and the divergences
in the subdiagrams, as well as corresponding subtraction diagrams. Actually,
the simplicity of the dimensional regularization is at the cost of three
essential limitations: (i) the definition of $\gamma _{5}$ in theories
beyond 4 dimension, (ii) the requirement of exact dimension of original
theories, (iii) preservation of quadratic divergence in original theories.
To overcome such limitations in the dimensional regularization is our main
purpose to look for a possible alternative consistent regularization scheme,
which will be helpful for understanding deeply the applicability and
consistency of QFTs. In this sense, the LORE method has been a step forward,
as already shown in \cite{Cui:2008uv, Cui:2008bk, Ma:2005md, Ma:2006yc,
cui:2011, DW,HW, Tang:2008ah, Tang:2010cr,Tang2011} at one loop level. It is
worthwhile to go further and make an explicit check at higher loop level,
which is our main goal in the present paper. Of course, for QFTs without
three limitations in principle mentioned, the dimensional regularization
scheme remains a powerful and simple one for a practical calculation.

The paper is organized as follows. In Sec. II, we briefly outline the LORE
method and the concept of ILIs at one loop level. In Sec. III, a particular
contribution of two-loop vacuum polarization diagram in QED is examined and
show the general structure of overlapping divergences. It is then
unavoidable to encounter the UV divergences hidden in the UVDP parameter
space. In Sec. IV, we apply the LORE method to the general $\alpha \beta
\gamma $ integrals with $\alpha =\gamma =1$, $\beta =2$, and explicitly show
how the LORE method can appropriately regularize the UV divergences either
from the original loop momenta or from the UVDP parameters. In Sec. V, we
show how the evaluation of ILIs and UVDP parametrization naturally merges
with the Bjorken-Drell's electric circuits analogy. In Sec. VI, The
Bjorken-Drell's electric circuit analogy of Feynman diagrams allows
us to analyze the origin of UV divergences contained in the UVDP parameter
space, and to figure out the one-to-one correspondence of divergences
between subdiagrams and UVDP parameters. In particular, the divergences in
Feynman diagrams is shown to correspond to infinite conductances or zero
resistances in electric circuits analogy. In Sec. VII, the LORE method combining
with the Bjorken-Drell's analogy shows the advantage in analyzing a
complicated overlapping divergence structure of Feynman diagrams. As an
explicit illustration, the case with $\alpha =\beta =\gamma =1$ of the
general $\alpha \beta \gamma $ integral is discussed in detail and all the
harmful divergences cancel exactly. As a practical application of all the
machinery privously introduced, we carry out in Sec. VIII a detailed
calculation of two loop contributions in the massive scalar $\phi ^{4}$
theory. Some additional quadratic corrections to the scalar mass are
obtained, and leads to a power-law running. Based on the general analysis
and explicit calculations, we arrive at in Sec. IX the general procedure of
applying the LORE method to high-loop calculations. Our conclusions and
remarks are presented in Sec. X.

\section{Concept of ILIs and Brief Outline on the LORE Method}

We start from the fact that all Feynamn integrals from the one-particle
irreducible (1PI) graphs in 1-loop can be written, by using Feynman
parametrization, in terms of the following sets of loop integrals,
\begin{equation*}
I_{-2\alpha }=\int \frac{d^{4}k}{(2\pi )^{4}}\ \frac{1}{(k^{2}-\mathcal{M}%
^{2})^{2+\alpha }}\ ,\qquad \alpha =-1,0,1,2,\cdots ,
\end{equation*}%
for scalar type integrals and
\begin{eqnarray}
&&I_{-2\alpha \ \mu \nu }=\int \frac{d^{4}k}{(2\pi )^{4}}\ \frac{k_{\mu
}k_{\nu }}{(k^{2}-\mathcal{M}^{2})^{3+\alpha }}\ ,  \notag \\
&&I_{-2\alpha \ \mu \nu \rho \sigma }=\int \frac{d^{4}k}{(2\pi )^{4}}\ \frac{%
k_{\mu }k_{\nu }k_{\rho }k_{\sigma }}{(k^{2}-\mathcal{M}^{2})^{4+\alpha }}\
,\qquad \alpha =-1,0,1,2,\cdots ,
\end{eqnarray}%
for tensor type integrals, where the number ($-2\alpha $) in the subscript
labels the dimension of power counting of energy momentum in the integrals.
Thus two special cases $\alpha =-1$ and $\alpha =0$ correspond to the
quadratic divergent integrals ($I_{2}$, $I_{2\mu \nu \cdots }$) and the
logarithmic divergent integrals ($I_{0}$, $I_{0\mu \nu \cdots }$). Note that
the mass factor $\mathcal{M}^{2}$ is in general a function of the Feynman
parameters and the external momenta $p_{i}$, namely $\mathcal{M}^{2}=%
\mathcal{M}^{2}(m_{1}^{2},p_{1}^{2},\cdots )$.

The above loop integrals are the so-called one-fold irreducible loop
integrals (ILIs)\cite{wu1}, which can be generalized to the n-fold ILIs
evaluated from n-loop overlapping Feynman integrals of loop momenta $k_{i}$ (%
$i=1,2,\cdots n$). In general, the n-fold ILIs are defined as the loop
integrals in which the overlapping momentum factor $(k_{i}-k_{j}+p_{ij})^{2}$
$(i\neq j)$ originally appearing in the overlapping Feynman integrals has
already been eliminated. It has been shown that any loop integrals can be
evaluated into the corresponding ILIs by using both the Feynman
parametrization and the UVDP parametrization methods\cite{wu1}. Note that in
the procedure of evaluating the ILIs, the algebraic computing for multi-$%
\gamma $ matrices involving loop momentum $k\hspace{-0.17cm}\slash$ such as $%
k\hspace{-0.17cm}\slash\gamma _{\mu }k\hspace{-0.17cm}\slash$ should be
carried out first and expressed in terms of the independent components: $%
\gamma _{\mu }$, $\sigma _{\mu \nu }$, $\gamma _{5}\gamma _{\mu }$, $\gamma
_{5}$.

The concept of ILIs is crucial in the LORE method. To see that, let us
briefly examine the vacuum polarization in the non-abelian gauge theory. We
begin with the following lagrangian in $R_{\xi }$ gauge,
\begin{equation}
\mathcal{L}=\bar{\psi}_{n}(i\gamma ^{\mu }D_{\mu }-m)\psi _{n}-\frac{1}{4}%
F_{\mu \nu }^{a}F_{a}^{\mu \nu }-\frac{1}{2\xi }(\partial ^{\mu }A_{\mu
}^{a})^{2}+\partial ^{\mu }\eta _{a}^{\ast }D_{\mu }\eta ^{a},
\end{equation}%
with
\begin{eqnarray}
&&F_{\mu \nu }^{a}=\partial _{\mu }A_{\nu }^{a}-\partial _{\nu }A_{\mu
}^{a}-gf_{abc}A_{\mu }^{b}A_{\nu }^{c} \nonumber \\
&&D_{\mu }\psi _{n}=(\partial _{\mu }+igT^{a}A_{\mu }^{a})\psi _{n},
\end{eqnarray}%
where $\xi $ is a gauge parameter. $\psi _{n}$, $A_{\mu }$ and $\eta $ are
fermions, gauge bosons and ghost fields, respectively. $T^{a}$ are the
generators of gauge group and $f_{abc}$ the structure constants of the gauge
group with $[T^{a},\ T^{b}]=if_{abc}T^{c}$. The vacuum polarization
corresponds to the self-energy diagrams of gauge boson, which contains the
quadratically divergent integrals, the most divergent behavior in all of the
Green functions in one-loop. Here we give the final results carried out by
using the usual Feynman rules in the general $\xi $ gauge. The details of
the calculation can be found in ref.\cite{wu1}. The explicit expressions for
the gauge boson self-energy diagrams are given, in terms of the ILIs, as
follows:
\begin{eqnarray*}
\Pi _{\mu \nu }^{(f)ab} &=&-g^{2}4N_{f}C_{2}\delta _{ab}\ \int_{0}^{1}dx\ [\
(2a_{2}-1)I_{2}(m)g_{\mu \nu } \\
&&+2x(1-x)(p^{2}g_{\mu \nu }-p_{\mu }p_{\nu })I_{0}(m)\ ],
\end{eqnarray*}%
for the fermion loop contribution to the gauge self-energy diagram, and
\begin{eqnarray}
&&\Pi _{\mu \nu }^{(g)ab}=g^{2}C_{1}\delta _{ab}(p^{2}g_{\mu \nu }-p_{\mu
}p_{\nu })\ \int_{0}^{1}dx\ \{\ [1+4x(1-x)]\ I_{0}\   \notag \\
&&+\frac{1}{2}\lambda \ [\ \left( \ 1+6x(1-x)(a_{0}+2)-3a_{0}\right) I_{0}\
-2x(1-x)\left( \ 1+12x(1-x)\ \right) p^{2}\ I_{-2}\ ]  \notag \\
&&+\frac{3}{4}\lambda ^{2}\ a_{-2}\ x(1-x)\ p^{2}\ I_{-2}\ \}  \notag \\
&&+g^{2}C_{1}\delta _{ab}\ \int_{0}^{1}dx\ \{\ 2(\ 2a_{2}-1\ )I_{2}\ g_{\mu
\nu }+\lambda (a_{0}-1)\ p_{\mu }p_{\nu }\ x(1-x)\ p^{2}\ I_{-2}\ \},
\end{eqnarray}%
for the gauge boson and ghost loop contributions to the gauge self-energy
diagram, where $p$ is the momentum of the external gauge boson, $N_{f}$ the
number of fermions flavors, $\lambda =1-\xi $, $f_{acd}f_{bcd}=C_{1}\delta
_{ab}$ and $tr\left( T^{a}T^{b}\right) =C_{2}\delta _{ab}$. We have also
used the following definitions from the general Lorentz decomposition
\begin{equation} \label{relations}
I_{2\mu \nu }=a_{2}\ I_{2}\ g_{\mu \nu },\qquad I_{0\mu \nu }=\frac{1}{4}%
a_{0}\ I_{0}\ g_{\mu \nu },\qquad I_{-2\mu \nu }=\frac{1}{4}a_{-2}\ I_{-2}\
g_{\mu \nu },
\end{equation}%
where $I_{-2}$ and $I_{-2\mu \nu }$ are convergent integrals with $%
a_{-2}=2/3 $. Note that $\Pi _{\mu \nu }^{(g)ab}$ depends on the gauge
parameter $\xi $. This is because the Green's functions are gauge dependent
while only the S-matrix elements are gauge independent. However, current
conservation implies that $\Pi _{\mu \nu }^{(f)ab}$ and $\Pi _{\mu \nu
}^{(g)ab}$ have to satisfy the Ward Identities $p^{\mu }\Pi _{\mu \nu
}^{(f)ab}=0$ and $p^{\mu }\Pi _{\mu \nu }^{(g)ab}=0$. Notice that the first
line of $\Pi _{\mu \nu }^{(f)ab}$ and in the last line of $\Pi _{\mu \nu
}^{(g)ab}$ contain both quadratically and logarithmically divergent
integrals which might violate gauge invariance. Only with the following
consistency conditions
\begin{equation}
a_{2}=1/2,\qquad a_{0}=1  \label{consistency}
\end{equation}%
then the gauge invariance can be preserved.

Nevertheless, from the naive analysis of Lorentz decomposition and tensor
manipulation, one gets by multiplying $g^{\mu \nu }$ on both sides of Eq.(%
\ref{relations}),
\begin{eqnarray}
&&g^{\mu \nu }I_{2\mu \nu }=I_{2}+\mathcal{M}^{2}I_{0}=4a_{2}I_{2},\quad %
\mbox{i.e.}\quad a_{2}=1/4+\mathcal{M}^{2}I_{0}/I_{2},  \notag \\
&&g^{\mu \nu }I_{0\mu \nu }=I_{0}+\mathcal{M}^{2}I_{-2}=a_{0}\ I_{0},\quad %
\mbox{i.e.}\quad a_{0}=1+\mathcal{M}^{2}I_{-2}/I_{0},  \notag
\end{eqnarray}%
which leads, without using any regularization schemes, to the following
relations
\begin{eqnarray}
&&I_{2\mu \nu }=\frac{1}{4}g_{\mu \nu }\ I_{2}+\frac{1}{4}g_{\mu \nu }%
\mathcal{M}^{2}\ I_{0},  \notag  \label{NR} \\
&&I_{0\mu \nu }=\frac{1}{4}g_{\mu \nu }\ I_{0}+g_{\mu \nu }\mathcal{M}%
^{2}I_{-2}=\frac{1}{4}g_{\mu \nu }\ I_{0}-\frac{i}{32\pi ^{2}}g_{\mu \nu }.
\end{eqnarray}%
Clearly, the above naive relations for the divergent ILIs will destroy the
gauge invariance. The reason is that in the divergent integrals which are
generally not mathematically well defined without using proper
regularization scheme, the tensor manipulation and integration do not
commute with each other, so the result for divergent integration is not
consistent in general. Thus in order to obtain a consistent result, one has
to adopt a regularization scheme to make the divergent integrals
well-defined. To see this, consider the time-time component on both sides of
the relation for the quadratic divergent ILIs in Eq.(\ref{relations})
\begin{equation}
I_{2\ 00}=a_{2}\ I_{2}\ g_{00}.
\end{equation}%
The Wick rotation will turn the four-dimensional energy momentum into
Euclidean space and integrating over the zero component of energy momentum $%
k_{0}$ on both sides, we get
\begin{eqnarray}
I_{2} &=&-i\int \frac{d^{4}k}{(2\pi )^{4}}\ \frac{1}{k^{2}+\mathcal{M}^{2}}%
=-i\int \frac{d^{3}k}{(2\pi )^{4}}\ \int dk_{0}\frac{1}{k_{0}^{2}+\mathbf{{k}%
^{2}+\mathcal{M}^{2}}}  \notag \\
&=&-i\int \frac{d^{3}k}{(2\pi )^{4}}\ 2\frac{1}{\sqrt{\mathbf{{k}^{2}+%
\mathcal{M}^{2}}}}\tan ^{-1}\left( k_{0}/\sqrt{\mathbf{{k}^{2}+\mathcal{M}%
^{2}}}\right) |_{k_{0}=0}^{k_{0}=\infty }  \notag \\
&=&-i\int \frac{d^{3}k}{(2\pi )^{3}}\ \frac{1}{2\sqrt{\mathbf{{k}^{2}+%
\mathcal{M}^{2}}}}
\end{eqnarray}%
for the right-hand side, and
\begin{eqnarray}
I_{2\ 00} &=&-i\int \frac{d^{4}k}{(2\pi )^{4}}\ \frac{k_{0}^{2}}{(k^{2}+%
\mathcal{M}^{2})^{2}}=-i\int \frac{d^{3}k}{(2\pi )^{4}}\ \int dk_{0}\frac{%
k_{0}^{2}}{(k_{0}^{2}+\mathbf{{k}^{2}+\mathcal{M}^{2})^{2}}}  \notag \\
&=&-i\int \frac{d^{3}k}{(2\pi )^{4}}\ \int dk_{0}\left( \frac{1}{k_{0}^{2}+%
\mathbf{{k}^{2}+\mathcal{M}^{2}}}-\frac{\mathbf{{k}^{2}+\mathcal{M}^{2}}}{%
(k_{0}^{2}+\mathbf{{k}^{2}+\mathcal{M}^{2})^{2}}}\right)  \notag \\
&=&-i\int \frac{d^{3}k}{(2\pi )^{4}}\ \int dk_{0}\left( \frac{1}{k_{0}^{2}+%
\mathbf{{k}^{2}+\mathcal{M}^{2}}}-\frac{1}{2}\frac{1}{k_{0}^{2}+\mathbf{{k}%
^{2}+\mathcal{M}^{2}}}\right) -\frac{k_{0}}{k_{0}^{2}+\mathbf{{k}^{2}+%
\mathcal{M}^{2}}}|_{k_{0}=0}^{k_{0}=\infty }  \notag \\
&=&\frac{-i}{2}\int \frac{d^{3}k}{(2\pi )^{4}}\ 2\frac{1}{\sqrt{\mathbf{{k}%
^{2}+\mathcal{M}^{2}}}}\tan ^{-1}\left( k_{0}/\sqrt{\mathbf{{k}^{2}+\mathcal{%
M}^{2}}}\right) |_{k_{0}=0}^{k_{0}=\infty }  \notag \\
&=&\frac{-i}{2}\int \frac{d^{3}k}{(2\pi )^{3}}\ \frac{1}{2\sqrt{\mathbf{{k}%
^{2}+\mathcal{M}^{2}}}}=\frac{1}{2}\ I_{2}\ g_{00}
\end{eqnarray}%
for the left-hand side. Note that the above integration over $k_{0}$ is
convergent, and should be safe for any algebraic manipulation. When
comparing the results with both left and right hand sides, we obtain $%
a_{2}=1/2$ which agrees with the consistency condition for gauge invariance.
We then come to the conclusion that the general relation between the
tensor-type and scalar-type quadratically divergent ILIs with $a_{2}=1/2$
must be the exact consistency condition.

We would like to emphasize that the above demonstration for obtaining the
consistency condition $a_{2}=1/2$ between the quadratically divergent ILIs
has nothing to do with any regularization schemes. Nevertheless, the
drawback here is that it is obtained only for one of the Lorentz components
rather than for the whole covariant Lorentz tensor. Thus it is necessary to
look for a proper regularization scheme which can realize the consistency
condition in a covariant way with the well-defined divergent integrals.
Meanwhile, it should also preserve the original divergent behavior for both
quadratical and logarithmic divergent integrals. Actually, it has explicitly
been proved\cite{wu1} that the LORE method does lead to the consistency
conditions with $a_{2}=1/2$ and $a_{0}=1$. A simple regularization
prescription operating on the ILIs has been realized in four dimensional
spacetime to satisfy the criteria mentioned in the introduction.

The regularization prescription of the LORE method is as follows: Firstly
rotating the momentum to the four dimensional Euclidean space, then
replacing the loop integrating variable $k^2$ and the loop integrating
measure $\int{d^4k}$ of the ILIs by the corresponding regularized ones $%
[k^2]_l$ and $\int[d^4k]_l$:
\begin{eqnarray}
& & \quad k^2 \rightarrow [k^2]_l \equiv k^2+M^2_l\ ,  \notag \\
& & \int{d^4k} \rightarrow \int[d^4k]_l \equiv \lim_{N,
M_l^2}\sum_{l=0}^{N}c_l^N\int{d^4k},
\end{eqnarray}
where $M_l^2$ ($l= 0,1,\ \cdots $) may be regarded as the regulator masses
for the ILIs. The coefficients $c_l^N$ and the regulator masses are chosen
to satisfy the following conditions:
\begin{eqnarray}
\lim_{N, M_l^2}\sum_{l=0}^{N}c_l^N(M_l^2)^n = 0 \quad (n= 0, 1, \cdots),
\label{cl conditions}
\end{eqnarray}
where the notation $\lim_{N, M_l^2}$ denotes the limiting case $\lim_{N,
M_R^2\rightarrow \infty}$. The initial conditions $M_0^2 = 0$ and $c_0^N = 1$
are taken to recover the original integrals in the limit $M_l^2 \to \infty$ (%
$l=1,2,\cdots$ ).

With the above regularization prescription, we have shown that the
regularized 1-fold ILIs satisfy the following consistency conditions\cite%
{wu1}:
\begin{eqnarray}  \label{CC}
& & I_{2\mu\nu}^R = \frac{1}{2} g_{\mu\nu}\ I_2^R, \quad
I_{2\mu\nu\rho\sigma }^R = \frac{1}{8} (g_{\mu\nu}g_{\rho\sigma} +
g_{\mu\rho}g_{\nu\sigma} + g_{\mu\sigma}g_{\rho\nu})\ I_2^R ,  \notag \\
& & I_{0\mu\nu}^R = \frac{1}{4} g_{\mu\nu} \ I_0^R, \quad
I_{0\mu\nu\rho\sigma }^R = \frac{1}{24} (g_{\mu\nu}g_{\rho\sigma} +
g_{\mu\rho}g_{\nu\sigma} + g_{\mu\sigma}g_{\rho\nu})\ I_0^R .
\end{eqnarray}
which are actually the necessary and sufficient conditions to preserve the
gauge symmetry in QFTs. Here the superscript ``R" denotes the regularized
ILIs. Note that the dimensional regularization scheme also leads to $a_2
=1/2 $ for $\mathcal{M}^2 \neq 0$ and $a_0 =1$, while the resulting $I^{R}_2$
is suppressed to be a logarithmic divergence multiplying by the mass scale $%
\mathcal{M}^2$, thus it goes to vanish $I^{R}_2=0$ when $\mathcal{M}^2 =0$.
This is the well-known fact that the dimensional regularization does not
preserve the quadratic divergent behavior of the original loop integrals.

As the simplest solution of Eq. (\ref{cl conditions}), take the string-mode
regulators
\begin{equation}
M_{l}^{2}=\mu _{s}^{2}+lM_{R}^{2},
\end{equation}%
with $l=1,2,\cdots $, then the coefficients $c_{l}^{N}$ are completely
determined to be
\begin{equation}
c_{l}^{N}=(-1)^{l}\frac{N!}{(N-l)!l!},  \label{mus}
\end{equation}%
where $M_{R}$ may be regarded as a basic mass scale of loop regulator. When
applying the above prescription and solution to the ILIs, the regularized
ILIs in the Euclidean space-time are generally expressed as follows:
\begin{eqnarray}
I_{-2\alpha }^{R} &=&i(-1)^{\alpha
}\lim_{N,M_{l}^{2}}\sum_{l=0}^{N}c_{l}^{N}\int \frac{d^{4}k}{(2\pi )^{4}}%
\frac{1}{(k^{2}+M^{2}+M_{l}^{2})^{2+\alpha }},  \notag \\
I_{-2\alpha \ \mu \nu }^{R} &=&-i(-1)^{\alpha
}\lim_{N,M_{l}^{2}}\sum_{l=0}^{N}c_{l}^{N}\int \frac{d^{4}k}{(2\pi )^{4}}%
\frac{k_{\mu }k_{\nu }}{(k^{2}+M^{2}+M_{l}^{2})^{3+\alpha }},\hspace{8mm}%
\alpha =-1,0,1,2,...  \notag \\
I_{-2\alpha \ \mu \nu \rho \sigma }^{R} &=&i(-1)^{\alpha
}\lim_{N,M_{l}^{2}}\sum_{l=0}^{N}c_{l}^{N}\int \frac{d^{4}k}{(2\pi )^{4}}%
\frac{k_{\mu }k_{\nu }k_{\rho }k_{\sigma }}{(k^{2}+M^{2}+M_{l}^{2})^{4+%
\alpha }}
\end{eqnarray}%
For the regularized quadratically and logarithmically divergent ILIs $%
I_{2}^{R}$ and $I_{0}^{R}$, we have shown that they have the following
explicit experssions\cite{wu1}:
\begin{eqnarray}
I_{2}^{R} &=&\frac{-i}{16\pi ^{2}}\{M_{c}^{2}-\mu ^{2}[\ln \frac{M_{c}^{2}}{%
\mu ^{2}}-\gamma _{w}+1+y_{2}(\frac{\mu ^{2}}{M_{c}^{2}})]\}  \notag \\
I_{0}^{R} &=&\frac{i}{16\pi ^{2}}[\ln \frac{M_{c}^{2}}{\mu ^{2}}-\gamma
_{w}+y_{0}(\frac{\mu ^{2}}{M_{c}^{2}})]
\end{eqnarray}%
with $\mu ^{2}=\mu _{s}^{2}+M^{2}$, and
\begin{eqnarray}
&&\gamma _{w}\equiv \lim_{N}\{\ \sum_{l=1}^{N}c_{l}^{N}\ln l+\ln [\
\sum_{l=1}^{N}c_{l}^{N}\ l\ln l\ ]\}=\gamma _{E}=0.5772\cdots ,  \notag
\label{y} \\
&&y_{0}(x)=\int_{0}^{x}d\sigma \frac{1-e^{-\sigma }}{\sigma },\quad y_{1}(x)=%
\frac{e^{-x}-1+x}{x}  \notag \\
&&y_{2}(x)=y_{0}(x)-y_{1}(x),\quad \lim_{x\rightarrow 0}y_{i}(x)\rightarrow
0,\ i=0,1,2 \\
&&M_{c}^{2}\equiv \lim_{N,M_{R}}M_{R}^{2}\sum_{l=1}^{N}c_{l}^{N}(l\ln
l)=\lim_{N,M_{R}}M_{R}^{2}/\ln N  \notag
\end{eqnarray}%
which indicates that the $\mu _{s}$ sets an IR `cutoff' at $M^{2}=0$ and $%
M_{c}$ provides an UV `cutoff'. For renormalizable QFTs, $M_{c}$ can be
taken to be infinity $(M_{c}\rightarrow \infty )$. In a theory without
infrared divergence, $\mu _{s}$ can safely be taken as $\mu _{s}=0$. In
fact, by taking $M_{c}\rightarrow \infty $ and $\mu _{s}=0$, we recover the
initial integral of the theory. Also by taking $M_{R}$ and $N$ to infinity,
we arrive at a regularized theory which becomes independent of the
regularization prescription. Note that the function $y_{0}(x)$ with $x=\mu
^{2}/M_{c}^{2}$ is actually the incomplete gamma function, which has the
property: $y_{0}(x)\rightarrow 0$ at $x\rightarrow 0$ (i.e., in the limit $%
M_{c}\rightarrow \infty $). In comparison with the dimensional
regularization, there is a correspondence: $\ln \frac{M_{c}^{2}}{\mu ^{2}}%
\rightarrow \frac{2}{\varepsilon }$ with $M_{c}\rightarrow \infty $ and $%
\varepsilon \rightarrow 0$, which indicates that the function $y_{0}(x)$
approaches to zero much faster than the polynomial of $\varepsilon $ in the
dimensional regularization. This can be seen explicitly from the expression:
$y_{0}(x)\simeq x\sim e^{-\frac{2}{\varepsilon }}\rightarrow 0$ in the limit
$M_{c}\rightarrow \infty $ and $\varepsilon \rightarrow 0$.

We would like to point out that the prescription in the LORE method looks
very similar to the Pauli-Villars prescription. Nevertheless, the basic
concept is quite different as the prescription in the LORE method is acting
on the ILIs rather than on the propagators in the Pauli-Villars scheme. This
is why the LORE method can preserve non-Abelian gauge symmetry, while the
Pauli-Villars regularization can not. In this sense, we would like to
emphasize that the concept of ILIs is a crucial point in the LORE method to
realize the interesting symmetry-preserving regularization scheme. In
particular, the introduction of two intrinsic energy scales without spoiling
symmetries of original theory is an advantage in the LORE method to avoid
the infinities of divergent Feynman integrals \cite%
{wu1,Wu:2003dd,Cui:2008uv,Cui:2008bk,Ma:2005md,Ma:2006yc,cui:2011,DW,HW,
Tang:2008ah, Tang:2010cr,Tang2011}. For the effective theories, the
intrinsic UV `cutoff' scale $M_c$ plays the role as the characteristic
energy scale below which the physics can be well described by the effective
quantum field theory\cite{DW,HW, Tang:2008ah, Tang:2010cr,Tang2011}.

\section{Overlapping Divergences and UVDP Parametrization}

It is well-known that for Feynman diagrams beyond one-loop order, a new
feature, \emph{overlapping} divergences. It happens when two divergent loops
share a common propagator. To illustrate this, consider one particular
contribution to the photon vacuum polarization at two-loop order of quantum
electrodynamics (QED) (see Fig. 1)
\begin{figure}[th]
\begin{center}
\includegraphics[scale=0.9]{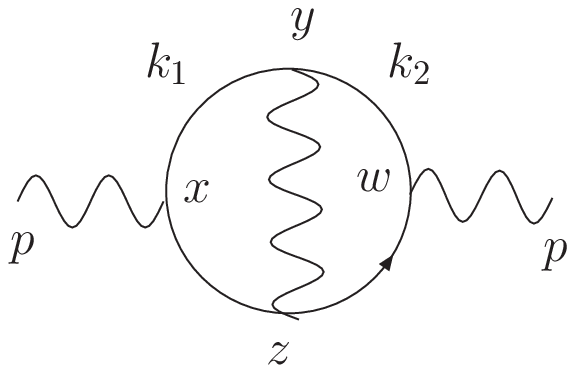}
\end{center}
\caption{{}}
\label{qedB1}
\end{figure}

\begin{figure}[th]
\begin{center}
\includegraphics[scale=0.9]{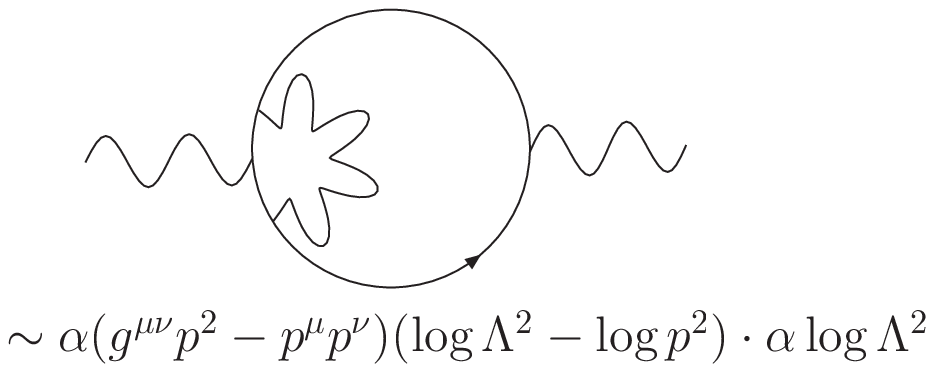}\\[0pt]
\end{center}
\caption{{}}
\label{vertex insertion}
\end{figure}

Here we may follow the argument in the textbook\cite{Peskin:1995ev}. As
discussed in the usual texts of QFTs, the UV divergences in Fig. \ref{qedB1}
can arise from several regions of momentum spaces. One divergent
contribution comes from the region where there is a large momentum passing
through the left subdiagram. This means that the three points $x,y,$ and $z$
in position space are very close together, while the point $w$ is farther
away. In this region one can think that the virtual photon gives the
corrections to the vertex $x$. Plug the divergent part of the one-loop
vertex corrections into the rest of diagram and integrate over $k_{1}$. We
get expression identical to the one-loop photon vacuum polarization
correction multiplied by the additional logarithmic divergence, as shown in
Fig. \ref{vertex insertion}. Obviously, a similar divergent contribution to
the diagram in Fig.\ref{qedB1} arises from the region with a large momentum
passing through the right subdiagram as shown in Fig. \ref{vertex insertion}%
.
It is manifest that the $\log ^{2}\Lambda ^{2}$ term comes from the region
where both $k_{1}$ and $k_{2}$ are large. While the $\log p^{2}\log \Lambda
^{2}$ term results from the region where $k_{2}$ is large but $k_{1}$ is
small, another such a term arises from the region where $k_{1}$ is large but
$k_{2}$ is small. The terms like $\log p^{2}\log \Lambda ^{2}$ are called
\emph{nonlocal} or \emph{harmful} divergences as such terms cannot be
canceled by the ordinary substraction scheme by introducing the
corresponding two loop counterterms in the Lagrangian.

\begin{figure}[ht]
\begin{center}
\includegraphics{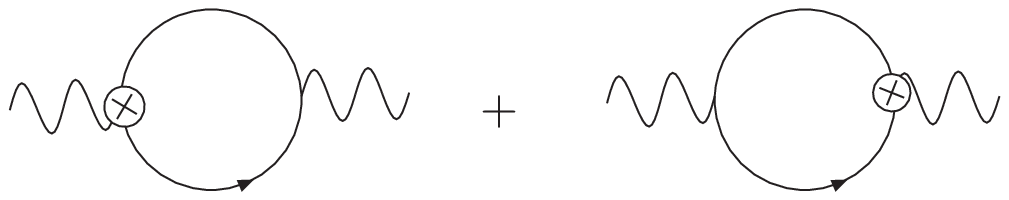}\\[0pt]
\end{center}
\caption{}
\label{counterinsertion}
\end{figure}

\begin{figure}[ht]
\begin{center}
\includegraphics[scale=0.8]{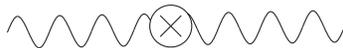}
\end{center}
\caption{local counter term}
\label{counterlocal}
\end{figure}

It is then expected that these harmful divergences are canceled by two types
of counterterm diagrams. First, we can build diagrams of order $\alpha^2$ by
inserting the order-$\alpha$ counterterm vertex into the one-loop vacuum
polarization diagram (see Fig. \ref{counterinsertion}).
Such two diagrams should cancel the harmful divergences as shown in Fig.\ref%
{vertex insertion}. Once these counterterm diagrams are added, the only
divergence left is exactly local and can be canceled by the two-loop overall
counterterm, which is diagrammatically represented in Fig. \ref{counterlocal}%
.

The discussion given above is a general description in the ordinary
textbooks and there is no problem in principle. However, in the practical
calculation, one actually meets some conceptional problems. There is no
doubt that we have to integrate over two loop momentums $k_{1}$ and $k_{2}$
one by one. Suppose that we first integrate over the loop momentum $k_{1}$,
which means that we integrate over the left subdiagram corresponding to the
left vertex insertion. Then we integrate over the loop momentum $k_{2}$,
which corresponds to the overall divergence of the whole diagram as its
divergent behavior is easily found to be quadratic from the simple power
counting. But the problem arises when we look into which loop momentum
integral represents the right subdiagram corresponding correction to the
right vertex. This is because we have already integrated over both loop
momenta in the diagram with the above procedure. It appears that we have
nothing to do with it. Actually, when carrying out the calculations by using
the Feynman parametrization and UVDP parametrization to combine the momenta
in the denominator, we will find that, besides of the divergences coming
from the integral of the two loop momenta $k_{1}$ and $k_{2}$, the
integrations over the UVDP parameters are also logarithmically divergent,
which exactly reproduce the divergence behavior of the vertex correction at
one-loop order. This observation makes it clear that the integration of
right subdiagram is \textquotedblleft hidden" in or transformed into the
parameter space with the usual procedure of dealing with the two-loop
overlapping diagrams.

Thus the next immediate question is whether, given a divergence in the UVDP
parameter space, we can find out the origin of this divergence in the
original Feynman diagrams. Our answer is positive. This is actually the main
purpose in our present paper. We shall show that there is an exact
correspondence between the UVDP parameter integrals and those from the
original loop momenta. The key conceptual tool for arriving at this
conclusion is the observation of the Bjorken-Drell's analogy between the
Feynman diagrams and electrical circuits, which will be demonstrated below.

Before proceeding, it is interesting to note that all the overlapping
divergent integrals (including scalar-type and tensor-type) of two-loop
Feynman diagrams in QED can be reduced to the following two types of
integrals by adopting the Feynman parametrization:
\begin{eqnarray}
I_{111} &=&\int \frac{d^{4}k_{1}}{(2\pi )^{4}}\int \frac{d^{4}k_{2}}{(2\pi
)^{4}}\frac{1}{%
(k_{1}^{2}-m_{1}^{2})(k_{2}^{2}-m_{2}^{2})[(k_{1}-k_{2}+p)^{2}-m_{3}^{2}]},
\\
I_{121} &=&\int \frac{d^{4}k_{1}}{(2\pi )^{4}}\int \frac{d^{4}k_{2}}{(2\pi
)^{4}}\frac{1}{%
(k_{1}^{2}-m_{1}^{2})(k_{2}^{2}-m_{2}^{2})^{2}[(k_{1}-k_{2}+p)^{2}-m_{3}^{2}]%
},
\end{eqnarray}%
where $m_{i}^{2}$ are in general the functions of the external momenta $p$
and Feynman parameters. Such integrals are actually the two special cases of
the general $\alpha \beta \gamma $ integals\cite{'tHooft:1972fi}. Therefore,
it is useful to make a general discussion and analysis on the regularization
and renormalization for the general $\alpha \beta \gamma $ integrals.

In order to avoid the complication involving the reducible loop integrals
and tensor-type integrals, we may consider only scalar-type ILIs. As pointed
out in Ref. \cite{'tHooft:1972fi} by 't Hooft and Veltman, a general
two-loop order Feynman diagram can be reduced to the general $\alpha \beta
\gamma $ integrals of the form:

\begin{equation}  \label{alpha-beta-gamma exp}
I_{\alpha\beta\gamma}=\int\frac{d^4k_1}{(2\pi)^4}\int\frac{d^4k_2}{(2\pi)^4}
\frac{1}{(k_1^2-m_1^2)^\alpha(k_2^2-m_2^2)^\beta[(-k_1-k_2+p)^2-m_3^2]^\gamma%
} .
\end{equation}

We shall focus on the problem how to disentangle the overlapping divergences
with the LORE method. Especially, we will show how to deal with the
divergences contained in the UVDP parameter space caused by the overlapping
structure. With the advantage of the UVDP parametrization in evaluating the
ILIs from Feynman diagrams, we naturally arrive at the Bjorken-Drell's
analogy between Feynman diagrams and electrical circuit diagrams. This
powerful tool gives us exact one-to-one correspondence of the divergences
between the parameter space and the subdiagrams, which can explicitly be
demonstrated. As a consequence, it straightforwardly leads to the important
theorem on the cancelation of harmful divergences. To realize those goals,
it is enough to keep track of only the overlapping divergences, such as the
terms $M_c^2\cdot\log\frac{M_c^2}{-p^2}$ and $\log \frac{M_c^2}{-p^2}\cdot
\log\frac{M_c^2}{-p^2}$. For the harmless divergences and finite terms, they
can be either absorbed into the two-loop overall counterterms or kept in the
final expression.

\begin{figure}[ht]
\begin{center}
\includegraphics{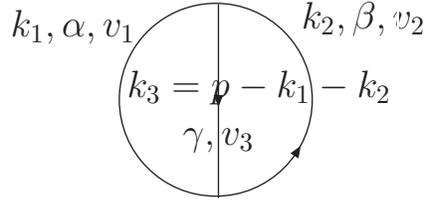}\\[0pt]
\end{center}
\caption{$\protect\alpha\protect\beta\protect\gamma$ diagram}
\label{abcoverall}
\end{figure}

\begin{figure}[ht]
\begin{center}
\includegraphics{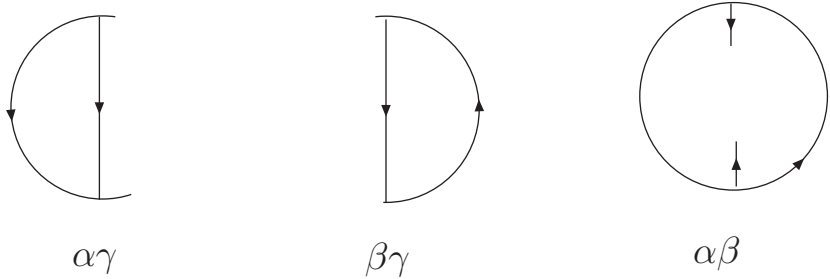}\\[0pt]
\end{center}
\caption{subdivergences}
\label{subdiag}
\end{figure}

\begin{figure}[ht]
\begin{center}
\includegraphics{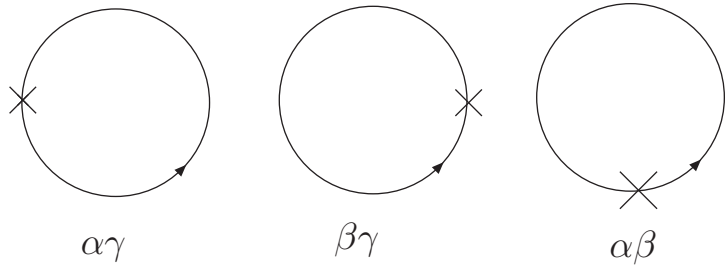}\\[0pt]
\end{center}
\caption{counterterm diagrams}
\label{stdiag}
\end{figure}

From the general form of Eq.(\ref{alpha-beta-gamma exp}), one can easily
recognize that there are in general one overall integral $\alpha\beta\gamma$
and three subintegrals ($\alpha\beta$, $\beta\gamma$ and $\gamma\alpha$),
represented diagrammatically as the following three corresponding
subdiagrams (See Fig.(\ref{abcoverall})and Fig.(\ref{subdiag})). The
corresponding counterterm diagrams are shown in Fig.(\ref{stdiag}), which
are generally needed for the cancelation of the harmful divergences.

By power counting, it is easy to see from Eq.(\ref{alpha-beta-gamma exp})
that there are only two cases which involve the overlapping divergences,
i.e., (1) $\alpha+\beta+\gamma=4$, and (2) $\alpha+\beta+\gamma=3$. Other
cases with $\alpha+\beta+\gamma>4$ contain only harmless divergences and the
overall integral is convergent. Thus we shall only discuss these two cases.

As the first step, we shall write the general $\alpha \beta \gamma $
integral given in Eq.(\ref{alpha-beta-gamma exp}) into the ILIs. With the
standard manipulations, such as combining factors in the denominator with
the UVDP parametrization and making translation of loop momenta, we can then
get rid of the cross terms of momenta in the denominator. Some useful
formula and further discussions on the UVDP parametrization method are given
in Appendix B, which enables us to reexpress Eq.(\ref{alpha-beta-gamma exp})
into the following form
\begin{eqnarray}
I_{\alpha \beta \gamma } &=&\int \frac{d^{4}k_{1}}{(2\pi )^{4}}\int \frac{%
d^{4}k_{2}}{(2\pi )^{4}}\frac{\Gamma (\alpha +\beta +\gamma )}{\Gamma
(\alpha )\Gamma (\beta )\Gamma (\gamma )}\int_{0}^{\infty }\prod_{i=1}^{3}%
\frac{dv_{i}}{(1+v_{i})^{2}}\delta (1-\sum_{i=1}^{3}\frac{1}{1+v_{i}})
\notag \\
&&\frac{\frac{1}{(1+v_{1})^{\alpha -1}}\frac{1}{(1+v_{2})^{\beta -1}}\frac{1%
}{(1+v_{3})^{\gamma -1}}}{\{\frac{1}{1+v_{1}}(k_{1}^{2}-m_{1}^{2})+\frac{1}{%
1+v_{2}}(k_{2}^{2}-m_{2}^{2})+\frac{1}{1+v_{3}}%
[(-k_{1}-k_{2}+p)^{2}-m_{3}^{2}]\}^{\alpha +\beta +\gamma }}  \notag \\
&=&\frac{\Gamma (\alpha +\beta +\gamma )}{\Gamma (\alpha )\Gamma (\beta
)\Gamma (\gamma )}\int_{0}^{\infty }\prod_{i=1}^{3}\frac{dv_{i}}{%
(1+v_{i})^{1+\alpha _{i}}}\delta (1-\sum_{j=1}^{3}\frac{1}{1+v_{j}})\int
\frac{d^{4}k_{1}^{\prime }}{(2\pi )^{4}}\int \frac{d^{4}k_{2}^{\prime }}{%
(2\pi )^{4}}  \notag \\
&&\frac{1}{[(\frac{1}{1+v_{1}}+\frac{1}{1+v_{3}}){k^{\prime }}_{1}^{2}+\frac{%
3+v_{1}+v_{2}+v_{3}}{(2+v_{1}+v_{2})(1+v_{3})}{k^{\prime }}_{2}^{2}+\frac{1}{%
3+v_{1}+v_{2}+v_{3}}p^{2}-\sum_{j=1}^{3}\frac{m_{j}^{2}}{1+v_{j}}]^{\alpha
+\beta +\gamma }},  \notag \\
&&
\end{eqnarray}%
where we have used $\alpha _{i}$ (i=1,2,3) to denote $\alpha $, $\beta $, $%
\gamma $ respectively, and made the following momentum translation
\begin{eqnarray}
{k^{\prime }}_{1} &=&k_{1}+\frac{1+v_{1}}{2+v_{1}+v_{3}}(p-k_{2}), \\
{k^{\prime }}_{2} &=&k_{2}+\frac{1+v_{2}}{3+v_{1}+v_{2}+v_{3}}p.
\end{eqnarray}%
Below we shall drop the prime on ${k^{\prime }}_{i}$ for simplicity. It is
seen that the cross term of momentum is eliminated.

In general, the power indices $\alpha ,\beta ,\gamma $ are positive
integers, so that we have $\alpha +\beta +\gamma \geq 3$. Thus the above
integral is convergent with respect to one of loop momentum $k_{i}$s. From
the general structure of Eq.(\ref{alpha-beta-gamma exp}), it is clear that
the final result is independent of the integration order over $k_{1}$ and $%
k_{2}$. Without lost of generality, we can first integrate over $k_{1}$ and
explicitly obtain the expression:
\begin{eqnarray}
I_{\alpha \beta \gamma } &=&\frac{i}{16\pi ^{2}}\frac{\Gamma (\alpha +\beta
+\gamma -2)}{\Gamma (\alpha )\Gamma (\beta )\Gamma (\gamma )}%
\int_{0}^{\infty }\prod_{i=1}^{3}dv_{i}\delta (1-\sum_{j=1}^{3}\frac{1}{%
1+v_{j}})F(v_{k})  \notag  \label{abc w/o k1} \\
&&\int \frac{d^{4}k_{2}}{(2\pi )^{4}}\frac{1}{[k_{2}^{2}-\mathcal{M}%
^{2}(p^{2},m_{k}^{2},v_{k})]^{\alpha +\beta +\gamma -2}},
\end{eqnarray}%
with
\begin{eqnarray}
\mathcal{M}^{2} &=&-\frac{(2+v_{1}+v_{3})(1+v_{2})}{(3+v_{1}+v_{2}+v_{3})^{2}%
}p^{2}+\frac{(2+v_{1}+v_{3})(1+v_{2})}{3+v_{1}+v_{2}+v_{3}}\sum_{j=1}^{3}%
\frac{m_{j}^{2}}{1+v_{j}}, \\
F(v_{k}) &=&\frac{(2+v_{1}+v_{3})^{\alpha +\beta +\gamma
-4}(1+v_{1})^{1-\alpha }(1+v_{3})^{1-\gamma }(1+v_{2})^{\alpha +\gamma -3}}{%
(3+v_{1}+v_{2}+v_{3})^{\alpha +\beta +\gamma -2}}.
\end{eqnarray}%
The above result is symmetric under the interchange between $v_{1}$ ($m_{1}$,%
$\alpha _{1}=\alpha $) and $v_{3}$($m_{3}$,$\alpha _{3}=\gamma $). In fact,
the original expression is also symmetric under the permutations among $%
v_{1} $ ($m_{1}$,$\alpha _{1}=\alpha $), $v_{2}$($m_{2}$,$\alpha _{2}=\beta $%
) and $v_{3}$($m_{3}$,$\alpha _{3}=\gamma $). Making the following scaling
transformation for the momentum
\begin{equation}
k_{2}^{2}=\frac{(2+v_{1}+v_{3})(1+v_{2})}{3+v_{1}+v_{2}+v_{3}}l_{+}^{2},
\label{scaling}
\end{equation}%
we then obtain the following more symmetric expression
\begin{eqnarray}
I_{\alpha \beta \gamma } &=&\frac{i}{16\pi ^{2}}\frac{\Gamma (\alpha +\beta
+\gamma -2)}{\Gamma (\alpha )\Gamma (\beta )\Gamma (\gamma )}%
\int_{0}^{\infty }\prod_{i=1}^{3}\frac{dv_{i}}{(1+v_{i})^{2}}\delta
(1-\sum_{j=1}^{3}\frac{1}{1+v_{j}})F(v_{k})  \notag  \label{abc int_ed0} \\
&&\int \frac{d^{4}l_{+}}{(2\pi )^{4}}\frac{1}{[l_{+}^{2}-\mathcal{M}%
^{2}(p^{2},m_{k}^{2},v_{k})]^{\alpha +\beta +\gamma -2}},
\end{eqnarray}%
with
\begin{equation*}
F(v_{k})=\frac{(1+v_{1})^{3-\alpha }(1+v_{2})^{3-\beta }(1+v_{3})^{3-\gamma }%
}{(3+v_{1}+v_{2}+v_{3})^{2}},
\end{equation*}%
\begin{equation*}
\mathcal{M}^{2}=\sum_{j=1}^{3}\left( \frac{m_{j}^{2}}{1+v_{j}}\right) -\frac{%
1}{3+v_{1}+v_{2}+v_{3}}p^{2}.
\end{equation*}%
In the subsequent sections, we will show how this formula can naturally be
obtained when merging the UVDP parametrization and the evaluation of ILIs
with the Bjorken-Drell's circuit analogy.

To go further, we need to consider some explicit values of $\alpha,
\beta,\gamma$. As mentioned above, the only cases involving overlapping
divergences are (1) $\alpha+\beta+\gamma=4$ and (2) $\alpha+\beta+\gamma=3$.
Up to the field redefinition, we can always take the corresponding cases to
be (1) $\alpha=\gamma=1, ~\beta=2$ and (2) $\alpha=\beta=\gamma=1$. We will
consider these two cases separately in detail.

\section{Treatment of Divergences in the UVDP parameter Space}

Let us first consider the simpler case with $\alpha =\gamma =1,~\beta =2$,
where the general form of $\alpha \beta \gamma $ integral Eq.(\ref{abc w/o
k1}) can be simplified into the following form
\begin{eqnarray}
I_{121} &=&\frac{i}{16\pi ^{2}}\int_{0}^{\infty }\prod_{i=1}^{3}dv_{i}\delta
(1-\sum_{j=1}^{3}\frac{1}{1+v_{j}})\frac{1}{%
(3+v_{1}+v_{2}+v_{3})^{2}(1+v_{2})}  \notag \\
&&\int \frac{d^{4}k_{2}}{(2\pi )^{4}}\frac{1}{[k_{2}^{2}-\mathcal{M}%
(p^{2},m_{k}^{2},v_{k})]^{2}}  \notag \\
&\rightarrow &-\frac{1}{(16\pi ^{2})^{2}}\int_{0}^{\infty
}\prod_{i=1}^{3}dv_{i}\delta (1-\sum_{j=1}^{3}\frac{1}{1+v_{j}})\frac{1}{%
(3+v_{1}+v_{2}+v_{3})^{2}(1+v_{2})}  \notag \\
&&(\ln \frac{M_{c}^{2}}{\mathcal{M}_{s}^{2}}-\gamma _{\omega }+y_{0}(\frac{%
\mathcal{M}_{s}^{2}}{M_{c}^{2}})).
\end{eqnarray}%
The integral over the loop momentum $k_{2}$ above is logarithmically
divergent, which represents the overall divergence. When carrying out the
integration over the loop momentum $k_{2}$, we have applied the LORE method
to regularize it, where $\mathcal{M}_{s}^{2}=\mathcal{M}^{2}+\mu _{s}^{2}$
with $\mu _{s}^{2}$ playing the role of IR divergence cut-off. In the
following, we are always working in the massive theory, so there is no IR
problem and the scale $\mu _{s}$ can safely be set to $\mu _{s}^{2}=0$.
Also, in the limit $M_{c}^{2}\rightarrow \infty $, $y_{i}(\frac{\mathcal{M}%
_{s}^{2}}{M_{c}^{2}})\simeq O(\frac{\mathcal{M}_{s}^{2}}{M_{c}^{2}}%
)\rightarrow 0$, so $y_{i}$'s vanish identically. By power counting, the
only contribution from $y_{i}$'s may arise when the overall quadratic
divergence is multiplied by $y_{0}(\frac{\mathcal{M}_{s}^{2}}{M_{c}^{2}})$.
Such a contribution is finite and does not disturb the divergent terms.
Nevertheless, in the present paper, we only focus on the divergent part to
show the consistency of the LORE method. Thus, the limit $%
M_{c}^{2}\rightarrow \infty $ is always taken and all the terms $y_{i}(\frac{%
\mathcal{M}^{2}}{M_{c}^{2}})$ are dropped below to simplify our expressions.

It is not difficult to see that there exists a divergence in the region of
UVDP parameter space with $v_{1},v_{3}\rightarrow \infty $, which reflects
the divergence of subdiagram $\alpha \gamma $. To extract the divergence, we
may focus on the region where $v_{1},v_{3}>V$ with $V\gg 1$. In such a
region, $v_{2}\rightarrow 0$, which is ensured by the delta function, so the
domain of the integration is transformed into $\int_{V}^{\infty
}dv_{1}\int_{V}^{\infty }dv_{3}\int_{0}^{\infty }dv_{2}$. With such a
treatment, it is easy to check that $\mathcal{M}\rightarrow m_{2}^{2}$ and
some terms small in comparison with $v_{1}$ and $v_{3}$ can be neglected.
Thus the integral $I_{121}$ is simplified into the following form

\begin{eqnarray}
I_{121} &\simeq &-\frac{1}{(16\pi ^{2})^{2}}\int_{V}^{\infty
}dv_{1}\int_{V}^{\infty }dv_{3}\int_{0}^{\infty }dv_{2}\delta (1-\frac{1}{%
1+v_{2}})  \notag \\
&&\frac{1}{(v_{1}+v_{3})^{2}}(\ln \frac{M_{c}^{2}}{m_{2}^{2}}-\gamma
_{\omega })  \notag \\
&=&-\frac{1}{(16\pi ^{2})^{2}}(\ln \frac{M_{c}^{2}}{m_{2}^{2}}-\gamma
_{\omega })\int_{V}^{\infty }dv_{1}\frac{1}{v_{1}+V},
\end{eqnarray}%
where we integrate over $v_{2}$ and $v_{3}$ as they are convergent. The
remaining integration over $v_{1}$ is divergent, which has to be regularized
appropriately. The LORE method has been shown to be more suitable in this
situation\cite{wu1,Wu:2003dd}, because such a divergence is a kind of scalar
type divergent ILIs, which is the object that can be regularized in the LORE
method, rather than other physical objects, such as propagators or the
dimension of the theory. To regularize the UVDP parameter integral, it is
more useful to transform it into a manifest ILI. For that, one just needs to
multiply a free mass-squared scale $q_{o}^{2}$ to $v_{1}$, which will be
determined by a suitable criterion. Eventually, it will cause the harmful
divergences of different diagrams to be canceled. In general, such a scale
can be the function of the intrinsic quantities in the theory, such as
masses of particles or external momenta.
\begin{eqnarray}
\int_{V}^{\infty }d(q_{o}^{2}v_{1})\frac{1}{q_{o}^{2}v_{1}+q_{o}^{2}V}
&=&\int_{(q_{o}^{2})V}^{\infty }dq_{1}^{2}\frac{1}{q_{1}^{2}+q_{o}^{2}V}
\notag  \label{treatment} \\
&=&\ln \frac{M_{c}^{2}}{2q_{o}^{2}V}-\gamma _{\omega },
\end{eqnarray}%
where we have defiened $q_{1}^{2}\equiv q_{o}^{2}v_{1}$. In the following,
we will frequently encounter similar divergent integrals in the UVDP
parameter space. Unless specified explicitly, we shall always use this
prescription to deal with them. Here we would like to emphasize that the
above prescription is the only one consistent within the framework of the
LORE method. For other regularization schemes, such an approach cannot give
consistent results, such as the Pauli-Villars regularization in which the
regularized objects are the propagators of internal particles.

It can be shown that in other regions of parameter space, there are no
further divergences. Namely, besides the overlapping divergence given above,
they contains only the harmless overall divergence from the integration of $%
k_{2}$. Thus the general form of overlapping divergence in the integral $%
I_{121}$ can be written as:
\begin{equation}
I_{121}\simeq -\frac{1}{(16\pi ^{2})^{2}}(\ln \frac{M_{c}^{2}}{m_{2}^{2}}%
-\gamma _{\omega })\cdot (\ln \frac{M_{c}^{2}}{2q_{o}^{2}V}-\gamma _{\omega
})  \label{I_121}
\end{equation}%
In order to show the exact cancelation of harmful divergences for $I_{121}$,
it is necessary to calculate its corresponding counterterm diagram ($\alpha
\gamma $). (see Fig. (\ref{stdiag})) 
\begin{equation}
I_{121}^{(c)(\alpha \gamma )}=-\int \frac{d^{4}k_{2}}{(2\pi )^{4}}\frac{1}{%
(k_{2}^{2}-m_{2}^{2})^{2}}\textsc{DP}\{\int \frac{d^{4}k_{1}}{(2\pi )^{4}}%
\frac{1}{(k_{1}^{2}-m_{1}^{2})[(k_{1}-k_{2}+p)^{2}-m_{3}^{2}]}\}
\label{I_121 alpha-gamma}
\end{equation}%
where $\textsc{DP}\{\}$ denotes the divergent part. Such a
counterterm integral can be easily computed,
\begin{equation}
I_{121}^{(c)(\alpha \gamma )}=+\frac{1}{(16\pi ^{2})^{2}}(\ln \frac{M_{c}^{2}%
}{m_{2}^{2}}-\gamma _{\omega })\cdot (\ln \frac{M_{c}^{2}}{\mu ^{2}}-\gamma
_{\omega })
\end{equation}%
where the second factor comes from the subintegral $(\alpha \gamma
)$ part contained in $\textsc{DP}\{\}$ and the first one from the
integration of internal loop momentum $k_{2}$.

It is obvious that there is an exact correspondence between the factors in
each expression. When taking the free scale to be $\mu ^{2}=2q_{o}^{2}V$,
the two divergent terms cancel each other exactly. Here the divergence
contained in the UVDP parameter space in the region $v_{1},v_{3}\rightarrow
\infty $ reproduces that of subintegral $(\alpha \gamma )$, namely the
integration over $k_{1}$. We also notice that the divergences of $I_{121}$
are factorizable and can be written as the product of two divergent
integrals, i.e., one from the integral $k_{2}$ for the overall divergence
and the other from the subintegral $k_{1}$ $(\alpha \gamma )$ for the
sub-divergence, which is transformed into and represented in the UVDP
parameter integral of the region $v_{1},v_{3}\rightarrow \infty $. This is
the general feature when using the LORE method to disentangle the two-loop
overlapping divergences. We will make a more explicit demonstration on this
feature below by merging with the Bjorken-Drell's analogy between Feynman
diagrams and electrical circuits.

\section{ Evaluation of ILIs and Bjorken-Drell's Analogy Between Feynman
Diagrams and Electrical Circuit Diagrams}


In order to generalize the correspondence between the divergences in the
UVDP parameter space and those in the subintegrals to more complicated
cases, it is interesting to observe that the UVDP parametrization and the
evaluation of ILIs in the LORE method naturally merge with the
Bjorken-Drell's analogy between the general Feynman diagrams and the
electrical circuits. A detailed description for such the analogy is referred
to the book by Bjorken and Drell\cite{bjor & drel}. It was originally
motivated for discussing the analyticity properties of Feynman diagrams from
the causality requirement. Here let us first establish such the analogy by
developing a standard procedure and notation following Bjorken and Drell,
and then apply it to the general $\alpha\beta\gamma$ integrals by merging it
to the LORE method.

For a general connected Feynman diagram, we shall always denote the external
momenta of the diagram by $p_{1},...,p_{m}$ with the direction of coming
into the diagram. Thus, according to overall momentum conservation, we have:
\begin{equation}
\sum_{s=1}^{m}p_{s}=0.  \label{pcons}
\end{equation}%
%
%
To each internal line we assign a momentum $k_{j}$ with a specified
direction and a mass $m_{j}$. At each vertex, we have a law of momentum
conservation of the form
\begin{equation}
\sum_{j=1}^{n}\epsilon _{ij}k_{j}+\sum_{s=1}^{m}\bar{\epsilon}_{is}p_{s}=0,
\label{kcons}
\end{equation}%
where $\epsilon _{ij}$ is chosen to be $+1$ if internal line $j$ enters
vertex $i$, while $-1$ if internal line $j$ leaves vertex $i$, otherwise $%
\epsilon _{ij}$ is defined to be 0. $\bar{\epsilon}_{is}$ has the similar
definition for the external lines which, by convention, are always taken to
enter vertices.

Each diagram has a definite number $k$ of internal loops. However, we have
the freedom to choose the concrete internal loops and assign each loop a
momentum $l_r$ which are going to be integrated out along the loop. Thus,
for each internal line $j$, we have the following decomposition:
\begin{equation}  \label{decomp}
k_j=q_j+\sum^k_{r=1}\eta_{jr}l_r ,
\end{equation}
where $\eta_{jr}$ is chosen to be 1 if the $j$th internal line lies on the $%
r $th loop and the momenta $k_j$ and $l_r$ are parallel, and -1 if the $j$th
line lies on the $r$th loop but $k_j$ and $l_r$ are antiparallel, otherwise $%
\eta_{jr}$ is 0. Notice that here we introduce another kind of internal
momentum $q_j$, which will be determined after we adopt the UVDP
parametrization for combining denominators to evaluate the ILIs. From the
decomposition Eq.(\ref{decomp}), we can immediately obtain the following
momentum conservation law for each vertex in terms of $q_j$:
\begin{equation}  \label{Vertex q}
\sum^n_{j=1}\epsilon_{ij}q_j+\sum^m_{s=1}\bar{\epsilon}_{is}p_s=0 ,
\end{equation}
which follows from Eq.(\ref{kcons}) and
\begin{equation}
\sum^n_{j=1}\epsilon_{ij}\eta_{jr}=0 ,
\end{equation}
which is a consequence of the definitions of $\epsilon_{ij}$ and $\eta_{jr}$
given in Eqs. (\ref{kcons}) and (\ref{decomp}).

The general structure of the Feynman integral can be written as follows:
\begin{equation}
I(p_{1},...,p_{m})=\int d^{4}l_{1}...d^{4}l_{k}\frac{N}{%
(k_{1}^{2}-m_{1}^{2})^{\alpha _{1}}...(k_{n}^{2}-m_{n}^{2})^{\alpha _{n}}},
\end{equation}%
where $N$ represents the numerator of a general matrix element, which can be
the products of external momenta, internal momenta, spin matrices, wave
functions and so on. By adopting the UVDP parametrization, the above
integral can be written as:
\begin{eqnarray}
I(p_{1},...,p_{m}) &=&\int d^{4}l_{1}...d^{4}l_{k}\frac{\Gamma
(\sum_{j=1}^{n}\alpha _{j})}{\Gamma (\alpha _{1})...\Gamma (\alpha _{n})}%
\int_{0}^{\infty }\prod_{i=1}^{n}\frac{dv_{i}}{(1+v_{i})^{\alpha _{i}+1}}%
\delta (1-\sum_{j=1}^{n}\frac{1}{1+v_{j}})  \notag  \label{genI} \\
&&\frac{N}{[\sum_{j=1}^{n}\frac{k_{j}^{2}-m_{j}^{2}}{1+v_{j}}%
]^{\sum_{j=1}^{n}\alpha _{j}}}  \notag \\
&=&\frac{\Gamma (\sum_{j=1}^{n}\alpha _{j})}{\Gamma (\alpha _{1})...\Gamma
(\alpha _{n})}\int d^{4}l_{1}...d^{4}l_{k}\ \int_{0}^{\infty }\prod_{i=1}^{n}%
\frac{dv_{i}}{(1+v_{i})^{\alpha _{i}+1}}\delta (1-\sum_{j=1}^{n}\frac{1}{%
1+v_{j}})  \notag \\
&&\frac{N}{[\sum_{j=1}^{n}\frac{q_{j}^{2}-m_{j}^{2}}{1+v_{j}}+2\sum_{j,r}%
\frac{q_{j}\eta _{jr}l_{r}}{1+v_{j}}+\sum_{j,r,r^{\prime }}\frac{\eta
_{jr}\eta _{jr^{\prime }}l_{r}l_{r^{\prime }}}{1+v_{j}}]^{\sum_{j=1}^{n}%
\alpha _{j}}}.
\end{eqnarray}%
In order to obtain the required ILIs, we need to eliminate the cross terms
in the denominator which implies
\begin{equation}
\sum_{j=1}^{n}\frac{\eta _{jr}q_{j}}{1+v_{j}}=0  \label{Loop q}
\end{equation}%
for each loop $r=1,...,k$. Now we have the enough conditions Eqs. (\ref%
{Vertex q}) and (\ref{Loop q}) to determine the momenta $q_{j}$ for each
diagram. The above procedure is essentially equivalent to the usual way of
shifting the loop momenta for completing the square in the denominator. Our
next task is to diagonalize the momentum integration variables so that we
can integrate over each momentum integrals separately in Eq. (\ref{genI}).

Before doing the calculation, let us try to understand Eqs.(\ref{Vertex q})
and (\ref{Loop q}) from an alternative interesting perspective. First we put
them into a more heuristic form:
\begin{eqnarray}
\sum_{q_{j}~in~loop~r}\frac{q_{j}}{1+v_{j}} &=&0,  \label{KVolt} \\
\sum_{q_{j},~p_{s}~entering~vertex~i}~(q_{j}+p_{s}) &=&0,  \label{KCurr}
\end{eqnarray}%
\begin{figure}[th]
\includegraphics[scale=0.6]{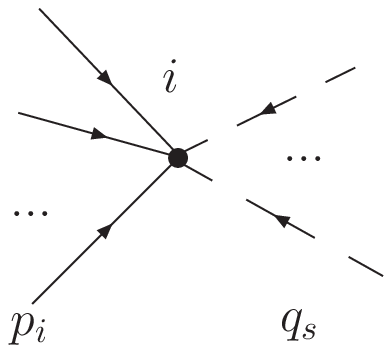}\label{Kvertex}~~~~~~~~~~~ %
\includegraphics[scale=0.8]{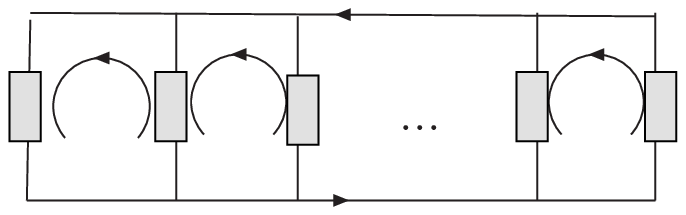}\newline
\caption{Left: Current conservation for each vertex i; Right: Conservation
of the voltage in any loop r}
\label{Kloop}
\end{figure}
we then arrive at a complete analogy between the Feynman diagrams and
electrical circuits. Specifically, we can think of the Feynman diagram as an
electrical circuit and associate the momenta with the currents. Thus $q_{j}$
are the internal currents flowing along the circuit and $p_{s}$ the external
currents entering it. When associating the parameters $\frac{1}{1+v_{j}}$
with the resistance of the $j$th line (so $v_{j}$ can be regarded as the
conductance of the $j$th line), we see explicitly that Eqs. (\ref{KVolt})
and (\ref{KCurr}) simply become the Kirchhoff's laws in this circuit
analogy. Eq. (\ref{KVolt}) shows that the sum of \textquotedblleft voltage
drop" around any closed loop is zero, and Eq. (\ref{KCurr}) indicates that
the sum of \textquotedblleft currents" flowing a vertex is zero.

Moreover, if we associate the voltage with the coordinate $x_{\mu }$ of the
vertex, we can even inquire the physical meaning of Ohm's law:
\begin{equation}
V=IR
\end{equation}%
to be the following relation by translating it into the language of Feynman
diagrams:
\begin{equation}
\Delta x_{j}^{\mu }=\frac{q_{j}^{\mu }}{1+v_{j}},  \label{EOM}
\end{equation}%
\begin{figure}[th]
\begin{center}
\includegraphics{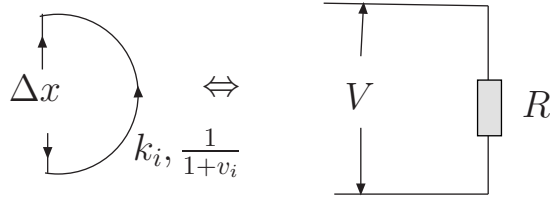}\\[0pt]
\end{center}
\caption{Ohm's Law}
\end{figure}
where $q_{j}$, $v_{j}$ are the momentum and UVDP parameter carried on the
internal line, and $\Delta x_{j}$ is the displacement between two points
connected by this line. In fact, Eq. (\ref{EOM}) is just the equation of
motion for a free particle, which becomes more apparent in terms of the
component forms,
\begin{equation}
\Delta \overrightarrow{x_{j}}=\overrightarrow{q_{j}}\cdot \frac{1}{1+v_{j}}%
,~~\Delta t_{j}=q_{j}^{0}\cdot \frac{1}{1+v_{j}},~~\frac{\Delta
\overrightarrow{x_{j}}}{\Delta t_{j}}=\frac{\overrightarrow{q_{j}}}{q_{j}^{0}%
}.  \label{eom comp}
\end{equation}%
As the parameter $v_{i}$ is positive definite, the causal propagation of the
particle is guaranteed:
\begin{equation}
\frac{\Delta t_{j}}{q_{j}^{0}}=\frac{1}{1+v_{j}}>0.
\end{equation}%
As the particle goes in the $\overrightarrow{q_{j}}$ direction according to (%
\ref{eom comp}), it moves either forward or backward in time depending on
whether the sign of the energy $q_{i}^{0}$ is positive or negative. This
agrees with the interpretation of causality of Feynman propagator in QFT.

The above description provides us a physical picture of the circuit analogy
which can be summarized as follows
\begin{eqnarray}
Feynman~ diagrams &\Leftrightarrow& Electrical~ Circuit~ diagrams \\
Displacement~ \Delta x_j &\Leftrightarrow& Voltage \\
UVDP~ parameter~ v_i &\Leftrightarrow& Conductance~ \geq 0 \\
Free~particle~equation~of~motion &\Leftrightarrow& Ohm^{\prime }s ~law \\
Cross~ term~ cancelation~ condition~for~ ILIs &\Leftrightarrow&
Kirchhoff^{\prime }s~ Law
\end{eqnarray}
while the positivity of the UVDP parameter $v_i$ as the ``conductance" is
related to the causality of propagation for the free particles.\newline
\begin{figure}[ht]
\begin{center}
\includegraphics{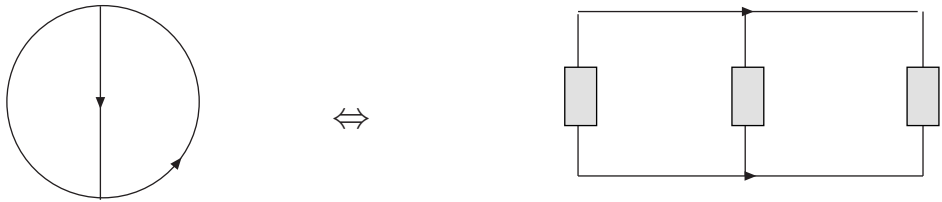}\\[0pt]
\end{center}
\caption{Analogue of Feynman diagrams and electrical circuit diagrams}
\label{}
\end{figure}

In order to carry out the integral over $l_{r}$ in Eq.(\ref{genI}), it is
useful to make the quadratic terms of the momentum $l_{r}$ diagonal. First
write it in terms of the matrix form
\begin{equation}
\sum_{j,r,r^{\prime }}\frac{\eta _{jr}\eta _{jr^{\prime }}l_{r}l_{r^{\prime
}}}{1+v_{j}}=\sum_{r,r^{\prime }}l_{r}M_{rr^{\prime }}l_{r^{\prime }}\equiv
L^{T}ML,\qquad M_{rr^{\prime }}=\sum_{j}\frac{\eta _{jr}\eta _{jr^{\prime }}%
}{1+v_{j}},
\end{equation}%
where $L^{T}=(l_{1},\cdots ,l_{k})$ is the transpose of the vector $L$ and $%
M_{rr^{\prime }}$ is a symmetric matrix. We then diagonalize the matrix $M$
by an orthogonal transformation $O$ with
\begin{equation}
L=OL^{\prime },\qquad O^{T}MO=diag(\lambda _{1},\cdots ,\lambda _{k})\equiv
diag(\lambda _{+},\lambda _{-(1)},\cdots ,\lambda _{-(k-1)}),
\end{equation}%
where $\lambda _{r}$ ($r=1,\cdots ,k$) or $\lambda _{+}$, $\lambda _{-(r)}$ $%
(r=1,\cdots ,k-1)$ are the eigenvalues of the matrix $M$, corresponding to
the eigenvectors $L^{\prime }=(l_{1}^{\prime },\cdots ,l_{k}^{\prime
})^{T}\equiv (l_{+}^{\prime },l_{-(1)}^{\prime },\cdots ,l_{-(k-1)}^{\prime
})^{T}$. As the transformation matrix $O$ is orthogonal, the integration
measure remains unchanged $d^{4}l_{1}^{\prime }\cdots d^{4}l_{k}^{\prime
}=d^{4}l_{1}\cdots d^{4}l_{k}$. Thus, the integral Eq.(\ref{genI}) can be
simplified to:
\begin{eqnarray}
I(p_{1},...,p_{m}) &=&\frac{\Gamma (\sum_{j=1}^{n}\alpha _{j})}{\Gamma
(\alpha _{1})...\Gamma (\alpha _{n})}\int_{0}^{\infty }\prod_{i=1}^{n}\frac{%
dv_{i}}{(1+v_{i})^{\alpha _{i}+1}}\delta (1-\sum_{j=1}^{n}\frac{1}{1+v_{j}})
\notag  \label{genI2} \\
&&\int d^{4}l_{1}...d^{4}l_{k}\frac{N}{[\sum_{j=1}^{n}\frac{%
k_{j}^{2}-m_{j}^{2}}{1+v_{j}}]^{\sum_{j=1}^{n}\alpha _{j}}}  \notag \\
&=&\frac{\Gamma (\sum_{j=1}^{n}\alpha _{j})}{\Gamma (\alpha _{1})...\Gamma
(\alpha _{n})}\int_{0}^{\infty }\prod_{i=1}^{n}\frac{dv_{i}}{%
(1+v_{i})^{\alpha _{i}+1}}\delta (1-\sum_{j=1}^{n}\frac{1}{1+v_{j}})  \notag
\\
&&\int d^{4}l_{1}^{\prime }...d^{4}l_{k}^{\prime }\frac{N}{[\sum_{j=1}^{n}%
\frac{q_{j}^{2}-m_{j}^{2}}{1+v_{j}}+\sum_{r}\lambda _{r}l_{r}^{^{\prime
}2}]^{\sum_{j=1}^{n}\alpha _{j}}}
\end{eqnarray}%
For a generic k-loop integrals where $k\geq 2$ and $n>k$, we have the
inequality $\sum_{j=1}^{n}\alpha _{j}\geq \frac{k(k+1)}{2}\geq 2k-1$.%
\footnote{%
The first inequality comes from the fact that in order that the expression
is generic, we need to consider every type of internal momentum combinations
in the denominator, such as $(l_{j}^{\prime }-p_{j})^{2},(l_{i}^{\prime
}+l_{i+1}^{\prime }-p_{i(i+1)})^{2},...,(l_{1}^{\prime }+l_{2}^{\prime
}+...+l_{k}^{\prime }-p_{12...k})^{2}$. The total number of the combinations
is $k+(k-1)+...+1=\frac{k(k+1)}{2}$. If all the types of combinations appear
in the denominator, then the inequality holds.} Thus we can explicitly
integrate out the loop momenta, as these integrals are already convergent.
In particular, when the numerator $N$ contains no $l_{i}^{^{\prime }}$
terms, we can integrate out the last $(k-1)$ internal loop momenta, say $%
l_{2}^{\prime },l_{3}^{\prime },...,l_{k}^{\prime }$.
\begin{eqnarray}
I(p_{1},...,p_{m}) &=&\frac{\Gamma (\sum_{j=1}^{n}\alpha _{j}-2k+2)}{\Gamma
(\alpha _{1})...\Gamma (\alpha _{n})}\int_{0}^{\infty }\prod_{j=1}^{n}\frac{%
dv_{j}}{(1+v_{j})^{\alpha _{j}+1}}\delta (1-\sum_{j=1}^{n}\frac{1}{1+v_{j}}%
)\prod_{r=1}^{k-1}\frac{1}{\lambda _{-(r)}^{2}}  \notag \\
&&\int d^{4}l_{+}^{\prime }\frac{1}{[\sum_{j=1}^{n}\frac{q_{j}^{2}-m_{j}^{2}%
}{1+v_{j}}+\lambda _{+}l_{+}^{^{\prime }2}]^{\sum_{j=1}^{n}\alpha
_{j}-2(k-1)}}.
\end{eqnarray}%
By a rescaling $l_{+}\rightarrow \sqrt{\lambda _{+}}l_{+}^{\prime }$, we
then obtain the following form:
\begin{eqnarray}
I(p_{1},...,p_{m}) &=&\frac{\Gamma (\sum_{j=1}^{n}\alpha _{j}-2k+2)}{\Gamma
(\alpha _{1})...\Gamma (\alpha _{n})}\int_{0}^{\infty }\prod_{i=1}^{n}\frac{%
dv_{i}}{(1+v_{i})^{\alpha _{i}+1}}\delta (1-\sum_{j=1}^{n}\frac{1}{1+v_{j}})
\notag  \label{genI3} \\
&&\frac{1}{(\det |M|)^{2}}\int d^{4}l_{+}\frac{1}{[\sum_{j=1}^{n}\frac{%
q_{j}^{2}-m_{j}^{2}}{1+v_{j}}+l_{+}^{2}]^{\sum_{j=1}^{n}\alpha _{j}-2(k-1)}},
\end{eqnarray}%
with the definition of the determinant for the matrix $M$
\begin{equation}
\det |M|=\prod_{r=1}^{k}\lambda _{r}\equiv \lambda
_{+}\prod_{r=1}^{k-1}\lambda _{-(r)}.
\end{equation}%
The above expression is the required form of ILIs, where the ILIs for the
momentum integral on $l_{+}$ reflects the overall divergence of the Feynman
diagram. From the above expression, it is clear that the UV divergences
contained in the loop momentum integrals over $l_{-(r)}^{\prime }$ $%
(r=1,\cdots ,k-1)$ for the original loop subdiagrams are now characterized
by the possible zero eigenvalues $\lambda _{-(r)}\rightarrow 0$ $(r=1,\cdots
,k-1)$ of the matrix $M$ in the allowed regions of the parameters $v_{i}$ $%
(i=1,\cdots ,n)$. Namely, each zero eigenvalue $\lambda _{-(r)}\rightarrow 0$
resulted from some infinities of parameters $v_{i}$ in the UVDP parameter
space leads to a singularity for the parameter integrals, which corresponds
to the divergence of subdiagram in the relevant loop momentum integral.

By applying the general LORE formulae to the above integration over the
momentum $l_{+}$, we get:
\begin{eqnarray}
I(p_{1},...,p_{m}) &=&\frac{\Gamma (\sum_{j=1}^{n}\alpha _{j}-2k+2)}{\Gamma
(\alpha _{1})...\Gamma (\alpha _{n})}\int_{0}^{\infty }\prod_{i=1}^{n}\frac{%
dv_{i}}{(1+v_{i})^{\alpha _{i}+1}}\delta (1-\sum_{j=1}^{n}\frac{1}{1+v_{j}})%
\frac{1}{(\det |M|)^{2}}  \notag  \label{genI4} \\
&&\lim_{N,M_{l}^{2}}\sum_{l}^{N}c_{l}^{N}\int d^{4}l_{+}\frac{%
i(-1)^{\sum_{j=1}^{n}\alpha _{j}}}{[\sum_{j=1}^{n}\frac{q_{j}^{2}-m_{j}^{2}}{%
1+v_{j}}+l_{+}^{2}+M_{l}^{2}]^{\sum_{j=1}^{n}\alpha _{j}-2(k-1)}}  \notag \\
&=&\frac{\Gamma (\sum_{j=1}^{n}\alpha _{j}-2k+2)}{\Gamma (\alpha
_{1})...\Gamma (\alpha _{n})}\int_{0}^{\infty }\prod_{i=1}^{n}\frac{dv_{i}}{%
(1+v_{i})^{\alpha _{i}+1}}\delta (1-\sum_{j=1}^{n}\frac{1}{1+v_{j}})  \notag
\\
&&\frac{1}{(\det |M|)^{2}}\ I_{-2\alpha }^{R}(\mathcal{M}^{2})
\end{eqnarray}%
with
\begin{equation}
\alpha =\sum_{j=1}^{n}\alpha _{j}-2k,\qquad \mathcal{M}^{2}=%
\sum_{j=1}^{n}(m_{j}^{2}-q_{j}^{2})/(1+v_{j})
\end{equation}%
where $I_{-2\alpha }^{R}(\mathcal{M}^{2})$ is the regularized 1-fold ILI for
the possible overall divergence of the Feynman diagram.

In general, there are $(k-1)$ zero eigenvalues $\lambda _{-(r)}\rightarrow 0$
$(r=1,\cdots ,k-1)$ in the UVDP parameter space for the k-rank matrix $M$,
and they correspond to the divergences of the $(k-1)$ loop subdiagrams in
the momentum space. In principle, to arrive at the k-fold ILIs for the
k-loop Feynman diagrams, one may perform $(n-k-1)$ integrations in the UVDP
parameter space. It requires us to appropriately analyze the zero
eigenvalues of the matrix $M$ associated with the corresponding regions of
the UVDP parameters. Alternatively, one may make an appropriate parameter
transformation, so that the integrations on the $(n-k-l)$ parameters become
convergent for the considered regions of parameters in a new UVDP parameter
space, thus they can be integrated safely. As a consequence, we obtain the
desired k-fold ILIs. We shall illustrate in detail its consistency and
advantage by applying it to the general $\alpha \beta \gamma $ integrals in
the $\phi ^{4}$ scalar theory. So far it becomes apparent that the above
general procedure explicitly realizes the UVDP parametrization and
systematically obtains the ILIs and this shows the powerful advantage when
merging the LORE method with the Bjorken-Drell analogy between Feynman
diagrams and electrical circuit diagrams.

In order to demonstrate explicitly the correspondence between two kinds of
divergences in the UVDP parameter space and in the momentum space, we are
going to apply the above general procedure to the $\alpha\beta\gamma$
integral in next section.

\section{ Divergence Correspondence between subdiagrams and UVDP parameters}

The corresponding Feynman diagram for $\alpha \beta \gamma $ integral is
already shown in Fig. (\ref{abcoverall}). With the internal momenta $k_{j}$
and the particular choice of loops defined therein, we can rewrite the $%
\alpha \beta \gamma $ integral as follows
\begin{eqnarray}
I_{\alpha \beta \gamma } &=&\int \frac{d^{4}k_{1}}{(2\pi )^{4}}\int \frac{%
d^{4}k_{2}}{(2\pi )^{4}}\frac{1}{(k_{1}^{2}-m_{1}^{2})^{\alpha
}(k_{2}^{2}-m_{2}^{2})^{\beta }(k_{3}^{2}-m_{3}^{2})^{\gamma }}  \notag
\label{abc in BJ} \\
&=&\int \frac{d^{4}k_{1}}{(2\pi )^{4}}\int \frac{d^{4}k_{2}}{(2\pi )^{4}}%
\frac{\Gamma (\alpha +\beta +\gamma )}{\Gamma (\alpha )\Gamma (\beta )\Gamma
(\gamma )}\int_{0}^{\infty }\prod_{i=1}^{3}\frac{dv_{i}}{(1+v_{i})^{\alpha
_{i}+1}}\delta (1-\sum_{j=1}^{3}\frac{1}{1+v_{j}})  \notag \\
&&\frac{1}{[\frac{k_{1}^{2}-m_{1}^{2}}{1+v_{1}}+\frac{k_{2}^{2}-m_{2}^{2}}{%
1+v_{2}}+\frac{k_{3}^{2}-m_{3}^{2}}{1+v_{3}}]^{\alpha +\beta +\gamma }},
\end{eqnarray}%
where we have introduced a new notation $\alpha _{i}$ (i=1,2,3)
corresponding to $\alpha ,\beta ,\gamma $ in the second line. According to
the diagram, we have the momentum conservation, either for overall diagram
or for both vertices:
\begin{equation*}
p_{1}=-p_{2}\equiv p,
\end{equation*}%
and
\begin{subequations}
\begin{eqnarray}
&&p_{1}-k_{1}-k_{2}-k_{3}=0, \\
&&p_{2}+k_{1}+k_{2}+k_{3}=0.
\end{eqnarray}

Following Eq. (\ref{decomp}), we decompose the internal momenta $k_{j}$ into
two parts: one represents the loop momentum flowing along the line $j$, and
the other for the external one carried by $j$
\end{subequations}
\begin{subequations}
\label{dcabc}
\begin{eqnarray}
k_{1} &=&q_{1}+l_{1}, \\
k_{2} &=&q_{2}+l_{2}, \\
k_{3} &=&q_{3}-l_{1}-l_{2},
\end{eqnarray}%
We then arrive at the momentum conservation laws for either vertex in terms
of $q_{j}$
\end{subequations}
\begin{equation}
p_{1}=q_{1}+q_{2}+q_{3}=-p_{2}=p.  \label{momcons abc}
\end{equation}%
Replacing the $k_{j}$ with $q_{j}$ and $l_{r}$ in Eq. (\ref{abc in BJ}) and
changing the integral variables to $l_{r}$ give us
\begin{equation*}
I_{\alpha \beta \gamma }=\int \frac{d^{4}l_{1}}{(2\pi )^{4}}\int \frac{%
d^{4}l_{2}}{(2\pi )^{4}}\frac{\Gamma (\alpha +\beta +\gamma )}{\Gamma
(\alpha )\Gamma (\beta )\Gamma (\gamma )}\int_{0}^{\infty }\prod_{i=1}^{3}%
\frac{dv_{i}}{(1+v_{i})^{\alpha _{i}+1}}\delta (1-\sum_{j=1}^{3}\frac{1}{%
1+v_{j}})\frac{1}{D^{\alpha +\beta +\gamma }},
\end{equation*}%
with
\begin{equation}
D=\sum_{j=1}^{3}\frac{q_{j}^{2}-m_{j}^{2}}{1+v_{j}}+2(\frac{q_{1}}{1+v_{1}}-%
\frac{q_{3}}{1+v_{3}})l_{1}+2(\frac{q_{2}}{1+v_{2}}-\frac{q_{3}}{1+v_{3}}%
)l_{2}+L^{T}ML,  \label{abc denom}
\end{equation}%
where we have introduced the definitions:
\begin{equation*}
L\equiv \left(
\begin{array}{c}
l_{1} \\
l_{2}%
\end{array}%
\right) ,\quad M\equiv \left(
\begin{array}{cc}
\frac{1}{1+v_{1}}+\frac{1}{1+v_{3}} & \frac{1}{1+v_{3}} \\
\frac{1}{1+v_{3}} & \frac{1}{1+v_{2}}+\frac{1}{1+v_{3}}%
\end{array}%
\right) .
\end{equation*}%
The elimination of the cross terms in the denominator $D$ requires that
\begin{subequations}
\label{voltcons abc}
\begin{eqnarray}
\frac{q_{1}}{1+v_{1}}-\frac{q_{3}}{1+v_{3}} &=&0, \\
\frac{q_{2}}{1+v_{2}}-\frac{q_{3}}{1+v_{3}} &=&0.
\end{eqnarray}%
These two formula explicitly illustrate the Kirchhoff's law for two loops in
the electrical circuit analogy of . By taking into account Eqs. (\ref%
{voltcons abc}) and (\ref{momcons abc}) together, we obtain the solutions:
\end{subequations}
\begin{subequations}
\begin{eqnarray}
q_{1} &=&\frac{1+v_{1}}{3+v_{1}+v_{2}+v_{3}}p, \\
q_{2} &=&\frac{1+v_{2}}{3+v_{1}+v_{2}+v_{3}}p, \\
q_{3} &=&\frac{1+v_{3}}{3+v_{1}+v_{2}+v_{3}}p.
\end{eqnarray}%
In order to perform the integral over $l_{r}$, we may first diagonalize the
matrix $M$ by a $2\times 2$ orthogonal matrix transformation $O$, so that
\end{subequations}
\begin{equation*}
L=OL^{\prime },\qquad O^{T}MO=\left(
\begin{array}{cc}
\lambda _{1} & 0 \\
0 & \lambda _{2}%
\end{array}%
\right) ,
\end{equation*}%
with $\lambda _{1,2}=\lambda _{+,-}$ given by
\begin{eqnarray}
\lambda _{\pm } &=&\frac{(1+\frac{1}{1+v_{3}})\pm \sqrt{(1+\frac{1}{1+v_{3}}%
)^{2}+4\Delta }}{2} \\
\Delta =\det |M| &=&\frac{1}{(1+v_{1})(1+v_{2})}+\frac{1}{(1+v_{2})(1+v_{3})}%
+\frac{1}{(1+v_{3})(1+v_{1})},
\end{eqnarray}%
which are the two eigenvalues of the matrix $M$ corresponding to two
eigenvectors $L^{\prime }=(l_{1}^{\prime },\ l_{2}^{\prime })$. Since the
transformation matrix $O$ is orthogonal, the integration measure remains the
same $d^{4}l_{1}^{\prime }d^{4}l_{2}^{\prime }=d^{4}l_{1}d^{4}l_{2}$. Thus,
the $\alpha \beta \gamma $ integral can be reexpressed as:
\begin{eqnarray}
I_{\alpha \beta \gamma } &=&\int \frac{d^{4}l_{1}^{\prime }}{(2\pi )^{4}}%
\int \frac{d^{4}l_{2}^{\prime }}{(2\pi )^{4}}\frac{\Gamma (\alpha +\beta
+\gamma )}{\Gamma (\alpha )\Gamma (\beta )\Gamma (\gamma )}\int_{0}^{\infty
}\prod_{i=1}^{3}\frac{dv_{i}}{(1+v_{i})^{\alpha _{i}+1}}\delta
(1-\sum_{j=1}^{3}\frac{1}{1+v_{j}})  \notag  \label{abc med} \\
&&\frac{1}{[\sum_{j=1}^{3}\frac{q_{j}^{2}-m_{j}^{2}}{1+v_{j}}+\lambda
_{1}l_{1}^{\prime 2}+\lambda _{2}l_{2}^{\prime 2}]^{\alpha +\beta +\gamma }}
\notag \\
&=&\int \frac{d^{4}l_{+}^{\prime }}{(2\pi )^{4}}\int \frac{%
d^{4}l_{-}^{\prime }}{(2\pi )^{4}}\frac{\Gamma (\alpha +\beta +\gamma )}{%
\Gamma (\alpha )\Gamma (\beta )\Gamma (\gamma )}\int_{0}^{\infty
}\prod_{i=1}^{3}\frac{dv_{i}}{(1+v_{i})^{\alpha _{i}+1}}\delta
(1-\sum_{j=1}^{3}\frac{1}{1+v_{j}})  \notag \\
&&\frac{1}{[\sum_{j=1}^{3}\frac{q_{j}^{2}-m_{j}^{2}}{1+v_{j}}+\lambda
_{+}l_{+}^{\prime 2}+\lambda _{-}l_{-}^{\prime 2}]^{\alpha +\beta +\gamma }}.
\end{eqnarray}%
From this formalism, it can be shown that the integration over $%
l_{-}^{\prime }$ represents the integral over subdiagrams, while the one
over $l_{+}^{\prime }$ is an overall diagram. However, the matrix $M$ is not
always invertible, since the determinant of $M$ vanishes when any two of $%
v_{i}$s tend to $\infty $. More specifically, take $v_{1},v_{3}\rightarrow
\infty $ for example. In this case the eigenvalue $\lambda _{-}$ vanishes.
It is also noted that the combination $\lambda _{+}l_{+}^{\prime }$ and $%
\lambda _{-}l_{-}^{\prime }$ in Eq.(\ref{abc med}) are on equal footing in
the denominator, it is then expected that $\lambda _{+}l_{+}^{\prime }$ and $%
\lambda _{-}l_{-}^{\prime }$ approach to infinity at the same speed when
both $l_{\pm }^{\prime }\rightarrow \infty $. Thus, when considering $%
\lambda _{-}\rightarrow 0$ while keeping $\lambda _{+}$ finite, it requires
that the speed of $l_{-}^{\prime }$ tending to infinity is faster than that
of $l_{+}^{\prime }$ in order to keep the balance. Recall that in our
previous general discussion on the divergence behavior of overlapping
diagrams, one of the features for the subdivergences is that the integration
variables approach to infinity faster than the overall one.

Based on the above analysis, we may conclude that the integral over $%
l_{-}^{\prime }$ reflects the asymptotic behavior of subintegrals when the
corresponding UVDP parameters approach to infinity. Here we would like to
emphasize that the integration over $l_{-}^{\prime }$ does not correspond to
any particular loop in the original Feynman diagram. Rather, it represents
all subintegrals and is specified according to the asymptotic regions in the
UVDP parameter space. For instance, when the divergences in the UVDP
parameter space occur in other regions, such as $v_{1},v_{2}\rightarrow
\infty $, then $l_{-}^{\prime }$ reflects the loop composing of lines 1 and
2. The above explicit construction helps us to understand the intuitive
analogy between the Feynman diagrams and electrical circuits. Especially, it
illustrates why and how the divergences in the subdiagrams are transmitted
to the corresponding divergences in the UVDP parameter space. Let us further
demonstrate this point from another perspective. By explicitly integrating
over $l_{-}^{\prime }$, we obtain,
\begin{eqnarray}
I_{\alpha \beta \gamma } &=&\frac{i}{16\pi ^{2}}\int \frac{%
d^{4}l_{+}^{\prime }}{(2\pi )^{4}}\frac{\Gamma (\alpha +\beta +\gamma -2)}{%
\Gamma (\alpha )\Gamma (\beta )\Gamma (\gamma )}\int_{0}^{\infty
}\prod_{i=1}^{3}\frac{dv_{i}}{(1+v_{i})^{\alpha _{i}+1}}\delta
(1-\sum_{j=1}^{3}\frac{1}{1+v_{j}})\frac{1}{\lambda _{-}^{2}}  \notag \\
&&\frac{1}{[\sum_{j=1}^{3}\frac{q_{j}^{2}-m_{j}^{2}}{1+v_{j}}+\lambda
_{+}l_{+}^{\prime 2}]^{\alpha +\beta +\gamma -2}}
\end{eqnarray}%
which explicitly shows that when $\lambda _{-}$ goes to zero, that is, any
two of the three UVDP parameters $v_{i}$ approach to infinity, the integrand
becomes singular and the integrations over the UVDP parameters give some UV
divergences.

By defining a new integral loop momenta $l_{+}$ as
\begin{equation}
l_{+}\equiv \sqrt{\lambda _{+}}l_{+}^{\prime },
\end{equation}%
we can transform the $\alpha \beta \gamma $ integral into a more tractable
form:
\begin{eqnarray}
I_{\alpha \beta \gamma } &=&\frac{i}{16\pi ^{2}}\int \frac{d^{4}l_{+}}{(2\pi
)^{4}}\frac{\Gamma (\alpha +\beta +\gamma -2)}{\Gamma (\alpha )\Gamma (\beta
)\Gamma (\gamma )}\int_{0}^{\infty }\prod_{i=1}^{3}\frac{dv_{i}}{%
(1+v_{i})^{\alpha _{i}+1}}\delta (1-\sum_{j=1}^{3}\frac{1}{1+v_{j}})  \notag
\label{abc int_ed2} \\
&&\frac{1}{(\det |M|)^{2}}\frac{1}{[\sum_{j=1}^{3}\frac{q_{j}^{2}-m_{j}^{2}}{%
1+v_{j}}+l_{+}^{2}]^{\alpha +\beta +\gamma -2}}  \notag \\
&=&\frac{i}{16\pi ^{2}}\frac{\Gamma (\alpha +\beta +\gamma -2)}{\Gamma
(\alpha )\Gamma (\beta )\Gamma (\gamma )}\int_{0}^{\infty }\prod_{i=1}^{3}%
\frac{dv_{i}}{(1+v_{i})^{2}}\delta (1-\sum_{j=1}^{3}\frac{1}{1+v_{j}}%
)F(v_{k})  \notag \\
&&\int \frac{d^{4}l_{+}}{(2\pi )^{4}}\frac{1}{[l_{+}^{2}-\mathcal{M}%
^{2}(p^{2},m_{k}^{2},v_{k})]^{\alpha +\beta +\gamma -2}},
\end{eqnarray}%
where
\begin{eqnarray}
F(v_{k}) &=&\frac{(1+v_{1})^{1-\alpha }(1+v_{2})^{1-\beta
}(1+v_{3})^{1-\gamma }}{(\det |M|)^{2}}  \notag \\
&=&\frac{(1+v_{1})^{3-\alpha }(1+v_{2})^{3-\beta }(1+v_{3})^{3-\gamma }}{%
(3+v_{1}+v_{2}+v_{3})^{2}},
\end{eqnarray}%
\begin{equation*}
\mathcal{M}^{2}=\sum_{j=1}^{3}\frac{m_{j}^{2}-q_{j}^{2}}{1+v_{j}}%
=\sum_{j=1}^{3}\frac{m_{j}^{2}}{1+v_{j}}-\frac{1}{3+v_{1}+v_{2}+v_{3}}p^{2}.
\end{equation*}%
The above equation is equivalent to the form in Eq.(\ref{abc w/o k1}) with a
rescaling given in Eq.(\ref{scaling}). Nevertheless, the derivation here is
more general and systematic and it explicitly shows the advantage when
merging the UVDP parametrization and the evaluation of ILIs with the
Bjorken-Drell's electrical circuit analogy of the Feynman diagrams.

In general, the integration for the momentum $l_{+}$ is divergent and needs
to be regularized. By applying the LORE method to the momentum integral, we
obtain:
\begin{eqnarray}
I_{\alpha \beta \gamma } &=&\frac{i}{16\pi ^{2}}\frac{\Gamma (\alpha +\beta
+\gamma -2)}{\Gamma (\alpha )\Gamma (\beta )\Gamma (\gamma )}%
\int_{0}^{\infty }\prod_{i=1}^{3}\frac{dv_{i}}{(1+v_{i})^{2}}\delta
(1-\sum_{j=1}^{3}\frac{1}{1+v_{j}})F(v_{k})  \notag  \label{abc int_ed3} \\
&&\lim_{N,M_{l}^{2}}\sum_{l=1}^{N}c_{l}^{N}\int \frac{d^{4}l_{+}}{(2\pi )^{4}%
}\frac{i(-1)^{\alpha +\beta +\gamma }}{[l_{+}^{2}+M_{l}^{2}+\mathcal{M}%
^{2}(p^{2},m_{k}^{2},v_{k})]^{\alpha +\beta +\gamma -2}} \\
&=&\frac{i}{16\pi ^{2}}\frac{\Gamma (\alpha +\beta +\gamma -2)}{\Gamma
(\alpha )\Gamma (\beta )\Gamma (\gamma )}\int_{0}^{\infty }\prod_{i=1}^{3}%
\frac{dv_{i}}{(1+v_{i})^{2}}\delta (1-\sum_{j=1}^{3}\frac{1}{1+v_{j}}%
)F(v_{k})\ I_{-2(\alpha +\beta +\gamma -4)}^{R}(\mathcal{M}^{2}).  \notag
\end{eqnarray}%
When applying the above general formula to the case $\alpha =\gamma =1,\beta
=2$, with the similar calculation as the one in the previous section, the
result is the same due to the equivalence of Eq. (\ref{abc w/o k1}) and Eq. (%
\ref{abc int_ed2}),
\begin{eqnarray}
I_{121} &=&\frac{i}{16\pi ^{2}}\int_{0}^{\infty }\prod_{i=1}^{3}\frac{dv_{i}%
}{(1+v_{i})^{2}}\delta (1-\sum_{j=1}^{3}\frac{1}{1+v_{j}})\frac{%
(1+v_{1})^{2}(1+v_{2})(1+v_{3})^{2}}{(3+v_{1}+v_{2}+v_{3})^{2}}\ I_{0}^{R}(%
\mathcal{M}^{2})  \notag \\
&\rightarrow &-\frac{1}{(16\pi ^{2})^{2}}\int_{0}^{\infty
}\prod_{i=1}^{3}dv_{i}\delta (1-\sum_{j=1}^{3}\frac{1}{1+v_{j}})\frac{1}{%
(3+v_{1}+v_{2}+v_{3})^{2}(1+v_{2})}  \notag \\
&&(\ln \frac{M_{c}^{2}}{\mathcal{M}_{s}^{2}}-\gamma _{\omega }+y_{0}(\frac{%
\mathcal{M}_{s}^{2}}{M_{c}^{2}}))
\end{eqnarray}%
which shows that the singular behavior in the region $v_{1},v_{3}\rightarrow
\infty $ becomes obvious as $\det |M|=\Delta =0$ due to the zero eigenvalue $%
\lambda _{-}\rightarrow 0$. For the other two regions: $v_{1},v_{2}%
\rightarrow \infty $ and $v_{2},v_{3}\rightarrow \infty $, the additional
factor $\frac{1}{(1+v_{2})}$ in these two cases makes the integration
finite. In contrast, for the case $\alpha =\beta =\gamma =1$, there is no
such a factor, so that there are more UV divergent structures in all the
three regions andis going to be discussed in detail below.

\section{Treatment of Overlapping Divergence and Advantage of the LORE
Method Merging with Bjorken-Drell's Analogy}

This section shows that the LORE method merging with the Bjorken-Drell's
circuit analogy has the advantage in analyzing the more complicated and
challenging overlapping divergence structures of Feynman diagrams. For an
explicit demonstration, we are going to consider the case with $\alpha
=\beta =\gamma =1$ in the $\alpha \beta \gamma $ integral. The difficulty
lies not only in the quadratic divergence but also in the more complicated
overlapping divergence structure. It will be seen that the LORE method
merging with Bjorken-Drell's analogy is extremely powerful in unraveling the
overlapping divergences.

The general form of $\alpha \beta \gamma $ integral (Eq.\ref{abc int_ed2})
can be simplified to:
\begin{eqnarray}
I_{111} &=&\frac{i}{16\pi ^{2}}\int_{0}^{\infty }\prod_{i=1}^{3}\frac{dv_{i}%
}{(1+v_{i})^{2}}\delta (1-\sum_{j=1}^{3}\frac{1}{1+v_{j}})\frac{1}{(\det
|M|)^{2}}  \notag  \label{I_111 def} \\
&&\lim_{N,M_{l}^{2}}\sum_{l=1}^{N}c_{l}\int \frac{d^{4}l_{+}}{(2\pi )^{4}}%
\frac{-i}{\sum_{j=1}^{3}\frac{q_{j}^{2}-m_{j}^{2}}{1+v_{j}}%
+l_{+}^{2}+M_{l}^{2}}  \notag \\
&\rightarrow &\frac{1}{(16\pi ^{2})^{2}}\int_{0}^{\infty }\prod_{i=1}^{3}%
\frac{dv_{i}}{(1+v_{i})^{2}}\delta (1-\sum_{j=1}^{3}\frac{1}{1+v_{j}})\frac{%
\prod_{j=1}^{3}(1+v_{j})^{2}}{(3+v_{1}+v_{2}+v_{3})^{2}}  \notag \\
&&[M_{c}^{2}-\mathcal{M}^{2}(\ln \frac{M_{c}^{2}}{\mathcal{M}^{2}}-\gamma
_{\omega }+1)],  \label{I111}
\end{eqnarray}%
where we have regularized the overall quadratic divergence of loop momentum
integral by the LORE method. The mass factor $\mathcal{M}$ is given in Eq.(%
\ref{Mv}). The UVDP parameter integrals are more involved due to the
appearance of the overlapping divergences. From the expression of integral $%
I_{111}$, it is seen that the three subintegrals $\alpha \gamma $, $\beta
\gamma $, and $\alpha \beta $ are all divergent. With the analogy of
circuits, we have shown that the UV divergences arising from the large
internal loop momenta transmit to the asymptotic regions of UVDP parameter
space, where the divergent conductances correspond to the following
asymptotic regions in the circuits:
\begin{eqnarray}
subdivergence~in~\alpha \beta \gamma ~diagrams &\Leftrightarrow
&divergence~in~\mbox{UVDP parameter}~space  \notag \\
Circuit~1:~\alpha \gamma ~divergence &\Leftrightarrow &v_{1}\rightarrow
\infty ,v_{3}\rightarrow \infty ,v_{2}\rightarrow 0, \\
Circuit~2:~\beta \gamma ~divergence &\Leftrightarrow &v_{2}\rightarrow
\infty ,v_{3}\rightarrow \infty ,v_{1}\rightarrow 0, \\
Circuit~3:~\alpha \beta ~divergence &\Leftrightarrow &v_{1}\rightarrow
\infty ,v_{2}\rightarrow \infty ,v_{3}\rightarrow 0.
\end{eqnarray}%
This result can also be obtained by considering the singularities in the
determinant $\det |M|=\Delta $ as discussed in the previous section.

Note that Eq.(\ref{I111}) has a permutation $Z_{3}$ symmetry among the three
pairs of parameters $(v_{1},m_{1})$, $(v_{2},m_{2})$, $(v_{3},m_{3})$, so
the treatment on three asymptotic regions in the circuits is essentially the
same. Let us consider in detail the first case of the Circuit~1.

\textbf{Circuit 1:} $v_{1}\rightarrow \infty $, $v_{3}\rightarrow \infty $
and $v_{2}\rightarrow 0$. In such region, the integral domain can be written
as $\int_{V}^{\infty }dv_{1}\int_{V}^{\infty }dv_{3}$ with $\mathcal{M}%
^{2}\rightarrow m_{2}^{2}$ and $F(v_{j})\rightarrow \frac{%
(1+v_{1})^{2}(1+v_{3})^{2}}{(v_{1}+v_{3})^{2}}$. Thus the integration is
simplified to:
\begin{eqnarray}
I_{111}^{(0)(\alpha \gamma )} &\simeq &\frac{1}{(16\pi ^{2})^{2}}%
\int_{V}^{\infty }\frac{dv_{1}}{(1+v_{1})^{2}}\int_{V}^{\infty }\frac{dv_{3}%
}{(1+v_{3})^{2}}\frac{(1+v_{1})^{2}(1+v_{3})^{2}}{(v_{1}+v_{3})^{2}}  \notag
\\
&&[M_{c}^{2}-m_{2}^{2}(\ln \frac{M_{c}^{2}}{m_{2}^{2}}-\gamma _{\omega }+1)]
\notag \\
&=&\frac{1}{(16\pi ^{2})^{2}}[M_{c}^{2}-m_{2}^{2}(\ln \frac{M_{c}^{2}}{%
m_{2}^{2}}-\gamma _{\omega }+1)]\int_{V}^{\infty }dv_{1}\frac{1}{v_{1}+V}+...
\notag \\
&=&\frac{1}{(16\pi ^{2})^{2}}[M_{c}^{2}-m_{2}^{2}(\ln \frac{M_{c}^{2}}{%
m_{2}^{2}}-\gamma _{\omega }+1)](\ln \frac{M_{c}^{2}}{2q_{o}^{2}V}-\gamma
_{\omega })+....
\end{eqnarray}%
Note that in the last step, we have applied the LORE method with the
treatment discussed in Eq.(\ref{treatment}). The dots represent other terms,
such as single logarithmic divergent term and finite terms, which are
irrelevant to our discussions as our main purpose here is to check the
cancelation of the harmful divergences. Note that our result here is
factorizable.

In order to compare the above divergence structure with those contained in
the subdiagram $(\alpha\gamma)$, we calculate the counterterm diagram $%
I^{(c)(\alpha\gamma)}_{111}$:
\begin{eqnarray}
I^{(c)(\alpha\gamma)}_{111} &=& - \int\frac{d^4k_2}{(2\pi)^4}\frac{1}{%
k_2^2-m_2^2}\textsc{DP} \{\int\frac{d^4k_1}{(2\pi)^4} \frac{1}{%
(k_1^2-m_1^2)[k_3^2-m_3^2]}\}  \notag \\
&\to & -\frac{1}{(16\pi^2)^2} (\ln\frac{M_c^2}{\mu^2}-\gamma_\omega)
[M_c^2-m_2^2(\ln\frac{M_c^2}{m_2^2}-\gamma_\omega+1)],  \label{I111_1}
\end{eqnarray}
where \textsc{DP}\{...\} means the divergence part of the integral in the
bracket, and $\mu^2$ is the renormalization scale. It is then manifest that
when we choose $\mu^2=2q_o^2V$, the harmful divergence parts in the two
expressions cancel exactly.

With a similar discussion based on the permutation $Z_{3}$ symmetry, it is
easy to show that the harmful divergent parts in the Circuit~2 and Circuit~3
also cancel exactly. So far we prove that there is no harmful divergence for
the case $\alpha =\beta =\gamma =1$ when combining with the corresponding
counterterm diagrams. We would like to mention that this is different from
the dimensional regularization, that in that we have an extra term
corresponding to the quadratic divergence $M_{c}^{2}$. As emphasized in \cite%
{wu1}, this term is natural to maintain the correct divergent behavior of
the original diagram, which can play an important role in effective field
theory for obtaining the correct gap equation to describe the dynamically
generated spontaneous symmetry breaking\cite{DW}. It will explicitly be
shown below that the presence of this term prevents us from having a mass
independent renormalization scheme. Thus, a consistent renormalization with
a well-defined subtraction scheme must be proposed for the LORE method. We
propose the following subtraction scheme:

(i) For quadratic divergence $(M_c^2- \mathcal{M}^2) $, subtract $%
(M_c^2-\mu^2)$ and leave $(\mu^2- \mathcal{M}^2)$ in the finite expression;

(ii) For logarithmic divergence $(\log\frac{M_c^2}{\mathcal{M}^2}%
-\gamma_\omega)$, subtract $(\log\frac{M_c^2}{\mu^2}-\gamma_\omega)$ and
leave term $\log\frac{\mu^2}{\mathcal{M}^2}$ in the finite expression.

Such a scheme may be regarded as a kind of energy scale subtraction scheme
at $\mu ^{2}$ and is similar to the usual momentum subtraction. For the
logarithmic divergence, it appears to be a $\bar{MS}$-like scheme in the
dimensional regularization as it is associated with the Euler number $\gamma
_{w}=\gamma _{E}$. It is interesting to note that once the energy scale
subtraction scheme for both the quadratic and logarithmic terms is set up at
the one-loop level with a correlated form $(M_{c}^{2}-\mu ^{2})$ and $\ln
M_{c}^{2}/\mu ^{2}$ via a single subtracted energy scale $\mu ^{2}$, and
suppose that the correlated form with a single subtracted energy scale $\mu
^{2}$ is required to be maintained, thus either the rescaling $\mu
^{2}\rightarrow e^{\alpha _{0}}\mu ^{2}$ or shifting $\mu ^{2}\rightarrow
\mu ^{2}-\alpha _{0}m^{2}$ for the subtracted energy scale $\mu ^{2}$ will
not be allowed. As a consequence, the mass renormalization at higher loop
becomes well-defined through such an energy scale subtraction scheme at the
one loop level, namely fixing the correlated form for the quadratic and
logarithmic terms via a single subtracted energy scale $\mu ^{2}$.

Based on the above analysis and discussions, we arrive at the following
theorems:

\emph{Factorization Theorem for Overlapping Divergences}: Overlapping
divergences which contain divergences of subintegrals and overall one in the
general Feynman loop integrals become factorizable in the corresponding
asymptotic regions.

\emph{Substraction Theorem for Overlapping Divergences}: For general
scalar-type two-loop integral $I_{\alpha\beta\gamma}$, when combined with
the corresponding subtraction integrals (which is composed of divergent
subintegrals multiplied by an overall integral), the sum will only contain
harmless divergence.

For completeness, we have the following theorems for dealing with the
Feynman integrals which do not involve the overlapping divergence. They are
so obvious that the proofs are omitted here.

\emph{Harmless Divergence Theorem}: If the general loop integral contains no
divergent subintegrals, then it is only possible to contain a harmless
single divergence arising from the overall divergence.

\emph{Trivial Convergence Theorem}: If the general loop integral contains
neither the overall divergence nor the divergent subintegrals, then it is
convergent.

In summary, the LORE method can properly deal with the overlapping
divergences, especially the subdivergences which is transformed
appropriately into the divergences in the UVDP parameter space. To extract
them, we need to explore the integrals in different asymptotic regions of
the parameter space. Moreover, we demonstrate that these overlapping
divergences can well be treated by the LORE method when merging with the
Bjorken-Drell's analogy between general Feynman diagrams and electrical
circuits, especially the correspondence between the UVDP parameters and the
conductances of internal lines in the circuit analogy. By applying this
intuitive picture, we can immediately recognize how a divergence in the
region of UVDP parameter space corresponds to a certain original divergent
subintegral composed by the lines that the divergent UVDP parameters are
attached on. This correspondence also helps us to find the right counterterm
diagram to cancel the notorious harmful divergences. As a result, we are
left with only the finite terms and the harmless divergence which can be
absorbed into the overall counterterm at two-loop order. These results are
summarized in the four theorems presented above. It is interesting to note
that the extension of the LORE method to the calculations beyond two-loop
order is straightforward, although we are aware that the degree of
complication and difficulty increases dramatically with the increase of loop
orders as more and more Feynman diagrams are involved. Nevertheless, it is
clearly indicated that merging with the Bjorken-Drell's analogy between
Feynman diagrams and electrical circuit diagrams, the LORE method gets its
consistency and advantage in the multiloop calculations, especially with the
aid of computer. 

\section{Application to $\protect\phi^4$ Theory at Two-Loop Order}

The discussion and analysis in the previous sections on the general two-loop
integrals appear to be a little bit too abstract. In this section, we shall
take the simple scalar $\phi ^{4}$ theory as a concrete example to
illustrate the LORE method in a practical calculation, and leave another
application involving tensor-type integrals to a separate paper\cite%
{QED2loop}.

The Lagrangian density for $\phi ^{4}$ theory is:
\begin{equation}
\mathcal{L}=\frac{1}{2}\partial _{\mu }\phi \partial ^{\mu }\phi -\frac{1}{2}%
m^{2}\phi ^{2}-\frac{\lambda }{4!}\phi ^{4}.
\end{equation}%
Its Feynman rules may be found in the standard textbooks, such as \cite%
{Peskin:1995ev,Itzykson:1980rh}. Our main purpose here is to explicitly
calculate the two-loop contributions to the mass term and coupling constant
from the self-energy and vertex diagrams. From this practical calculation,
we will demonstrate in detail the consistency and advantage of the LORE
method when merging with the Bjorken-Drell's circuits analogy.

\subsection{ Renormalization At One-Loop Level}

Before proceeding to a detailed calculation at two-loop level, we need the
one-loop counterterms first. This is equivalent to specify the
renormalization condition in the LORE method, which is the main goal of this
subsection.

\begin{figure}[th]
\begin{center}
\includegraphics[scale=0.8]{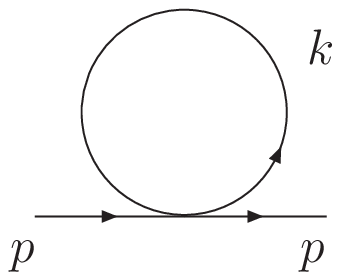}\\[0pt]
\end{center}
\caption{{}}
\label{1loop2}
\end{figure}

\begin{figure}[th]
\begin{center}
\includegraphics[scale=0.8]{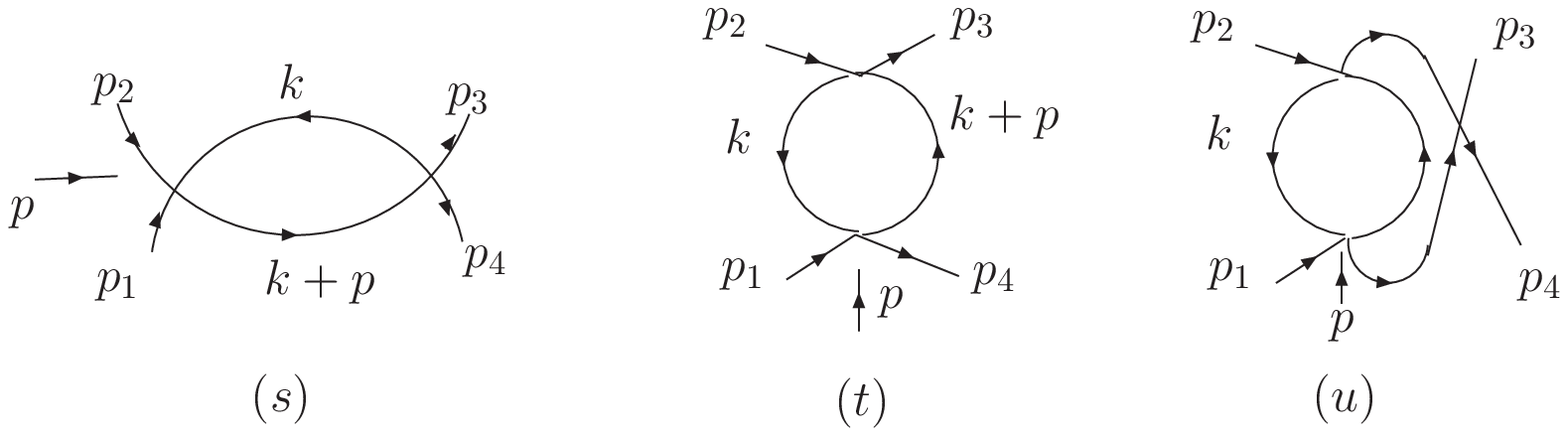}\\[0pt]
\end{center}
\caption{{}}
\label{1loop4}
\end{figure}

At the one loop level, there are two types of diagrams corresponding to the
self-energy correction and vertex correction respectively, as shown in Figs. (%
\ref{1loop2}) and (\ref{1loop4}).\newline
\newline
For the self-energy correction, the calculation is straightforward and the
result is given by
\begin{eqnarray}
-iM_{(1)}^{2} &=&-i\lambda \cdot \frac{1}{2}\int \frac{d^{4}k}{(2\pi )^{4}}%
\frac{i}{k^{2}-m^{2}}  \notag  \label{1_selfenergy} \\
&\rightarrow &-\frac{i\lambda }{2(4\pi )^{2}}[M_{c}^{2}-m^{2}(\ln \frac{%
M_{c}^{2}}{m_{s}^{2}}-\gamma _{\omega }+1+y_{2}(\frac{m_{s}^{2}}{M_{c}^{2}}%
))]  \notag \\
&\rightarrow &-\frac{i\lambda }{2(4\pi )^{2}}[(M_{c}^{2}-m^{2})-m^{2}(\ln
\frac{M_{c}^{2}}{m^{2}}-\gamma _{\omega })]
\end{eqnarray}%
where we have applied the LORE method to obtain the result in the second
line. The result in the last line is obtained by taking $\mu _{s}=0$ and $%
M_{c}\rightarrow \infty $. Under the energy scale $\mu ^{2}$ subtraction
scheme described in the previous section, the mass and wave function
counterterms take the following forms:
\begin{eqnarray}
-i\delta _{m^{2}}^{(1)} &=&\frac{i\lambda }{2(4\pi )^{2}}[(M_{c}^{2}-\mu
^{2})-m^{2}(\ln \frac{M_{c}^{2}}{\mu ^{2}}-\gamma _{\omega })],
\label{delta_m1} \\
i\delta _{Z}^{(1)} &=&0,  \label{delta_z1}
\end{eqnarray}%
and the finite term is found to be
\begin{equation}
-iM_{(1)}^{2}=-\frac{i\lambda }{2(4\pi )^{2}}[(\mu ^{2}-m^{2})-m^{2}\ln
\frac{\mu ^{2}}{m^{2}}].  \label{111}
\end{equation}%
Note that the result given in Eq.(\ref{111}) is different from the one
obtained by using the dimensional regularization method. The difference
arises from the quadratic behavior $\mu ^{2}$ in the renormalization
counterterm. This difference may have important physical implications: it
greatly changes the renormalization group\cite%
{Tang:2008ah,Tang:2010cr,Tang2011} with a power law running, and generates
the physically meaningful dynamical mass scales in the effective field theory%
\cite{wu1,DW}.

By a similar calculation of the one-loop four-point Green function, we
obtain the following vertex correction for s-channel:
\begin{eqnarray}
-i\Lambda^{(s)}_{(1)} &=& \frac{(-i\lambda)^2}{2} \int\frac{d^4 k}{(2\pi)^4}
\frac{i}{k^2-m^2}\frac{i}{(k+p)^2-m^2}  \notag \\
&=& \frac{\lambda^2}{2}\int\frac{d^4k}{(2\pi)^4}\int^1_0 dx \frac{1}{%
[xk^2+(1-x)(k+p)^2-m^2]^2}  \notag \\
&=& \frac{\lambda^2}{2}\int\frac{d^4k}{(2\pi)^4}\int^1_0 dx \frac{1}{%
[k^2+x(1-x)p^2-m^2]^2}  \notag \\
&\to & \frac{i\lambda^2}{2(4\pi)^2}\int^1_0 dx [\ln\frac{M_c^2}{m^2-x(1-x)p^2%
}-\gamma_\omega],
\end{eqnarray}
where $-p^2=-(p_1+p_2)^2\equiv s$. For other two channels (t- and
u-channels), we will obtain the same expression except for the definition of
$p^2$: $-p^2=-(p_1-p_4)^2\equiv t$ for t-channel and $-p^2=-(p_1-p_3)^2%
\equiv u$ for u-channel.

According to the renormalization scheme of the LORE method, we have the
following counterterm:
\begin{equation}  \label{delta_lmd1}
-i\delta^{(1)}_\lambda = -\frac{3i\lambda^2}{2(4\pi)^2} [\ln\frac{M_c^2}{%
\mu^2}-\gamma_\omega],
\end{equation}
where the factor 3 comes from three diagrams corresponding to the $s$, $t$, $%
u$-channels.

So the renormalized vertex correction is simply given by:
\begin{eqnarray}
-i\Lambda_{(1)} = \frac{i\lambda^2}{2(4\pi)^2}\int^1_0 dx [\ln\frac{\mu^2}{%
m^2+x(1-x)s} +\ln\frac{\mu^2}{m^2+x(1-x)t} +\ln\frac{\mu^2}{m^2+x(1-x)u}].
\end{eqnarray}

\subsection{Self-Energy Contribution at Two Loop}

\begin{figure}[th]
\begin{center}
\includegraphics[scale=0.8]{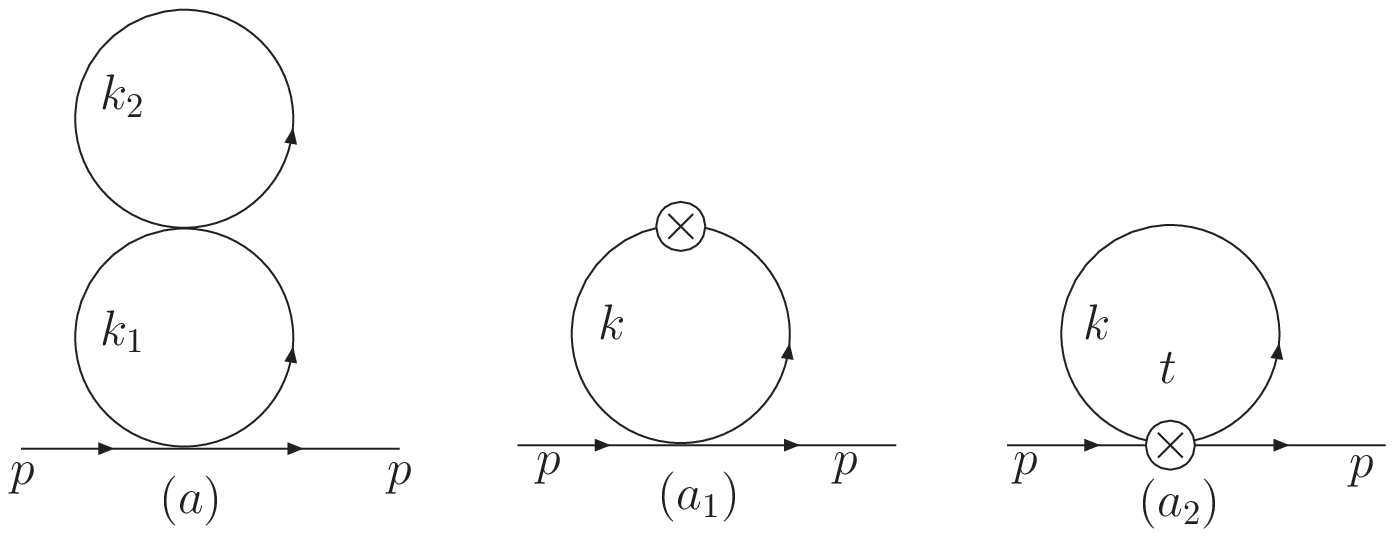}\\[0pt]
\end{center}
\caption{{}}
\label{groupA}
\end{figure}

\begin{figure}[th]
\begin{center}
\includegraphics[scale=0.7]{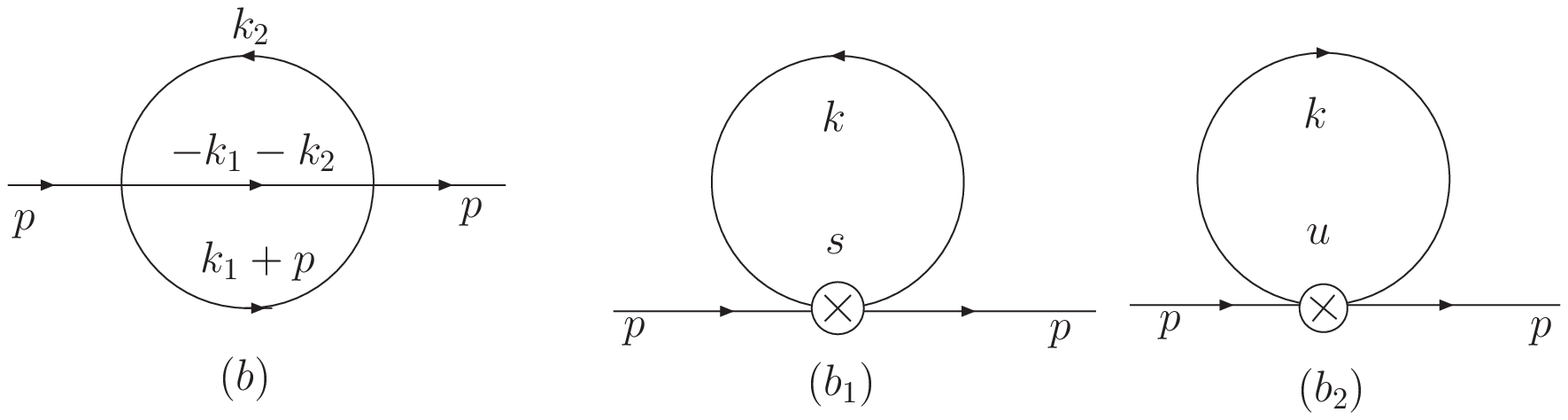}\\[0pt]
\end{center}
\caption{{}}
\label{groupB}
\end{figure}

There are two diagrams contributing to the two-loop self-energy corrections,
which are shown in Figs. (\ref{groupA}) and (\ref{groupB}).\newline

The calculation of diagram (a) is straightforward and the result is:
\begin{eqnarray}
-iM_{(2)}^{2(a)} &=&\frac{1}{4}(-i\lambda )^{2}\int \frac{d^{4}k_{1}}{(2\pi
)^{4}}\frac{i}{k_{1}^{2}-m^{2}}\frac{i}{k_{1}^{2}-m^{2}}\int \frac{d^{4}k_{2}%
}{(2\pi )^{4}}\frac{i}{k_{2}^{2}-m^{2}}  \notag \\
&\rightarrow &\frac{1}{4}\frac{i\lambda ^{2}}{(16\pi )^{2}}(\ln \frac{%
M_{c}^{2}}{m^{2}}-\gamma _{\omega })\cdot \lbrack M_{c}^{2}-m^{2}(\frac{%
M_{c}^{2}}{m^{2}}-\gamma _{\omega }+1)].
\end{eqnarray}%
The corresponding counterterm diagrams are shown in $(a_{1})$ and $(a_{2})$,
and their sum gives:
\begin{eqnarray}
-iM_{(2)}^{2(a1)+(a2)} &=&\frac{1}{2}(-i\lambda )(-i\delta _{m^{2}})\int
\frac{d^{4}k}{(2\pi )^{4}}\frac{i^{2}}{(k^{2}-m^{2})^{2}}+\frac{1}{2}%
(-i\delta _{\lambda }^{t})\int \frac{d^{4}k}{(2\pi )^{4}}\frac{i}{k^{2}-m^{2}%
}  \notag \\
&\rightarrow &-\frac{1}{4}\frac{i\lambda ^{2}}{(16\pi ^{2})^{2}}%
\{[(M_{c}^{2}-\mu ^{2})-m^{2}(\ln \frac{M_{c}^{2}}{\mu ^{2}}-\gamma _{\omega
})](\ln \frac{M_{c}^{2}}{m^{2}}-\gamma _{\omega })  \notag \\
&&+(\ln \frac{M_{c}^{2}}{\mu ^{2}}-\gamma _{\omega })\cdot \lbrack
M_{c}^{2}-m^{2}(\ln \frac{M_{c}^{2}}{m^{2}}-\gamma _{\omega }+1)]\},
\end{eqnarray}%
where $\delta _{m^{2}}$ is the one-loop mass counterterm defined in Eq. (\ref%
{delta_m1}) and $\delta _{\lambda }^{t}$ only the t-channel vertex
counterterm, which is represented in Fig.13($a_{2}$). Thus, the sum of $%
(a),(a_{1})$ and $(a_{2})$ gives:
\begin{eqnarray}
-iM_{(2)}^{2(a)+(a1)+(a2)} &=&\frac{1}{4}\frac{i\lambda ^{2}}{(16\pi
^{2})^{2}}\{-(\ln \frac{M_{c}^{2}}{\mu ^{2}}-\gamma _{\omega
})[(M_{c}^{2}-\mu ^{2})-m^{2}(\ln \frac{M_{c}^{2}}{\mu ^{2}}-\gamma _{\omega
})]  \notag  \label{A_2} \\
&&+[(\mu ^{2}-m^{2})-m^{2}\ln \frac{\mu ^{2}}{m^{2}}]\ln \frac{\mu ^{2}}{%
m^{2}}\}.
\end{eqnarray}%
According to our renormalization scheme proposed in the previous subsection,
the overall two-loop counterterm for diagram (a) is defined as:
\begin{equation}
-i\delta _{m^{2}}^{(a)}=\frac{1}{4}\frac{i\lambda ^{2}}{(16\pi ^{2})^{2}}%
(\ln \frac{M_{c}^{2}}{\mu ^{2}}-\gamma _{\omega })[(M_{c}^{2}-\mu
^{2})-m^{2}(\ln \frac{M_{c}^{2}}{\mu ^{2}}-\gamma _{\omega })],
\label{A_counter}
\end{equation}%
and the renormalized correction to the two-point function from this diagram
is:
\begin{equation}
-iM_{(2)R}^{2(a)}=\frac{1}{4}\frac{i\lambda ^{2}}{(16\pi ^{2})^{2}}[(\mu
^{2}-m^{2})-m^{2}\ln \frac{\mu ^{2}}{m^{2}}]\ln \frac{\mu ^{2}}{m^{2}}.
\label{A result}
\end{equation}

Let us now compute the most complicated diagrams in $\phi ^{4}$ theory at
two loop order, namely the sunrise diagram.
According to the internal momentum parameterizations shown in the diagram $%
(b)$ and the general Feynman rules, we can write the expression explicitly:
\begin{eqnarray}
-iM_{(2)}^{2(b)} &=&\frac{1}{6}(-i\lambda )^{2}\int \frac{d^{4}k_{1}}{(2\pi
)^{4}}\int \frac{d^{4}k_{2}}{(2\pi )^{4}}\frac{i}{(k_{1}+p)^{2}-m^{2}}\frac{i%
}{k_{2}^{2}-m^{2}}\frac{i}{(k_{1}+k_{2})^{2}-m^{2}}  \notag \\
&=&\frac{i\lambda ^{2}}{6}\int \frac{d^{4}k_{1}}{(2\pi )^{4}}\int \frac{%
d^{4}k_{2}}{(2\pi )^{4}}\frac{1}{%
[(k_{1}+p)^{2}-m^{2}](k_{2}^{2}-m^{2})[(k_{1}+k_{2})^{2}-m^{2}]}.
\end{eqnarray}%
Obviously, the above integral is just the special case of $\alpha \beta
\gamma $ integral  with $\alpha =\beta =\gamma =1$ and the same mass $%
m_{i}^{2}=m^{2}$. Thus we can apply the general result Eq. (\ref{I_111 def})
to this case
\begin{eqnarray}
-iM_{(2)}^{2(b)} &\rightarrow &\frac{i\lambda ^{2}}{6(16\pi ^{2})^{2}}%
\int_{0}^{\infty }\prod_{j=1}^{3}\frac{dv_{j}}{(1+v_{j})^{2}}\delta
(1-\sum_{j=1}^{3}\frac{1}{1+v_{j}})\frac{\prod_{j=1}^{3}(1+v_{j})^{2}}{%
(3+v_{1}+v_{2}+v_{3})^{2}}  \notag \\
&&[M_{c}^{2}-\mathcal{M}^{2}(\ln \frac{M_{c}^{2}}{\mathcal{M}^{2}}-\gamma
_{\omega }+1)],
\end{eqnarray}%
with
\begin{equation*}
\mathcal{M}^{2}=m^{2}-\frac{1}{3+v_{1}+v_{2}+v_{3}}p^{2}.
\end{equation*}

We now compute the contribution to the two-point Green function, rather than
just giving the asymptotic expression for showing the cancelation of harmful
divergences, which was already demonstrated in the previous section. For
this purpose, it is useful to introduce a new set of UVDP parameters $u,v,w$
via
\begin{subequations}
\label{uw transfer}
\begin{eqnarray}
\frac{1}{1+v_{1}} &\equiv &\frac{1}{(1+u)(1+w)}, \\
\frac{1}{1+v_{2}} &\equiv &\frac{u}{1+u}, \\
\frac{1}{1+v_{3}} &\equiv &\frac{1}{(1+u)(1+v)},
\end{eqnarray}%
so the integration measure is now
\end{subequations}
\begin{eqnarray}
&&\int_{0}^{\infty }\prod_{j=1}^{3}\frac{dv_{j}}{(1+v_{j})^{2}}\delta
(1-\sum_{j=1}^{3}\frac{1}{1+v_{j}})  \notag \\
&=&\int_{0}^{\infty }\frac{du}{(1+u)^{3}}\int_{0}^{\infty }\frac{dw}{%
(1+w)^{2}}\frac{dv}{(1+v)^{2}}\delta (1-\frac{1}{1+w}-\frac{1}{1+v}),
\end{eqnarray}%
with
\begin{eqnarray}
\mathcal{M}^{2} &=&m^{2}-\frac{u}{1+u}\frac{1}{u(1+w)(1+v)+1}p^{2}, \\
F &=&\frac{(1+u)^{4}}{[u+\frac{1}{(1+w)(1+v)}]^{2}}.
\end{eqnarray}%
With the above transformation, we finally arrive at:
\begin{eqnarray}
-iM_{(2)}^{2(b)} &=&\frac{i\lambda ^{2}}{6(16\pi ^{2})^{2}}\int_{0}^{\infty
}du\int_{0}^{\infty }\frac{dw}{(1+w)^{2}}\frac{dv}{(1+v)^{2}}\delta (1-\frac{%
1}{1+w}-\frac{1}{1+v})\frac{1+u}{[u+\frac{1}{(1+w)(1+v)}]^{2}}  \notag
\label{parts} \\
&&[M_{c}^{2}-\mathcal{M}^{2}(\ln \frac{M_{c}^{2}}{\mathcal{M}^{2}}-\gamma
_{\omega }+1)].
\end{eqnarray}%
For the quadratic divergence, we can carry out the integration:
\begin{eqnarray}
-iM_{(2)quad}^{2(b)} &=&\frac{i\lambda ^{2}}{6(16\pi ^{2})^{2}}%
M_{c}^{2}\int_{0}^{\infty }\frac{dw}{(1+w)^{2}}\frac{dv}{(1+v)^{2}}\delta (1-%
\frac{1}{1+w}-\frac{1}{1+v})  \notag  \label{B_quad} \\
&&\int_{0}^{\infty }du\frac{1+u}{[u+\frac{1}{(1+w)(1+v)}]^{2}}  \notag \\
&=&\frac{i\lambda ^{2}}{6(16\pi ^{2})^{2}}M_{c}^{2}\int_{0}^{\infty }\frac{dw%
}{(1+w)^{2}}\frac{dv}{(1+v)^{2}}\delta (1-\frac{1}{1+w}-\frac{1}{1+v})
\notag \\
&&\int_{0}^{\infty }du[\frac{1}{u+\frac{1}{(1+w)(1+v)}}+\frac{1}{[u+\frac{1}{%
(1+w)(1+v)}]^{2}}(1-\frac{1}{(1+w)(1+v)})]  \notag \\
&\rightarrow &\frac{i\lambda ^{2}}{6(16\pi ^{2})^{2}}M_{c}^{2}[3(\ln \frac{%
M_{c}^{2}}{q_{o}^{2}}-\gamma _{\omega })+1]
\end{eqnarray}%
which is local when we choose the free scale $q_{o}^{2}=\mu ^{2}$.

For the logarithmic divergence part, the result is given by:
\begin{eqnarray}
-iM_{(2)log}^{2(b)} &=&-\frac{i\lambda ^{2}}{6(16\pi ^{2})^{2}}%
\int_{0}^{\infty }\frac{dw}{(1+w)^{2}}\frac{dv}{(1+v)^{2}}\delta (1-\frac{1}{%
1+w}-\frac{1}{1+v})\int_{0}^{\infty }du\frac{1+u}{[u+\frac{1}{(1+w)(1+v)}%
]^{2}}  \notag \\
&&\{m^{2}-\frac{u}{(1+u)(1+w)(1+v)[u+\frac{1}{(1+w)(1+v)}]}p^{2}\}[\ln \frac{%
M_{c}^{2}}{\mathcal{M}^{2}}-\gamma _{\omega }+1]
\end{eqnarray}%
From the analysis and discussion presented in previous sections for the
overlapping divergences, there are three parameter regions which contain
divergent contributions. To extract them, we need to separate the general
expression into several parts, each of which may give an asymptotical result
in a single region. In terms of the new set of UVDP parameters $u,v,w$, the
situation becomes much simpler than the original parameters $%
v_{1},v_{2},v_{3}$. The coefficients of the logarithmic divergence in the
above expression can be separated into the following four parts:
\begin{eqnarray}
&&\frac{1+u}{[u+\frac{1}{(1+w)(1+v)}]^{2}}\{m^{2}-\frac{u}{(1+u)(1+w)(1+v)[u+%
\frac{1}{(1+w)(1+v)}]}p^{2}\}  \notag  \label{M2} \\
&=&\frac{1}{u+\frac{1}{(1+w)(1+v)}}m^{2}+\frac{1}{[u+\frac{1}{(1+w)(1+v)}%
]^{2}}(1-\frac{1}{(1+w)(1+v)})m^{2}  \notag \\
&&-\frac{u}{(1+w)(1+v)[u+\frac{1}{(1+w)(1+v)}]^{3}}p^{2}  \notag \\
&=&I+(II+III)+IV.
\end{eqnarray}%
These four parts are divergent in the following asymptotic UVDP
parameter regions:
\begin{eqnarray}
I &:&\quad \quad u\rightarrow \infty \,\quad \quad \quad vw=1;\qquad
v_{1}\rightarrow \infty \,\ v_{3}\rightarrow \infty \,\ v_{2}\rightarrow 0
\notag \\
II &:&\quad \quad v\rightarrow \infty \,\ u\rightarrow 0\,\ w\rightarrow
0;\qquad v_{2}\rightarrow \infty \,\ v_{3}\rightarrow \infty \,\
v_{1}\rightarrow 0  \notag \\
III &:&\quad \quad w\rightarrow \infty \,\ u\rightarrow 0\,\ v\rightarrow
0;\qquad v_{2}\rightarrow \infty \,\ v_{1}\rightarrow \infty \,\
v_{3}\rightarrow 0  \notag \\
IV &:&\quad \quad p^{2}\gg m^{2},  \notag
\end{eqnarray}%
where the second part in Eq.(\ref{M2}) contains two asymptotic regions II
and III symmetric under the exchange of parameters $v$ and $w$ or $v_{1}$
and $v_{3}$. In general, it is difficult to carry out the whole integration
and obtaining a complete result for $-iM_{(2)}^{2(b)}$ due to the complicity
of $\mathcal{M}^{2}$ in the logarithm. But it is sufficient for our present
purpose to obtain $-iM_{(2)}^{2(b)}$ by simplifying $\mathcal{M}^{2}$ in the
above asymptotic regions, allowing us to get the results up to logarithmic
divergence.

\textbf{Region (I)}: $u\rightarrow \infty$. In this region, $\mathcal{M}^2$
can be simplified to the form:
\begin{eqnarray}
\mathcal{M}^2 \simeq m^2.
\end{eqnarray}
So the approximate expression is given by:
\begin{eqnarray}
-iM^{2(b)(I)}_{(2)log} &\simeq & -\frac{i\lambda^2}{6(16\pi^2)^2}
\int^\infty_0 \frac{dw}{(1+w)^2} \frac{dv}{(1+v)^2} \delta(1-\frac{1}{1+w}-%
\frac{1}{1+v})  \notag \\
&& \int^\infty_0\frac{du}{u+\frac{1}{(1+w)(1+v)}}m^2 (\ln\frac{M_c^2}{m^2}%
-\gamma_\omega+1)  \notag \\
&\to & -\frac{i\lambda^2 m^2}{6(16\pi^2)^2} (\ln\frac{M^2_c}{q_o^2}%
-\gamma_\omega+2)(\ln\frac{M^2_c}{m^2}-\gamma_\omega+1).
\end{eqnarray}
Notice that in the case $u \to \infty$, the only relevant mass scale is $m^2$%
, so we can simply take $q_o^2 = m^2$. The final expression in this region
is found to be:
\begin{equation}  \label{B_I}
-iM^{2(b)(I)}_{(2)log} \simeq -\frac{i\lambda^2}{6(16\pi^2)^2} m^2 [(\ln%
\frac{M^2_c}{m^2}-\gamma_\omega)^2 +3(\ln\frac{M_c^2}{m^2}-\gamma_\omega)].
\end{equation}
\textbf{Region (II+III)}: $v \to \infty$ $or$ $w \to \infty$. It is
interesting to note that the integral is symmetric under the exchange of
parameters $v$ and $w$. Thus taking the limit $v\to \infty$, or $w\to\infty$%
, we shall arrive at the same results. In both cases, the asymptotic form of
$\mathcal{M}^2$ is given by:
\begin{equation}
\mathcal{M}^2 \simeq m^2
\end{equation}
\begin{eqnarray}
-iM^{2(b)(II+III)}_{(2)log} &\simeq & -\frac{i\lambda^2}{6(16\pi^2)^2} m^2
\int^\infty_0 \frac{dw}{(1+w)^2} \frac{dv}{(1+v)^2} \delta(1-\frac{1}{1+w}-%
\frac{1}{1+v})  \notag \\
&&[1-\frac{1}{(1+w)(1+v)}] \int^\infty_0 \frac{du}{[u+\frac{1}{(1+w)(1+v)}]^2%
} [\ln\frac{M_c^2}{m^2}-\gamma_\omega+1]  \notag \\
&=& -\frac{i\lambda^2}{6(16\pi^2)^2} m^2 \int^\infty_0 \frac{dw}{(1+w)^2}
\frac{dv}{(1+v)^2} \delta(1-\frac{1}{1+w}-\frac{1}{1+v})  \notag \\
&&[1-\frac{1}{(1+w)(1+v)}](1+w)(1+v)(\ln\frac{M_c^2}{m^2}-\gamma_\omega+1)
\notag \\
&=& -\frac{i\lambda^2}{6(16\pi^2)^2} m^2 [\int^\infty_0\frac{dw}{1+w}%
+\int^\infty_0 \frac{dv}{1+v}-1](\ln\frac{M_c^2}{m^2}-\gamma_\omega+1)
\notag \\
&\to & -\frac{i\lambda^2}{6(16\pi^2)^2} m^2 [2(\ln\frac{M_c^2}{q_o^2}%
-\gamma_\omega)-1](\ln\frac{M_c^2}{m^2}-\gamma_\omega+1),
\end{eqnarray}
where in the third equality, we have used the constraint in the delta
function $(1+w)(1+v)=(1+v)+(1+w)$ to simplify the integral into the form of
two 1-loop ILIs which can be regularized by the LORE method as shown
previously. Again the only mass scale in the limit $v\to \infty$ or $w\to
\infty$ is the mass of the particle $m^2$, so the scale $q_o^2$ can be fixed
to be $m^2$ and the result is given by:
\begin{equation}  \label{B_II}
-iM^{2(b)(II+III)}_{2log} \simeq -\frac{i\lambda^2}{6(16\pi^2)^2} m^2 [2(\ln%
\frac{M_c^2}{m^2}-\gamma_\omega)^2+(\ln\frac{M_c^2}{m^2}-\gamma_\omega)].
\end{equation}

Note that in $-iM_{(2)log}^{2(b)(I)}$ and $-iM_{(2)log}^{2(b)(II+III)}$,
there are three logarithmic divergences hidden in the UVDP parameter space,
and they reproduce the corresponding subdivergences in the subdiagrams of
Fig.13(b). This feature was already anticipated by the electric circuits
analogy of Feynman diagrams discussed in section III. However, when adopting
a different set of UVDP parameters $u,v,w$ transformed from the ones $%
v_{1},v_{2},v_{3}$, the divergence regions in the parameter space are also
changed correspondingly.

\textbf{Region (IV): $-p^2\gg m^2$}. In this region, we obtain the first
order correction to the wave function renormalization in the $\phi^4$
theory. Clearly, there is no harmful divergence in this region as all the
integrals of UVDP parameters are convergent. When $-p^2\gg m^2$, we can
ignore all the terms proportional to $m^2$ in $\mathcal{M}^2$ and the
integral can be simplified to:
\begin{eqnarray}  \label{B_III}
-iM^{2(b)(IV)}_{(2)log} &\simeq & -\frac{i\lambda^2}{6(16\pi^2)^2}(-p^2)
\int^\infty_0 \frac{dw}{(1+w)^3} \frac{dv}{(1+v)^3} \delta(1-\frac{1}{1+w}-%
\frac{1}{1+v})  \notag \\
&& \int^\infty_0 \frac{du u}{[u+\frac{1}{(1+w)(1+v)}]^3} [\ln\frac{M_c^2}{%
\frac{u}{(1+u)[u(1+w)(1+v)+1]}(-p^2)}-\gamma_\omega+1]  \notag \\
&=& \frac{i\lambda^2}{6(16\pi^2)^2}p^2[\frac{1}{2}(\ln\frac{M_c^2}{-p^2}%
-\gamma_\omega+1)  \notag \\
&&+\frac{1}{108}(-81-2~\psi^{(1)}(\frac{1}{6})-2~\psi^{(1)}(\frac{1}{3}%
)+2~\psi^{(1)}(\frac{2}{3}) +2~\psi^{(1)}(\frac{5}{6}))],
\end{eqnarray}
where $\psi^{(1)}(z)\equiv \frac{d^{2}}{d z^{2}}\ln \Gamma(z)$ is the
polygamma function of order 1.

Adding up all the contributions Eqs.(\ref{B_quad}),(\ref{B_I})(\ref{B_II}),(%
\ref{B_III}), we arrive at the final result for the divergent contributions
of the sunrise diagram Fig.13(b)
\begin{eqnarray}
-iM_{(2)}^{2(b)} &\simeq &\frac{i\lambda ^{2}}{6(16\pi ^{2})^{2}}\{[3(\ln
\frac{M_{c}^{2}}{\mu ^{2}}-\gamma _{\omega })+1]M_{c}^{2}-3m^{2}(\ln \frac{%
M_{c}^{2}}{m^{2}}-\gamma _{\omega })^{2}-4m^{2}(\ln \frac{M_{c}^{2}}{m^{2}}%
-\gamma _{\omega })  \notag  \label{B_2} \\
&&+\frac{1}{2}p^{2}(\ln \frac{M_{c}^{2}}{-p^{2}}-\gamma _{\omega })\}.
\end{eqnarray}%
The counterterm diagram for Fig.13(b) is shown in Fig. 13($b_{1,2}$) and the
result is simply given by:
\begin{eqnarray}
-iM_{(2)}^{2(b_{1})+(b_{2})} &=&\frac{1}{2}(-i\delta _{\lambda }^{s+u})\int
\frac{d^{4}k}{(2\pi )^{4}}\frac{i}{k^{2}-m^{2}}  \notag  \label{B_2(1)} \\
&\rightarrow &-\frac{i\lambda ^{2}}{2(16\pi ^{2})^{2}}(\ln \frac{M_{c}^{2}}{%
\mu ^{2}}-\gamma _{\omega })[M_{c}^{2}-m^{2}(\ln \frac{M_{c}^{2}}{m^{2}}%
-\gamma _{\omega }+1)].
\end{eqnarray}%
Note that we have used the vertex counterterm insertion of $s$ and $u$%
-channels, so there is a factor of 2 in the above calculation. By summing up
Figs. 13 (b) and ($b_{1}$), we obtain:
\begin{eqnarray}
-iM_{(2)}^{2(b)+(b_{1})+(b_{2})} &=&\frac{i\lambda ^{2}}{(16\pi ^{2})^{2}}\{[%
\frac{1}{6}(M_{c}^{2}-\mu ^{2})-\frac{1}{6}m^{2}(\ln \frac{M_{c}^{2}}{\mu
^{2}}-\gamma _{\omega })+\frac{1}{12}p^{2}(\ln \frac{M_{c}^{2}}{\mu ^{2}}%
-\gamma _{\omega })]  \notag \\
&&+[\frac{1}{6}(\mu ^{2}-m^{2})-\frac{1}{2}m^{2}(\ln \frac{\mu ^{2}}{m^{2}}%
)^{2}-\frac{2}{3}m^{2}\ln \frac{\mu ^{2}}{m^{2}}+\frac{1}{12}p^{2}\ln \frac{%
\mu ^{2}}{-p^{2}}]\}+...  \notag \\
&&
\end{eqnarray}%
As it is expected, the potentially harmful divergences $m^{2}(\ln \frac{%
M_{c}^{2}}{\mu ^{2}}-\gamma _{\omega })\ln \frac{\mu ^{2}}{m^{2}}$ cancel
exactly.

By considering the following overall counterterms for diagram (b):
\begin{equation} \label{B_counter}
i(p^{2}\delta _{Z}^{(b)}-\delta _{m^{2}}^{(b)})=-\frac{i\lambda ^{2}}{(16\pi
^{2})^{2}}[\frac{1}{6}(M_{c}^{2}-\mu ^{2})-\frac{1}{6}m^{2}(\ln \frac{%
M_{c}^{2}}{\mu ^{2}}-\gamma _{\omega })+\frac{1}{12}p^{2}(\ln \frac{M_{c}^{2}%
}{\mu ^{2}}-\gamma _{\omega })], 
\end{equation}%
we get the final contribution to the two-loop self-energy as shown in
Fig.13(b):
\begin{equation}
-iM_{(2)R}^{2(b)}=\frac{i\lambda ^{2}}{(16\pi ^{2})^{2}}[\frac{1}{6}(\mu
^{2}-m^{2})-\frac{1}{2}m^{2}(\ln \frac{\mu ^{2}}{m^{2}})^{2}-\frac{2}{3}%
m^{2}\ln \frac{\mu ^{2}}{m^{2}}+\frac{1}{12}p^{2}\ln \frac{\mu ^{2}}{-p^{2}}%
]+...
\end{equation}

We are now in the position to put all the results from diagrams Fig.13(a)
and Fig.13(b) together, and obtain the total contributions to the two-loop
self-energy,
\begin{eqnarray}
-iM_{(2)R}^{2} &=&\frac{i\lambda ^{2}}{(16\pi ^{2})^{2}}[\frac{1}{4}(\mu
^{2}-m^{2})\ln \frac{\mu ^{2}}{m^{2}}+\frac{1}{6}(\mu ^{2}-m^{2})  \notag \\
&&-\frac{3}{4}m^{2}(\ln \frac{\mu ^{2}}{m^{2}})^{2}-\frac{11}{12}m^{2}\ln
\frac{\mu ^{2}}{m^{2}}+\frac{1}{12}p^{2}\ln \frac{\mu ^{2}}{-p^{2}}]+...
\end{eqnarray}%
Considering the massless limit $m^{2}\rightarrow 0$ and ignoring the
quadratic contribution $\mu ^{2}\rightarrow 0$, one arrives at
\begin{equation}
-iM_{(2)R}^{2}=\frac{i\lambda ^{2}}{12(16\pi ^{2})^{2}}p^{2}\ln \frac{\mu
^{2}}{-p^{2}},
\end{equation}%
which agrees with the one obtained by using the standard dimensional
regularization method(see page 345 of the book\cite{Peskin:1995ev}).

\subsection{Vertex Contribution at Two Loop}

\begin{figure}[ht]
\begin{center}
\includegraphics[scale=0.8]{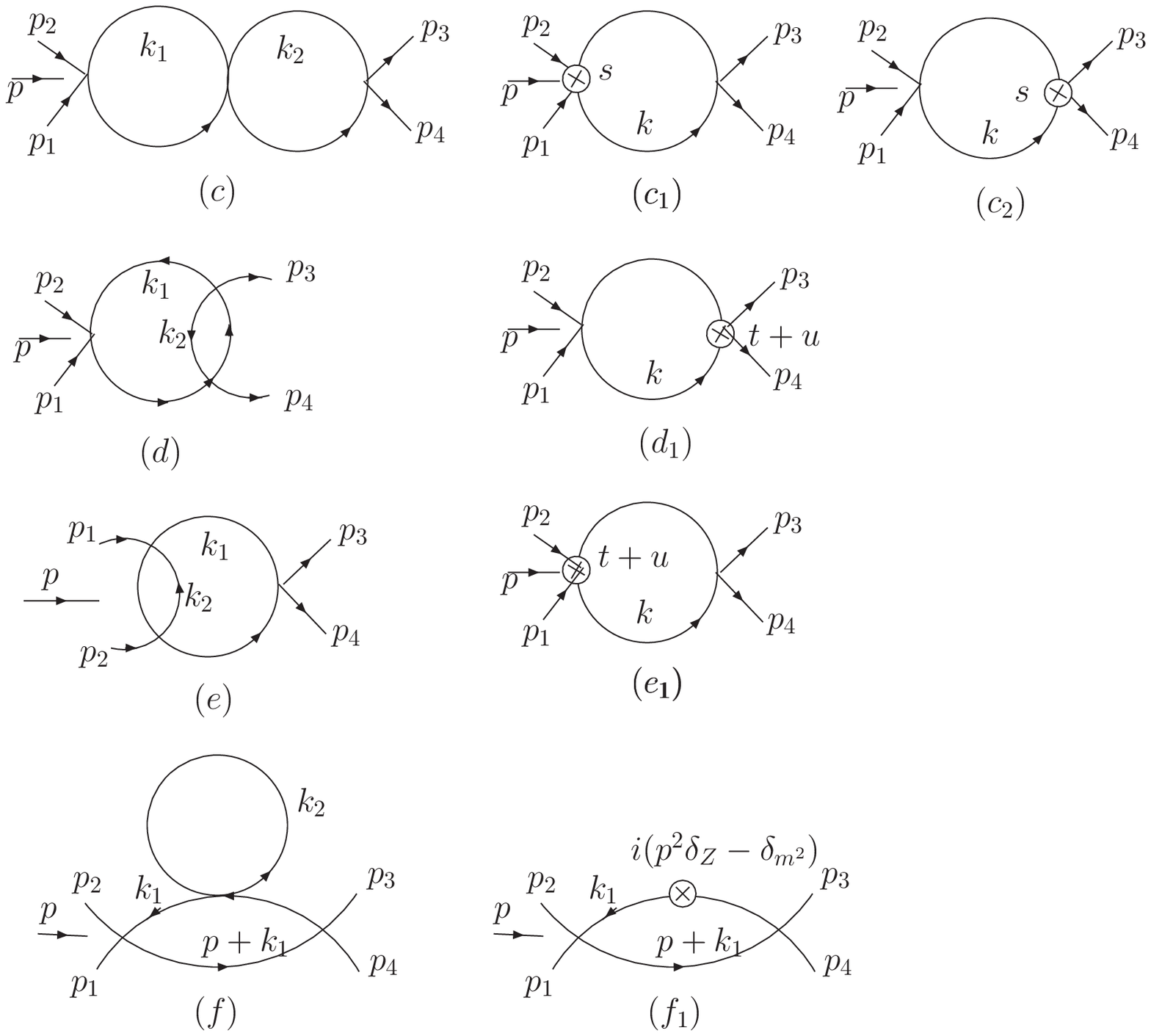}\\[0pt]
\end{center}
\caption{}
\label{cde}
\end{figure}

The two-loop vertex contribution for the s-channel in $\phi^4$ theory
involves four groups of diagrams as shown in Fig. (\ref{cde}).
It is expected that all harmful divergences cancel separately within each
group. As the groups of diagrams (d) and (e) are related by a simple
interchange of initial and final momenta, they should give the same results
and it only needs to calculate either group and then to multiply by a factor
of 2.

First, let us calculate the simplest two-loop diagram (f) and its
counterterm diagram (f1) in Fig.14. Since the only UV divergence in (f) can
be completely canceled by that in (f1), the group (f) does not require
two-loop overall counterterm.
\begin{eqnarray}
-i\Lambda _{(2)}^{(f)} &=&\frac{(-i\lambda )^{3}}{2}\int \frac{d^{4}k_{2}}{%
(2\pi )^{4}}\frac{i}{k_{2}^{2}-m^{2}}\int \frac{d^{4}k_{1}}{(2\pi )^{4}}%
\frac{i^{3}}{(k_{1}^{2}-m^{2})^{2}[(k_{1}+p)^{2}-m^{2}]}  \notag \\
&\rightarrow &\frac{i\lambda ^{3}}{2}\frac{-i}{16\pi ^{2}}%
[M_{c}^{2}-m^{2}(\ln \frac{M_{c}^{2}}{m^{2}}-\gamma _{\omega }+1)]\frac{%
\Gamma (3)}{\Gamma (2)\Gamma (1)}\int_{0}^{1}dx(1-x)  \notag \\
&&\int \frac{d^{4}k_{1}}{(2\pi )^{4}}\frac{1}{%
\{[(1-x)(k_{1}^{2}-m^{2})+x[(k_{1}+p)^{2}-m^{2}]\}^{3}}  \notag \\
&=&-\frac{i\lambda ^{3}}{2(16\pi ^{2})^{2}}[M_{c}^{2}-m^{2}(\ln \frac{%
M_{c}^{2}}{m^{2}}-\gamma _{\omega }+1)]\int_{0}^{1}dx\frac{1-x}{%
m^{2}-x(1-x)p^{2}},
\end{eqnarray}%
\begin{eqnarray}
-i\Lambda _{(2)}^{(f1)} &=&i(p^{2}\delta _{Z}^{(1)}-\delta
_{m^{2}}^{(1)})(-i\lambda )^{2}\int \frac{d^{4}k_{1}}{(2\pi )^{4}}\frac{i^{3}%
}{(k_{1}^{2}-m^{2})^{2}[(k_{1}+p)^{2}-m^{2}]}  \notag \\
&=&\frac{i\lambda ^{3}}{2(16\pi ^{2})^{2}}[(M_{c}^{2}-\mu ^{2})-m^{2}(\ln
\frac{M_{c}^{2}}{\mu ^{2}}-\gamma _{\omega })]\int_{0}^{1}dx\frac{1-x}{%
m^{2}-x(1-x)p^{2}},
\end{eqnarray}%
\begin{equation}
-i\Lambda _{(2)}^{(f)+(f1)}=-\frac{i\lambda ^{3}}{2(16\pi ^{2})^{2}}[(\mu
^{2}-m^{2})-m^{2}\ln \frac{\mu ^{2}}{m^{2}}]\int_{0}^{1}dx\frac{1-x}{%
m^{2}-x(1-x)p^{2}}.
\end{equation}

The computation of diagram (c) and its counterterm diagrams $(c1)$ and $(c2)$
in Fig.14 is also straightforward, as these diagrams can be factored into a
product of two one-loop integrals. The result is:
\begin{eqnarray}
-i\Lambda _{(2)}^{(c)} &=&\frac{1}{4}(-i\lambda )^{3}[\int \frac{d^{4}k}{%
(2\pi )^{4}}\frac{i}{k^{2}-m^{2}}\frac{i}{(k+p)^{2}-m^{2}}]^{2}  \notag \\
&=&\frac{i\lambda ^{3}}{4}[\int \frac{d^{4}k}{(2\pi )^{4}}\int_{0}^{1}dx%
\frac{1}{\{(1-x)(k^{2}-m^{2})+x[(k+p)^{2}-m^{2}]\}^{2}}]^{2}  \notag \\
&=&-\frac{i\lambda ^{3}}{4(16\pi ^{2})^{2}}[\int_{0}^{1}dx(\ln \frac{%
M_{c}^{2}}{m^{2}-x(1-x)p^{2}}-\gamma _{\omega })]^{2},
\end{eqnarray}%
and
\begin{eqnarray}
-i\Lambda _{(2)}^{(c1)+(c2)} &=&2\cdot \frac{1}{2}(-i\delta _{\lambda
}^{s})(-i\lambda )\int \frac{d^{4}k}{(2\pi )^{4}}\frac{i}{k^{2}-m^{2}}\frac{i%
}{(k+p)^{2}-m^{2}}  \notag \\
&=&\frac{i\lambda ^{3}}{2(16\pi ^{2})^{2}}(\ln \frac{M_{c}^{2}}{\mu ^{2}}%
-\gamma _{\omega })[\int_{0}^{1}dx(\ln \frac{M_{c}^{2}}{m^{2}-x(1-x)p^{2}}%
-\gamma _{\omega })],
\end{eqnarray}%
where in the calculation of counterterm diagrams, the factor 2 in the first
line accounts for the two equal diagrams (c1) and (c2). By adding up the
above results, we have:
\begin{equation*}
-i\Lambda _{(2)}^{(c)+(c1)+(c2)}=-\frac{i\lambda ^{3}}{4(16\pi ^{2})^{2}}%
[(\int_{0}^{1}dx\ln \frac{\mu ^{2}}{m^{2}-x(1-x)p^{2}})^{2}-(\ln \frac{%
M_{c}^{2}}{\mu ^{2}}-\gamma _{\omega })^{2}].
\end{equation*}%
By considering the two-loop overall counterterm for diagram (c):
\begin{equation}
-i\delta _{\lambda }^{(2)(c)}=-\frac{i\lambda ^{3}}{4(16\pi ^{2})^{2}}(\ln
\frac{M_{c}^{2}}{\mu ^{2}}-\gamma _{\omega })^{2},  \label{C_counter}
\end{equation}%
we obtain the renormalized contribution from the group (c) of diagrams:
\begin{equation}
-i\Lambda _{(2)R}^{(c)}=-\frac{i\lambda ^{3}}{4(16\pi ^{2})^{2}}%
[\int_{0}^{1}dx\ln \frac{\mu ^{2}}{m^{2}-x(1-x)p^{2}}]^{2}.  \label{C_result}
\end{equation}

For diagram (d) in Fig.(\ref{cde}), due to the complicated dependence on
external momenta, we shall focus on a simplified situation where only the
s-channel contributes. Also, we only keep the leading divergent
contributions, namely the $\log \cdot \log $ and $\log $ terms, because $%
\mathcal{M}^{2}$ can be simplified in such a subdivergence region. To see
the momentum dependence in Fig. (\ref{cde}), we write explicitly the
expression for diagram (d):
\begin{eqnarray}
-i\Lambda _{(2)}^{(d)} &=&\frac{1}{2}(-i\lambda )^{3}\int \frac{d^{4}k_{1}}{%
(2\pi )^{4}}\int \frac{d^{4}k_{2}}{(2\pi )^{4}}\frac{i}{k_{1}^{2}-m^{2}}%
\frac{i}{(k_{1}+p)^{2}-m^{2}}\frac{i}{k_{2}^{2}-m^{2}}\frac{i}{%
(k_{2}+k_{1}+p_{3})^{2}-m^{2}}  \notag \\
&=&\frac{i\lambda ^{3}}{2}\int \frac{d^{4}k_{1}}{(2\pi )^{4}}\int \frac{%
d^{4}k_{2}}{(2\pi )^{4}}\int_{0}^{1}dx\frac{1}{%
[x(k_{1}+p)^{2}+(1-x)k_{1}^{2}-m^{2}]^{2}}\cdot   \notag \\
&&\frac{1}{[k_{2}^{2}-m^{2}][(k_{2}+k_{1}+p_{3})^{2}-m^{2}]}  \notag \\
&=&\frac{i\lambda ^{3}}{2}\int_{0}^{1}dx\int \frac{d^{4}k_{1}}{(2\pi )^{4}}%
\int \frac{d^{4}k_{2}}{(2\pi )^{4}}\frac{1}{[k_{1}^{2}+x(1-x)p^{2}-m^{2}]^{2}%
}  \notag \\
&&\frac{1}{[k_{2}^{2}-m^{2}][(k_{2}+k_{1}+p_{3}-xp)^{2}-m^{2}]}.
\end{eqnarray}%
Note that in the process to obtain the second and third equalities, we have
transformed the original integration into the general $\alpha \beta \gamma $
integral with $\alpha =\gamma =1,\beta =2$ and $%
m_{1}^{2}=m_{3}^{2}=m^{2},m_{2}^{2}=m^{2}-x(1-x)p^{2}$ by adopting the usual
Feynman parametrization and making the translation of variable $%
k_{1}\rightarrow k_{1}-xp$. Thus, the general formulae Eq. (\ref{abc int_ed2}%
) gives,
\begin{eqnarray}
-i\Lambda _{(2)}^{(d)} &=&-\frac{\lambda ^{3}}{2\cdot 16\pi ^{2}}%
\int_{0}^{1}dx\int_{0}^{\infty }\prod_{j=1}^{3}dv_{j}\delta (1-\sum_{j=1}^{3}%
\frac{1}{1+v_{j}})  \notag \\
&&\frac{1}{(1+v_{2})(3+v_{1}+v_{2}+v_{3})^{2}}\int \frac{d^{4}l}{(2\pi )^{4}}%
\frac{1}{[l^{2}-\mathcal{M}^{2}]^{2}}  \notag \\
&\rightarrow &-\frac{i\lambda ^{3}}{2\cdot (16\pi ^{2})^{2}}%
\int_{0}^{1}dx\int_{0}^{\infty }\prod_{j=1}^{3}dv_{j}\delta (1-\sum_{j=1}^{3}%
\frac{1}{1+v_{j}})  \notag \\
&&\frac{1}{(1+v_{2})(3+v_{1}+v_{2}+v_{3})^{2}}(\ln \frac{M_{c}^{2}}{\mathcal{%
M}^{2}}-\gamma _{\omega }),
\end{eqnarray}%
with
\begin{equation*}
\mathcal{M}^{2}=m^{2}-\frac{x(1-x)p^{2}}{1+v_{2}}-\frac{(p_{3}-xp)^{2}}{%
3+v_{1}+v_{2}+v_{3}}.
\end{equation*}%
In order to carry out the above integral, we transform the UVDP parameters $%
v_{1},v_{2},v_{3}$ to the new set $u,v,w$ as shown in Eq. (\ref{uw transfer}%
), so that the form of $-i\Lambda _{(2)}^{(d)}$ is changed to:
\begin{eqnarray}
-i\Lambda _{(2)}^{(d)} &=&-\frac{i\lambda ^{3}}{2\cdot (16\pi ^{2})^{2}}%
\int_{0}^{1}dx\int_{0}^{\infty }\frac{dw}{(1+w)^{2}}\frac{dv}{(1+v)^{2}}%
\delta (1-\frac{1}{1+w}+\frac{1}{1+v})  \notag  \label{lmd2} \\
&&\int_{0}^{\infty }du\frac{u}{[u+\frac{1}{(1+w)(1+v)}]^{2}}[\ln \frac{%
M_{c}^{2}}{\mathcal{M}^{2}}-\gamma _{\omega }],
\end{eqnarray}%
where $\mathcal{M}^{2}$ is given by :
\begin{equation*}
\mathcal{M}^{2}=m^{2}-\frac{u}{u+1}x(1-x)p^{2}-\frac{u}{u+1}\cdot \frac{%
(p_{3}-xp)^{2}}{u(1+w)(1+v)+1}.
\end{equation*}%
Notice that in Eq. (\ref{lmd2}) only the integration over $u$ is
logarithmically divergent, while the ones over $w,v$ convergent, so we can
make the following approximation for $\mathcal{M}^{2}$ in the limit $%
u\rightarrow \infty $
\begin{equation}
\mathcal{M}^{2}\simeq m^{2}-x(1-x)p^{2},
\end{equation}%
and the integration can be performed as follows
\begin{eqnarray}
-i\Lambda _{(2)}^{(d)} &\simeq &-\frac{i\lambda ^{3}}{2\cdot (16\pi ^{2})^{2}%
}\int_{0}^{1}dx\int_{0}^{\infty }\frac{dw}{(1+w)^{2}}\frac{dv}{(1+v)^{2}}%
\delta (1-\frac{1}{1+w}+\frac{1}{1+v})  \notag \\
&&\int_{0}^{\infty }du\frac{u}{[u+\frac{1}{(1+w)(1+v)}]^{2}}[\ln \frac{%
M_{c}^{2}}{m^{2}-x(1-x)p^{2}}-\gamma _{\omega }]  \notag \\
&\rightarrow &-\frac{i\lambda ^{3}}{2\cdot (16\pi ^{2})^{2}}%
\int_{0}^{1}dx(\ln \frac{M_{c}^{2}}{q_{o}^{2}}-\gamma _{\omega }+1)(\ln
\frac{M_{c}^{2}}{m^{2}-x(1-x)p^{2}}-\gamma _{\omega }),
\end{eqnarray}%
where the mass scale $q_{o}^{2}$ is taken to be $m^{2}-x(1-x)p^{2}$ as the
only scale in the limit $u\rightarrow \infty $ is $\mathcal{M}^{2}\simeq
m_{2}^{2}=m^{2}-x(1-x)p^{2}$. Thus we can write down the regularized
expression for the diagram (d) as:
\begin{eqnarray}
-i\Lambda _{(2)}^{(d)} &\simeq &-\frac{i\lambda ^{3}}{2\cdot (16\pi ^{2})^{2}%
}\int_{0}^{1}dx[(\ln \frac{M_{c}^{2}}{m^{2}-x(1-x)p^{2}}-\gamma _{\omega
})^{2}  \notag  \label{D} \\
&&+(\ln \frac{M_{c}^{2}}{m^{2}-x(1-x)p^{2}}-\gamma _{\omega })].
\end{eqnarray}%
The counterterm diagram ($d_{1}$) in Fig. 14 can also be simply calculated:
\begin{eqnarray}
-i\Lambda _{(2)}^{(d1)} &=&\frac{1}{2}(-i\lambda )(-i\delta _{\lambda
}^{t+u})\int \frac{d^{4}k}{(2\pi )^{4}}\frac{i}{k^{2}-m^{2}}\frac{i}{%
(k+p)^{2}-m^{2}}  \notag  \label{D_1} \\
&=&\frac{i\lambda ^{3}}{2(16\pi ^{2})^{2}}\int_{0}^{1}dx(\ln \frac{M_{c}^{2}%
}{\mu ^{2}}-\gamma _{\omega })(\ln \frac{M_{c}^{2}}{m^{2}-x(1-x)p^{2}}%
-\gamma _{\omega }),
\end{eqnarray}%
where $\delta _{\lambda }^{t+u}$ accounts for the $t-$ and $u-$ channels in
the subdiagram. By combining the above two equations Eq.(\ref{D}) and (\ref%
{D_1}), we have
\begin{equation}
-i\Lambda _{(2)}^{(d)+(d1)}\simeq -\frac{i\lambda ^{3}}{2(16\pi ^{2})^{2}}%
\int_{0}^{1}dx[(\ln \frac{\mu ^{2}}{m^{2}-x(1-x)p^{2}})^{2}+(\ln \frac{%
M_{c}^{2}}{m^{2}-x(1-x)p^{2}}-\gamma _{\omega })].  \notag  \label{D+D_1}
\end{equation}%
Write the two-loop overall counterterm for the diagram (d) as
\begin{equation}
-i\delta _{\lambda }^{(2)(d)}=\frac{i\lambda ^{3}}{2(16\pi ^{2})^{2}}(\ln
\frac{M_{c}^{2}}{\mu ^{2}}-\gamma _{\omega }),  \label{D_counter}
\end{equation}%
then the two-loop vertex contribution of the diagram (d) is,
\begin{equation}
-i\Lambda _{(2)R}^{(d)}=-\frac{i\lambda ^{3}}{2(16\pi ^{2})^{2}}%
\int_{0}^{1}dx[(\ln \frac{\mu ^{2}}{m^{2}-x(1-x)p^{2}})^{2}+\ln \frac{\mu
^{2}}{m^{2}-x(1-x)p^{2}}]+...
\end{equation}%
Obviously, the diagram (e) and its counterterm diagram (e1) in Fig.14 gives
the same result as diagram (d) and (d1).

By summing up the renormalized results of the diagrams (c), (d), (e) and (f)
in Fig.(\ref{cde}), we finally obtain the s-channel two-loop correction for
the four-point function:
\begin{eqnarray}
-i\Lambda^{(s)}_{(2)R} &=& -\frac{i\lambda^3}{(16\pi^2)^2}\{\frac{1}{4}%
[\int^1_0 dx \ln\frac{\mu^2}{m^2-x(1-x)p^2}]^2  \notag \\
&& +\int^1_0dx[ (\ln\frac{\mu^2}{m^2-x(1-x)p^2})^2+\ln\frac{\mu^2}{%
m^2-x(1-x)p^2}]  \notag \\
&& +\frac{1}{2} [(\mu^2-m^2)-m^2\ln\frac{\mu^2}{m^2}]\int^1_0 dx \frac{1-x}{%
m^2-x(1-x)p^2}\},
\end{eqnarray}
where $-p^2=-(p_1+p_2)^2\equiv s$ for s-channel.

\begin{figure}[th]
\begin{center}
\includegraphics[scale=0.8]{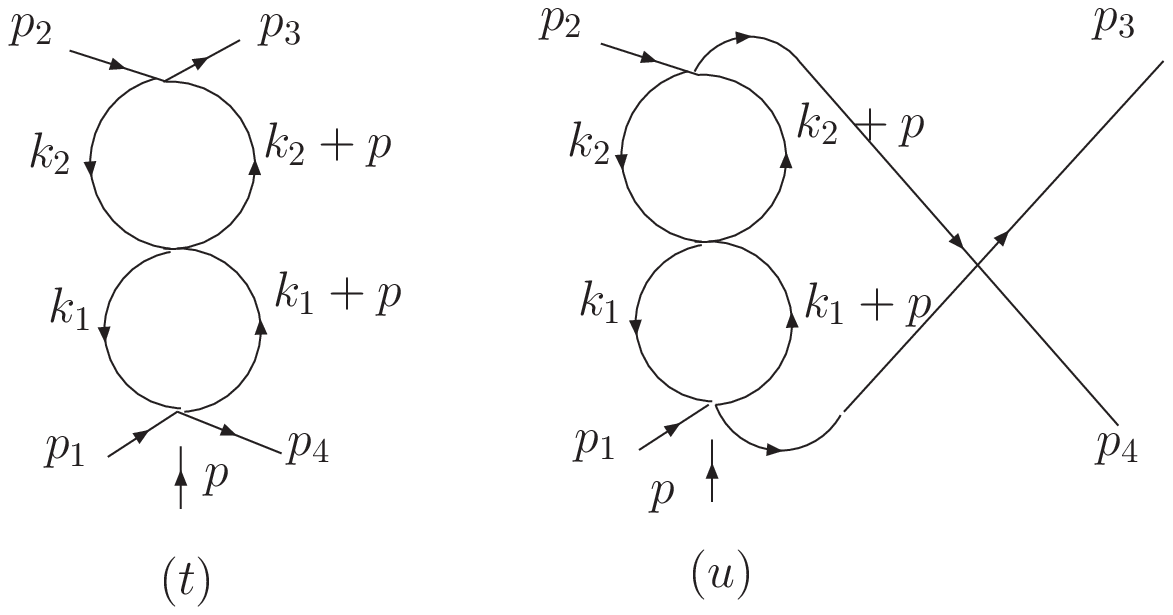}
\end{center}
\caption{{}}
\label{ctu}
\end{figure}

For completeness, we also have to consider $t$- and $u$-channel
contributions, which will give us the similar expressions except for the
substitution of $p^{2}$: $-p^{2}=-(p_{1}-p_{4})^{2}\equiv t$ for $t-$channel
and $-p^{2}=-(p_{1}-p_{3})^{2}\equiv u$ for u-channel. In Fig.(\ref{ctu}),
we only present counterpart of diagram (c) for t- and u-channels, and other
diagrams in groups (c), (d), (e) and (f) could be obtained by the same
change of external legs.

As a consistency check for the above results, let us consider the situation
with massless limit $m\rightarrow 0$ and $s\rightarrow \infty $ but keeping $%
t$ fixed. From the identity $s+t+u=0$, which indicates $u\simeq
-s\rightarrow -\infty $, one needs to consider u-channel as well. In this
case, the four-point vertex correction is found to be:
\begin{eqnarray}
-i\Lambda _{(2)R}^{(s)+(u)} &\simeq &-\frac{i\lambda ^{3}}{(16\pi ^{2})^{2}}[%
\frac{5}{4}(\ln \frac{\mu ^{2}}{s})^{2}+\frac{5}{4}(\ln \frac{\mu ^{2}}{u}%
)^{2}]  \notag \\
&\simeq &-\frac{5}{2}\cdot \frac{i\lambda ^{3}}{(16\pi ^{2})^{2}}(\ln \frac{%
\mu ^{2}}{s})^{2}
\end{eqnarray}%
where in the last equality we only keep the leading external momentum $%
s=-p^{2}\simeq u$ dependence when the external momentum $s$ is large. Such a
result agrees with the one given in the book\cite{Peskin:1995ev}(on page
345).

\subsection{Two Loop $\protect\beta$ Functions and Anomalous Mass Dimension
in $\protect\phi^4$ Theory}

Having calculated the divergence behavior of all the two-loop diagrams in
the $\phi^4$ theory, we are now ready to obtain two-loop $\beta$ functions.
From Eqs. (\ref{A_counter}) and (\ref{B_counter}), we can extract the
two-loop mass and wave function counterterms respectively:
\begin{eqnarray}
-i\delta^{(2)}_{m^2} &=& \frac{i\lambda^2}{(16\pi^2)^2}[ \frac{1}{4}%
(M_c^2-\mu^2)(\ln\frac{M_c^2}{\mu^2}-\gamma_\omega)-\frac{1}{6}(M_c^2-\mu^2)
\notag \\
&&-\frac{1}{4}m^2(\ln\frac{M_c^2}{\mu^2}-\gamma_\omega)^2 +\frac{1}{6}m^2(\ln%
\frac{M_c^2}{\mu^2}-\gamma_\omega)],  \label{delta_m2} \\
i\delta^{(2)}_{Z}&=& -\frac{i\lambda^2}{12(16\pi^2)^2}(\ln\frac{M_c^2}{\mu^2}%
-\gamma_\omega).  \label{delta_z2}
\end{eqnarray}
And the two-loop vertex counterterm can be extracted from Eqs. (\ref%
{C_counter}) and (\ref{D_counter}):
\begin{eqnarray}  \label{delta_lmd2}
-i\delta^{(2)}_{\lambda} &=&
3\cdot(-i\delta^{(c)}_\lambda-i\delta^{(d)}_\lambda-i\delta^{(e)}_\lambda)
\notag \\
&=& \frac{i\lambda^3}{(16\pi^2)^2}[-\frac{3}{4}(\ln\frac{M_c^2}{\mu^2}%
-\gamma_\omega)^2 +3(\ln\frac{M_c^2}{\mu^2}-\gamma_\omega)],
\end{eqnarray}
where the factor 3 in the first line accounts for the s, t, u-channels
respectively.

Recall the relation between renormalized coupling constant $\lambda $ and
the bare one $\lambda _{0}$ is:
\begin{eqnarray}
\lambda  &=&\lambda _{0}Z_{\phi }^{2}-\delta _{\lambda }=\lambda
_{0}(1+\delta _{Z})^{2}-\delta _{\lambda }  \notag \\
&\approx &\lambda _{0}(1+2\delta _{Z})-\delta _{\lambda },
\end{eqnarray}%
where $\delta _{Z}$ and $\delta _{\lambda }$ are the function of the bare
coupling $\lambda _{0}$, which is independent of scale $\mu $. In the
perturbative calculation of $\lambda $ at two-loop level, we have,
\begin{eqnarray}
\lambda  &\approx &\lambda _{0}-2\cdot \frac{\lambda _{0}^{3}}{12(16\pi
^{2})^{2}}(\ln \frac{M_{c}^{2}}{\mu ^{2}}-\gamma _{\omega })  \notag \\
&&-\frac{3\lambda _{0}^{2}}{2\cdot (16\pi ^{2})}(\ln \frac{M_{c}^{2}}{\mu
^{2}}-\gamma _{\omega })-\frac{\lambda _{0}^{3}}{(16\pi ^{2})^{2}}[\frac{3}{4%
}(\ln \frac{M_{c}^{2}}{\mu ^{2}}-\gamma _{\omega })^{2}-3(\ln \frac{M_{c}^{2}%
}{\mu ^{2}}-\gamma _{\omega })]  \notag \\
&=&\lambda _{0}-\frac{3\lambda _{0}^{2}}{2\cdot (16\pi ^{2})}(\ln \frac{%
M_{c}^{2}}{\mu ^{2}}-\gamma _{\omega })-\frac{\lambda _{0}^{3}}{(16\pi
^{2})^{2}}[\frac{3}{4}(\ln \frac{M_{c}^{2}}{\mu ^{2}}-\gamma _{\omega })^{2}-%
\frac{17}{6}(\ln \frac{M_{c}^{2}}{\mu ^{2}}-\gamma _{\omega })].
\end{eqnarray}%
Thus according to the definition of $\beta $-function which is supposed to
sum up all the leading logarithmic terms (ignoring the logarithmic-squared
term), we arrive at the $\beta $-function for the renormalized coupling
constant $\lambda $ as:
\begin{eqnarray}
\beta _{\lambda } &=&\mu \frac{d\lambda }{d\mu }  \notag \\
&=&\frac{3\lambda _{0}^{2}}{16\pi ^{2}}-\frac{2\lambda _{0}^{3}}{(16\pi
^{2})^{2}}\frac{17}{6}  \notag \\
&\approx &\frac{3\lambda ^{2}}{16\pi ^{2}}-\frac{17}{3}\frac{\lambda ^{3}}{%
(16\pi ^{2})^{2}},
\end{eqnarray}%
where the bare constant $\lambda _{0}$ has been replaced in the last line by
its renormalized one, leading to the standard result $\beta _{\lambda }.$
\cite{Brezin,Chetyrkin,Dittes:1977aq}.

Similarly, we can evaluate the anomalous mass dimension at two-loop level.
From the definition of the renormalized mass:
\begin{equation}
m^{2}=Z_{\phi }m_{0}^{2}-\delta _{m^{2}}=m_{0}^{2}+m_{0}^{2}\delta
_{Z}-\delta _{m^{2}},
\end{equation}%
we have the following approximate relation for the renormalized $m^{2}$
given in terms of bare mass $m_{0}^{2}$ and the bare coupling constant $%
\lambda _{0}$ at two-loop level:
\begin{eqnarray}
m^{2} &=&m_{0}^{2}+\frac{\lambda _{0}}{2(16\pi ^{2})}[(M_{c}^{2}-\mu
^{2})-m_{0}^{2}(\ln \frac{M_{c}^{2}}{\mu ^{2}}-\gamma _{\omega })]  \notag \\
&&+\frac{\lambda _{0}^{2}}{(16\pi ^{2})^{2}}[\frac{1}{4}(M_{c}^{2}-\mu
^{2})(\ln \frac{M_{c}^{2}}{\mu ^{2}}-\gamma _{\omega })-\frac{1}{6}%
(M_{c}^{2}-\mu ^{2})  \notag \\
&&-\frac{1}{4}m_{0}^{2}(\ln \frac{M_{c}^{2}}{\mu ^{2}}-\gamma _{\omega
})^{2}+\frac{1}{12}m_{0}^{2}(\ln \frac{M_{c}^{2}}{\mu ^{2}}-\gamma _{\omega
})],
\end{eqnarray}%
which is different from the result obtained by using the dimensional
regularization approach due to the appearance of the quadratic terms. The
anomalous mass dimension where we sum up all the leading quadratic and
logarithmic terms (i.e., not considering the logarithmic-squared term and
quadratic-logarithmic cross term) is given by:
\begin{eqnarray}
\gamma _{\phi ^{2}} &=&\frac{\mu ^{2}}{m^{2}}\frac{dm^{2}}{d\mu ^{2}}  \notag
\\
&=&-\frac{\lambda _{0}}{(16\pi ^{2})m^{2}}(\frac{1}{2}\mu ^{2}-\frac{1}{2}%
m_{0}^{2})+\frac{\lambda _{0}^{2}}{(16\pi ^{2})^{2}m^{2}}(\frac{1}{6}\mu
^{2}-\frac{1}{12}m_{0}^{2})  \notag \\
&\approx &-\frac{\lambda }{16\pi ^{2}}(\frac{1}{2}\frac{\mu ^{2}}{m^{2}}-%
\frac{1}{2})+\frac{\lambda ^{2}}{(16\pi ^{2})^{2}}(\frac{1}{6}\frac{\mu ^{2}%
}{m^{2}}-\frac{1}{12})  \notag \\
&=&\frac{1}{2}\frac{\lambda }{16\pi ^{2}}-\frac{1}{12}\left( \frac{\lambda }{%
16\pi ^{2}}\right) ^{2}-\frac{\mu ^{2}}{m^{2}}\left[ \frac{1}{2}\frac{%
\lambda }{16\pi ^{2}}-\frac{1}{6}\left( \frac{\lambda }{16\pi ^{2}}\right)
^{2}\right] ,
\end{eqnarray}%
where we have replaced in the third line the bare mass and coupling constant
with the renormalized ones.

Note that the resulting $\gamma _{\phi ^{2}}$ is different from that
obtained in ref.\cite{Kazakov} by using the dimensional regularization
approach with the $\bar{MS}$ subtraction scheme. The difference occurs in
both the power-law and the logarithmic running terms. For the power-law
running terms with the form $\mu ^{2}/m^{2}$ in $\gamma _{\phi ^{2}}$, it
reflects the fact that the LORE method maintains the original quadratic
divergence. For the logarithmic terms, the difference can be caused from the
well-known fact that the two-loop anomalous mass dimension in $\phi ^{4}$
theory is in general subtraction scheme dependent. This may be seen from the
rescaling $\mu ^{2}\rightarrow e^{\alpha _{0}}\mu ^{2}$, the resulting
leading logarithmic term at two loop level is changed by an additional
contribution from the logarithmic-squared term, thus the corresponding $%
\gamma _{\phi ^{2}}$ for the logarithmic running is changed to be
\begin{equation*}
\gamma _{\phi ^{2}}|_{log}=\frac{1}{2}\frac{\lambda }{16\pi ^{2}}-\frac{1}{12%
}(1+6\alpha _{0})\left( \frac{\lambda }{16\pi ^{2}}\right) ^{2}.
\end{equation*}%
As a consequence, both the $\mu ^{2}$-independent term and the quadratic $%
\mu ^{2}$-dependent terms also changed correspondingly. Similarly, when
shifting the scale $\mu ^{2}\rightarrow \hat{\mu}^{2}\equiv \mu ^{2}-\alpha
_{0}m^{2}$, the leading logarithmic term also receives an extra contribution
from the quadratic-logarithmic cross term, and the resulting $\gamma _{\phi
^{2}}$ for the logarithmic running in terms of the new subtraction energy
scale $\hat{\mu}^{2}$ is modified to be
\begin{equation*}
\gamma _{\phi ^{2}}|_{log}=\frac{1}{2}\frac{\lambda }{16\pi ^{2}}-\frac{1}{12%
}(1+3\alpha _{0})\left( \frac{\lambda }{16\pi ^{2}}\right) ^{2}.
\end{equation*}%
However, the quadratic-logarithmic cross term is now given in terms of two
energy scales $\mu ^{2}$ and $\hat{\mu}^{2}$ rather than a single one, i.e.,
$(M_{c}^{2}-\mu ^{2})(\ln M_{c}^{2}/\hat{\mu}^{2}-\gamma _{w})$. From the
above illustration, it is seen that either the rescaling or the shifting of
the subtracted energy scale $\mu ^{2}$ will change the initial correlative
form $(M_{c}^{2}-\mu ^{2})$ and $\ln M_{c}^{2}/\mu ^{2}$. Therefore, when
the quadratic terms are kept by using the LORE method, the arbitrariness
caused by the subtraction scheme for the scalar mass renormalization at high
loop order may be eliminated by requiring to maintain the correlative form $%
(M_{c}^{2}-\mu ^{2})$ and $\ln M_{c}^{2}/\mu ^{2}$ with a single subtracted
energy scale.

\section{ General Procedure of LORE Method}

With the explicit calculations of two-loop Feynman diagrams in the $\phi ^{4}
$ theory given above, it is useful to summariz the general procedure in
applying the LORE method to multi-loop calculations. It is expected that the
same procedure is applicable to higher-order calculations with similar
features when merging the LORE method with the Bjorken-Drell's analogy
between the Feynman diagrams and electrical circuits, though we have only
shown it in the two-loop calculations. The procedure may be wtated in the
following steps:

(i) Write down the corresponding Feynman integrals by using the Feynman
rules of the theory for any given Feynman diagrams.

(ii) Combine the denominators by using Feynman parameters to evaluate the
two-loop integrals into the sum of the $\alpha \beta \gamma $ integrals of
scalar-type and tensor-type. The use of the usual Feynman parametrization in
this step needs to be distinguished from the UVDP parametrization adopted
for the $\alpha \beta \gamma $ integrals. The latter may contain the UV
divergences, while the former is in general irrelevant to the UV divergences
but it can contain infrared (IR) divergences. From this point of view,
making distinction of Feynman parameters from UVDP parameters enables us to
separate IR divergences from UV divergences in two parameter spaces.

(iii) By applying the general formulae Eqs.(\ref{abc int_ed2}-\ref{abc
int_ed3}) for the ILIs of two-loop $\alpha\beta\gamma$ integrals to the
resulting $\alpha\beta\gamma$ integrals coming from a given Feynman
integrals, we can straightforwardly read off the final results for those
integrals. Alternatively, one may also adopt a practically useful procedure
by completing the squares of the factors in the denominator and evaluate the
$\alpha\beta\gamma$ integrals into the 2-fold ILIs as proposed in ref.\cite%
{wu1}, which shows that for each internal loop momentum, one can always
transform the integrals into the 1-fold ILIs with respect to it, and then
integrate out the ILIs by means of the LORE method. The two procedures are
actually equivalent. For tensor-type integrals, we need to apply the
consistency conditions Eq. (\ref{CC}) to transform them into the
corresponding scalar-type ones first.

(iv) When the integrals involve overlapping divergences, the above procedure
will transform the divergences appearing in the subdiagrams into the ones in
the UVDP parameter space. In order to identify those divergences, it is
helpful to use the advantage of the Bjorken-Drell's analogy between the
Feynman diagrams and electric circuits. To extract the UV divergence
behavior, it is useful to explore the possible divergence regions in the
UVDP parameter space.  Then apply the prescription described in Eq.(\ref%
{treatment}) through introducing a mass scale $q_{o}^{2}$ to transform the
integrals into the momentum-like ones, so that we can directly apply the
LORE method. The scale $q_{o}^{2}$ is in general taken to be the
renomalization scale or some intrinsic scales in the original Feynman
integrals, such as the masses of particles and/or the external momenta. The
explicit form of $q_{o}^{2}$ should be fixed by certain criteria, such as
the typical scale in the divergent regions of parameter space, so that the
harmful divergences cancel exactly.

It is interesting to notice that in the LORE method only the overall
divergence of the overlapping Feynman diagrams is expressed in the momentum
integration and the resulting functions $y_0(x)$ and $y_2(x)$ can depend on
the mass factor $\mathcal{M}^2$ through which a dependence on kinematic
invariants comes in. In contrast, all other divergences arising from the
subdiagrams are actually given in terms of the UVDP parameters and the
resulting function $y_0$ in the logarithmic divergence will be independent
of any kinematic invariants. This can be seen from the general scalar-type
integral Eq.(\ref{genI4}) which can arise from n-loop Feynman diagrams. The
more detailed evaluations are carried out for the so-called $%
\alpha\beta\gamma$ integral of two loop diagrams, which can explicitly be
seen in Eqs.(\ref{abc int_ed3}-\ref{I111_1}). As a consequence, the
regularized divergent quantity of subdiagrams involves only a
kinematic-independent scale $\mu$ via a polynomial of $\mu^2/M_c^2$ with $%
\mu^2/M_c^2\to 0$ at $M_c\to \infty$, thus the function $y_0(x)$ arising
from the subdiagrams is no longer a complicated function of kinematic
invariants.


\section{ Conclusions and Remarks}

We have explicitly shown how the loop regularization (LORE) method can be
consistently applied to two loop calculations of Feynman diagrams,
appropriately treating the overlapping divergences. The key concept of the
LORE method\cite{wu1,Wu:2003dd} is the introduction of the irreducible loop
integrals(ILIs), which are generally evaluated from the Feynman diagrams by
using the Feynman parametrization and the
ultraviolet-divergence-preserving(UVDP) parametrization. We have
demonstrated in this paper how the evaluation of ILIs and UVDP
parametrization naturally merges with the Bjorken-Drell's analogy between
Feynman diagrams and electric circuits. In particular, the UVDP parameters
can be regarded as the conductance or resistance in the electric circuit
analogy, and the sets of conditions required for evaluating the ILIs and the
momentum conservations have been found to associate with the conservations
of electric voltages in each loop and the conservations of electric currents
at each vertex respectively. As a consequence, the divergences in Feynman
diagrams correspond to infinite conductances or zero resistances in electric
circuits, and the LORE method merging with the Bjorken-Drell's analogy has
the advantage in analyzing the complicated overlapping divergence structure
of Feynman diagrams. Therefore, the Bjorken-Drell's circuit analogy allows
us to clarify the origin of UV divergences in the UVDP parameter space and
identify the correspondence of the divergences between subdiagrams and UVDP
parameters. From the explicit calculations of the case with $\alpha =\beta
=\gamma =1$ in the general $\alpha \beta \gamma $ integral, the divergences
arising from the subintegrals manifest themselves in the integration over
the corresponding asymptotic regions of the UVDP parameter space. The
calculations of the corresponding counterterm diagrams confirm our intuitive
picture that all the harmful divergences cancel exactly in the final result.
Although the procedures and calculations in the LORE method are not as
concise as the ones in the dimensional regularization, the overlapping
divergent structure and behavior as well as its treatment become more
physically clear in the LORE method.

As an interesting application, we have taken the massive scalar $\phi^4$
theory as an example and performed the detailed calculation of two loop
contributions by applying the general formalism of the LORE method. By
explicitly computing the two- and four-point functions at two-loop level and
carefully using the advantage of Bjorken-Drell's circuit analogy, all the
harmful divergences cancel exactly and the resulting two loop corrections
agree with the standard results for the logarithmic corrections. The
power-law running of mass is explicitly given at two loop level.

In this paper, we have only carried out two-loop calculations and explicitly
demonstrated the consistency of the LORE method at two-loop level, but it
can be shown that the general procedure of the LORE method shown in Eqs.(\ref%
{genI2}-\ref{genI4}) is applicable to even higher-loop calculations by
taking advantage of Bjorken-Drell's circuit analogy. Furthermore, we only
considered the scalar-type two-loop integrals. However, as shown in \cite%
{Wu:2003dd}, in order to ensure the gauge invariance, it is necessary to
keep the consistency conditions Eq.(\ref{CC}) which correctly transform the
tensor-type ILIs into the scalar-type ones. We shall demonstrate how these
consistency conditions in two-loop or even higher-loop order by an explicit
calculation\cite{QED2loop}, although it has already been demonstrated in a
general way in\cite{Wu:2003dd}. We would like to point out that the
advantage of merging the LORE method with Bjorken-Drell's circuit analogy
enables us to figure out a more general and rigorous proof for the validity
of the LORE method to all orders in the perturbation theory\cite{progress}.

\vspace{1 cm}

\centerline{{\bf Acknowledgement}}

\vspace{20 pt}

The authors would like to thank J.W. Cui and Y.B. Yang for useful
discussions and L.F. Li for helpfully reading the manuscript. This work was supported in part by the National Science
Foundation of China (NSFC) under Grant \#No. 10821504, 10975170 and the key
project of the Chinese Academy of Science. 
\appendix

\section{ Useful Formulae in UV Divergence Preserving (UVDP) Parametrization}

The introduction of UVDP parameters is to combine the various denominators
propagating factors, whose utility is similar to Feynman parameters. The
motivation to introduce a new UVDP parametrization method is to transform a
divergent integral in the UVDP parameter space into a ILI-like divergent
one, the object regularized by the LORE method. The simplest case is to
combine only two factors in the denominator by using the identity:
\begin{equation*}
\frac{1}{AB}=\int_{0}^{\infty }\frac{du}{(1+u)^{2}}\frac{dv}{(1+v)^{2}}%
\delta (1-\frac{1}{1+u}-\frac{1}{1+v})\frac{1}{[\frac{A}{1+u}+\frac{B}{1+v}%
]^{2}}.
\end{equation*}%
If one of the factors have more than one power, we can differentiate with
respect to $A$ or $B$ to get,
\begin{equation*}
\frac{1}{AB^{n}}=\int_{0}^{\infty }\frac{du}{(1+u)^{2}}\frac{dv}{(1+v)^{2}}%
\delta (1-\frac{1}{1+u}-\frac{1}{1+v})\frac{\frac{n}{(1+v)^{n-1}}}{[\frac{A}{%
1+u}+\frac{B}{1+v}]^{n+1}}.
\end{equation*}%
More general identity for more than two factors is:
\begin{equation*}
\frac{1}{A_{1}A_{2}\cdots A_{n}}=\int_{0}^{\infty }\prod_{i=1}^{n}\frac{%
dv_{i}}{(1+v_{i})^{2}}\delta (\sum_{i=1}^{n}\frac{1}{1+v_{i}}-1)\frac{(n-1)!%
}{[\sum_{i=1}^{n}\frac{A_{i}}{1+v_{i}}]^{n}}.
\end{equation*}%
Even more general form can be derived:
\begin{equation}
\frac{1}{A_{1}^{m_{1}}A_{2}^{m_{2}}\cdots A_{n}^{m_{n}}}=\int_{0}^{\infty
}\prod_{i=1}^{n}\frac{dv_{i}}{(1+v_{i})^{2}}\delta (\sum_{i=1}^{n}\frac{1}{%
1+v_{i}}-1)\frac{\prod_{i=1}^{n}\frac{1}{(1+v_{i})^{m_{i}-1}}}{%
[\sum_{i=1}^{n}\frac{A_{i}}{1+v_{i}}]^{\sum_{i=1}^{n}m_{i}}}.
\label{UVDP gen}
\end{equation}

Alternatively, we may also take another more useful form for the case of two
factors by just integrating out one of the parameters $u$ and $v$ by using
the delta function, which has been adopted in \cite{wu1}:
\begin{equation}
\frac{1}{AB^{n}}=\int_{0}^{\infty }du\frac{nu^{n-1}}{[A+uB]^{n+1}},
\label{simp UVDP}
\end{equation}%
but this form cannot be generalized to the more general case easily. \newline

>From the general identity Eq.(\ref{UVDP gen}), we notice that the relation
of the UVDP parameters $v_i$ to Feynman parameters $x_i$ is:
\begin{equation}
x_i=\frac{1}{1+v_i}.
\end{equation}
This identification allows us to transform a divergent integral with Feynman
parameters into the one with UVDP parameters, which can be further
transformed into a ILI-like integral by introducing a free mass scale and
being regularized in the framework of the LORE method. Such a trick is
discussed in Eq.(\ref{treatment}). 

\end{document}